\documentclass[letterpaper,11pt]{JHEP3}

\usepackage{graphicx}
\usepackage{amsmath}
\usepackage{amssymb}
\usepackage{xspace}
\usepackage[small]{subfigure}

\newcommand{\eq}[1]{Eq.~\eqref{eq:#1}}
\newcommand{\eqs}[2]{Eqs.~\eqref{eq:#1} and \eqref{eq:#2}}
\renewcommand{\sec}[1]{Sec.~\ref{sec:#1}}

\newcommand{\subsec}[1]{Sec.~\ref{subsec:#1}}
\renewcommand{\app}[1]{App.~\ref{app:#1}}
\newcommand{\fig}[1]{Fig.~\ref{fig:#1}}
\newcommand{\figs}[2]{Figs.~\ref{fig:#1} and \ref{fig:#2}}

\newcommand{\abs}[1]{\lvert#1\rvert}
\newcommand{\Abs}[1]{\bigl\lvert#1\bigr\rvert}
\newcommand{\ord}[1]{O(#1)}
\newcommand{\ORd}[1]{O\Bigl(#1\Bigr)}

\newcommand{\Vev}[1]{\bigl\langle #1 \bigr\rangle}
\newcommand{\mae}[3]{\langle#1\rvert#2\rvert#3\rangle}
\newcommand{\Mae}[3]{\bigl\langle#1\bigr\rvert#2\bigr\rvert#3\bigr\rangle}
\newcommand{\MAe}[3]{\Bigl\langle#1\Bigr\rvert#2\Bigr\rvert#3\Bigr\rangle}
\newcommand{\bra}[1]{\langle#1\rvert}
\newcommand{\ket}[1]{\lvert#1\rangle}

\newcommand{\df}{\mathrm{d}}

\newcommand{\img}{\mathrm{i}}

\renewcommand{\Im}{\mathrm{Im}}

\newcommand{\sdt}{\!\cdot\!}
\newcommand{\tr}{\textrm{tr}}

\newcommand{\bt}{\beta}
\newcommand{\ga}{\gamma}
\newcommand{\de}{\delta}
\newcommand{\eps}{\epsilon}

\newcommand{\la}{\lambda}
\newcommand{\w}{\omega}

\newcommand{\Ga}{\Gamma}

\newcommand{\cB}{{\mathcal B}}

\newcommand{\cI}{{\mathcal I}}
\newcommand{\cL}{{\mathcal L}}
\newcommand{\cP}{{\mathcal P}}

\newcommand{\tB}{\widetilde{B}}
\newcommand{\tS}{\widetilde{S}}
\newcommand{\tga}{ {\tilde{\gamma}} }

\newcommand{\hp}{\hat{p}}
\newcommand{\hJ}{\widehat{J}}

\newcommand{\tZ}{\widetilde{Z}}

\newcommand{\bn}{\bar{n}}
\newcommand{\bnP}{\overline {\mathcal P}}

\newcommand{\nslash}{n\!\!\!\slash}
\newcommand{\bnslash}{\bar{n}\!\!\!\slash}
\newcommand{\pslash}{p\!\!\!\slash}

\newcommand{\ellslash}{\ell\!\!\!\slash}

\newcommand{\TeV}{\,\mathrm{TeV}}

\newcommand{\nn}{\nonumber}

\newcommand{\lqcd}{\Lambda_\mathrm{QCD}}
\newcommand{\alem}{\alpha_\mathrm{em}}

\newcommand{\lp}{ {\tilde{p}} }         

\newcommand{\hemiin}{\mathrm{ihemi}}

\newcommand{\Ecm}{E_\mathrm{cm}}

\newcommand{\Disc}{\mathrm{Disc}}

\newcommand{\cut}{\mathrm{cut}}
\renewcommand{\max}{\mathrm{max}}

\newcommand{\cusp}{\mathrm{cusp}}
\newcommand{\bare}{\mathrm{bare}}

\newcommand{\zero}{{(0)}}
\newcommand{\one}{{(1)}}

\newcommand{\SCETa}{\ensuremath{{\rm SCET}_{\rm I}}\xspace}
\newcommand{\SCETb}{\ensuremath{{\rm SCET}_{\rm II}}\xspace}

\newcommand{\op}{{\mathcal{O}}}
\newcommand{\tiop}{\widetilde{\mathcal{O}}}
\newcommand{\oq}{{\mathcal{Q}}}

\allowdisplaybreaks[2]

\preprint{arXiv:1002.2213\\MIT--CTP 4097\\February 10, 2010}

\title{The Quark Beam Function at NNLL}

\author{Iain W.~Stewart, Frank J.~Tackmann and Wouter J.~Waalewijn\\
Center for Theoretical Physics, Massachusetts Institute of Technology,\\
Cambridge, MA~02139, U.S.A.}

\abstract{
In hard collisions at a hadron collider the most appropriate
description of the initial state depends on what is measured in the final
state. Parton distribution functions (PDFs) evolved to the hard collision
scale $Q$ are appropriate for inclusive observables, but not for measurements
with a specific number of hard jets, leptons, and photons.  Here the incoming
protons are probed and lose their identity to an incoming jet at a scale
$\mu_B\ll Q$, and the initial state is described by universal beam functions.
We discuss the field-theoretic treatment of beam functions, and show that the
beam function has the same RG evolution as the jet function to all orders in
perturbation theory. In contrast to PDF evolution, the beam function evolution
does not mix quarks and gluons and changes the virtuality of the colliding
parton at fixed momentum fraction.  At $\mu_B$, the incoming jet can be
described perturbatively, and we give a detailed derivation of the one-loop
matching of the quark beam function onto quark and gluon PDFs. We compute the
associated NLO Wilson coefficients and explicitly verify the cancellation of
IR singularities. As an application, we give an expression for the
next-to-next-to-leading logarithmic order (NNLL) resummed Drell-Yan beam
thrust cross section.
}


\begin{document}

\newpage
\section{Introduction}
\label{sec:Intro}

The primary goal of the experiments at the LHC and Tevatron is to search for the
Higgs particle and physics beyond the Standard Model through collisions at the
energy frontier. The fact that the short-distance processes of interest are
interlaced with QCD interactions complicates the search.  A schematic picture of
a proton-proton collision is shown in \fig{LHC}. A quark or gluon is extracted
from each proton (the red circles labeled $f$), and emits initial-state
radiation ($\cI$) prior to the hard short-distance collision (at $H$). The
hard collision produces strongly interacting partons which hadronize into
collimated jets of hadrons ($J_{1,2,3}$), as well as non-strongly
interacting particles (represented in the figure by the $\ell^+\ell^-$).
Finally, all the strongly interacting particles, including the
spectators in the proton, interact with soft low-momentum gluons and
can exchange perpendicular momentum by virtual Glauber gluons (both indicated
by the short orange lines labeled $S$).

\FIGURE[b]{%
\includegraphics[width=0.6\textwidth]{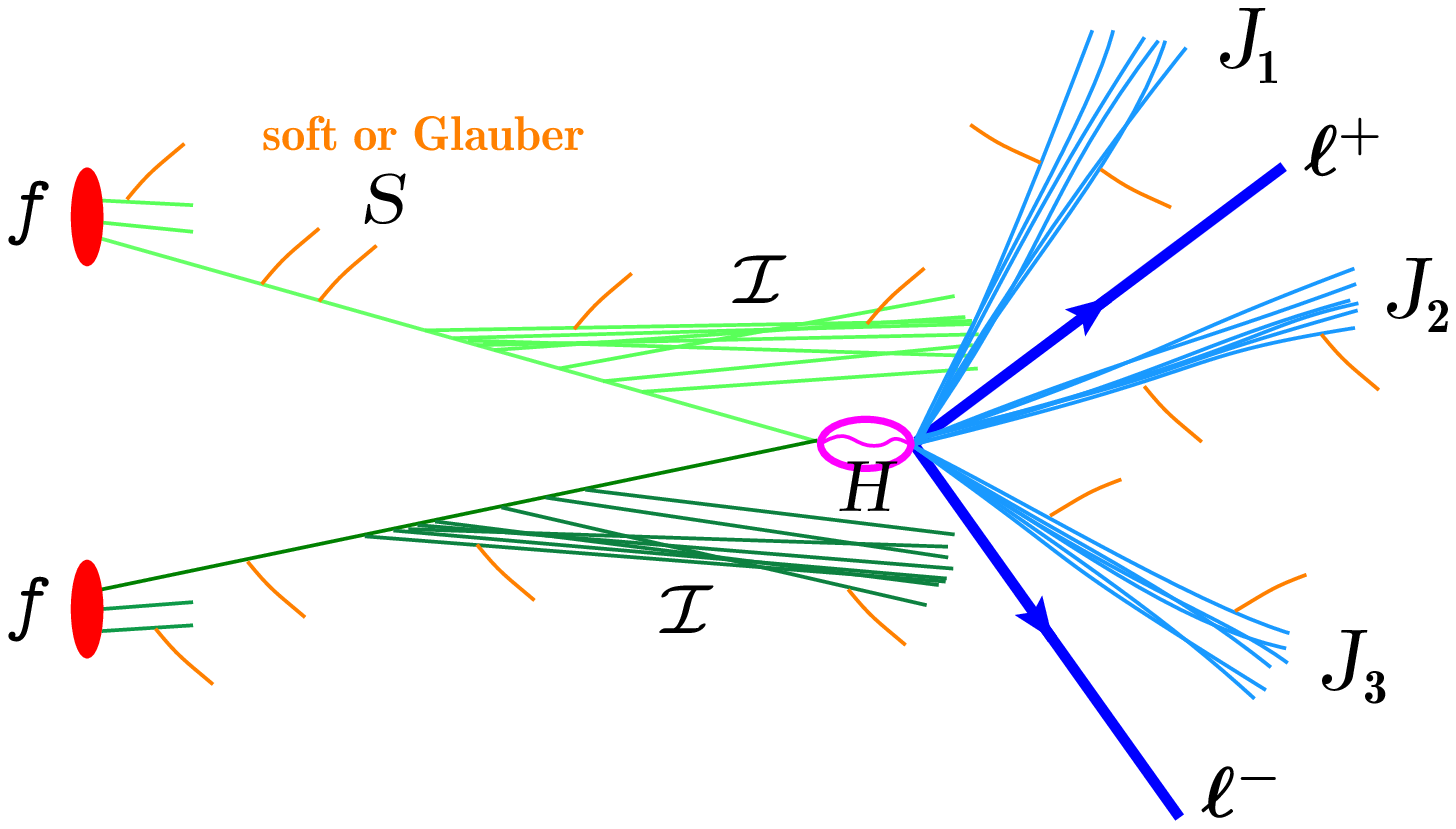}%
\caption{Schematic picture of a proton-(anti)proton collision at the LHC or Tevatron.}
\label{fig:LHC}}

The theoretical description of the collision is dramatically simplified for
inclusive measurements, such as $pp\to X\ell^+\ell^-$, where one does not restrict
the hadronic final state $X$. In this case, the cross section can be factorized
as $\df\sigma = H_\mathrm{incl} \otimes f\otimes f$, where each $f$ denotes a parton
distribution function (PDF), which gives the probability of extracting a parton
from the proton, while all other components of the collision are gathered together
in a perturbatively calculable function $H_\mathrm{incl}$. However, inclusive
measurements do not necessarily contain all the desired information.
Experimentally, identifying a certain hard-interaction process
requires distinguishing between events that have a specific number of hard jets,
leptons, or photons separated from each other and from the beam directions.
Such measurements introduce new low energy scales and perturbative
series with large double logarithms. For these situations it is necessary in
the theoretical description to distinguish more of the ingredients in \fig{LHC}, such as
$\cI$, $J_i$, and $S$.  Monte Carlo programs provide a widely used method to
model the ingredients in the full cross section, $\df\sigma = H\otimes f\otimes f \otimes \cI\otimes
\cI \otimes \prod_i J_i \otimes S$, using notions from QCD factorization and
properties of QCD in the soft and collinear limits. Monte Carlos have the virtue of
providing a general tool for any observable, but have the disadvantage of making
model-dependent assumptions to combine the ingredients and to calculate some of them.
For specific observables a better approach is to use factorization
theorems (when they are available), since this provides a rigorous method of
defining and combining the various ingredients.

Here we investigate so-called beam functions, $B= \cI \otimes f$, which describe
the part of \fig{LHC} associated with the initial state. They incorporate
PDF effects as well as initial-state radiation via functions $\cI$ that
can be computed in perturbation theory~\cite{Stewart:2009yx}.  Below we will
describe a particular class of measurements, which correspond to
$pp \to L + 0$ jets with $L$ a non-hadronic final state such as $Z\to \ell^+\ell^-$
or $h\to \gamma\gamma$. For these measurements, a rigorous factorization
theorem has been proven that involves beam functions ~\cite{Stewart:2009yx}.
We start by describing the general physical picture associated with beam functions, which suggests that
they will have a wider role in describing cross sections for events with any number of
distinguished jets, e.g. $pp\to W/Z + n$ jets. That is, the beam functions
have a more universal nature than what has been proven explicitly so far for the $0$-jet
case.

\begin{figure}
\subfigure[]{\includegraphics[width=0.5\textwidth]{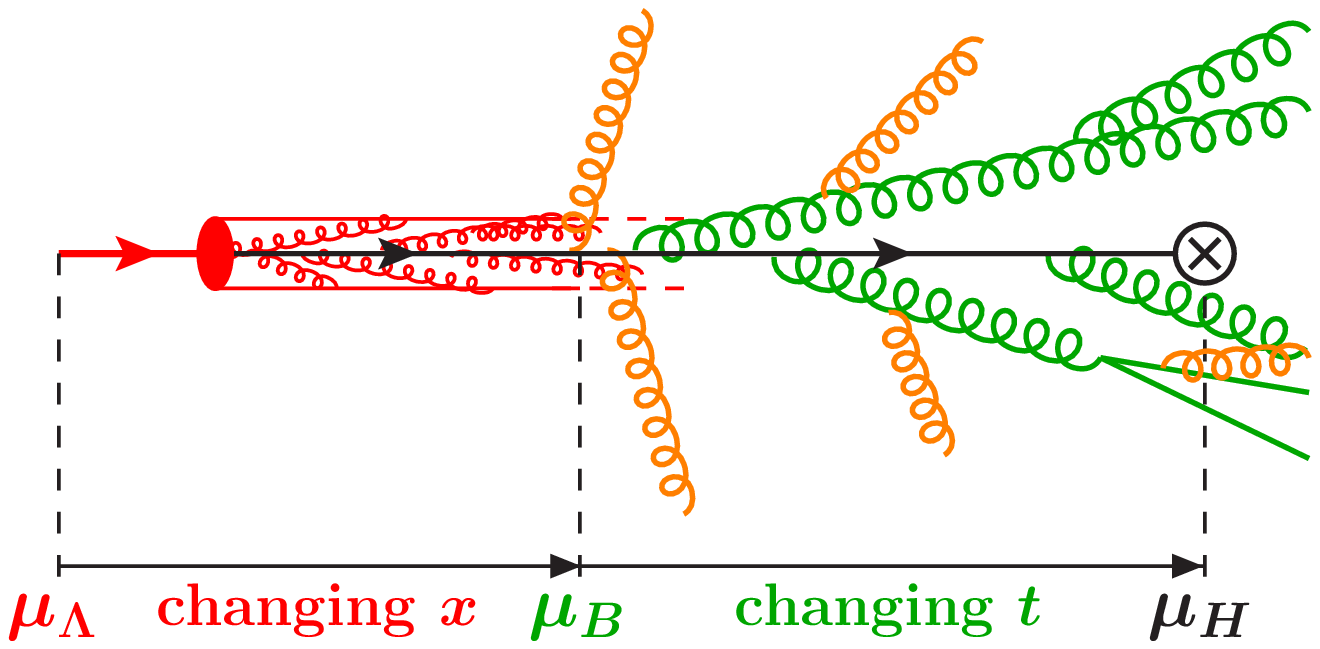}\label{fig:beam}}%
\hfill%
\subfigure[]{\includegraphics[width=0.45\textwidth]{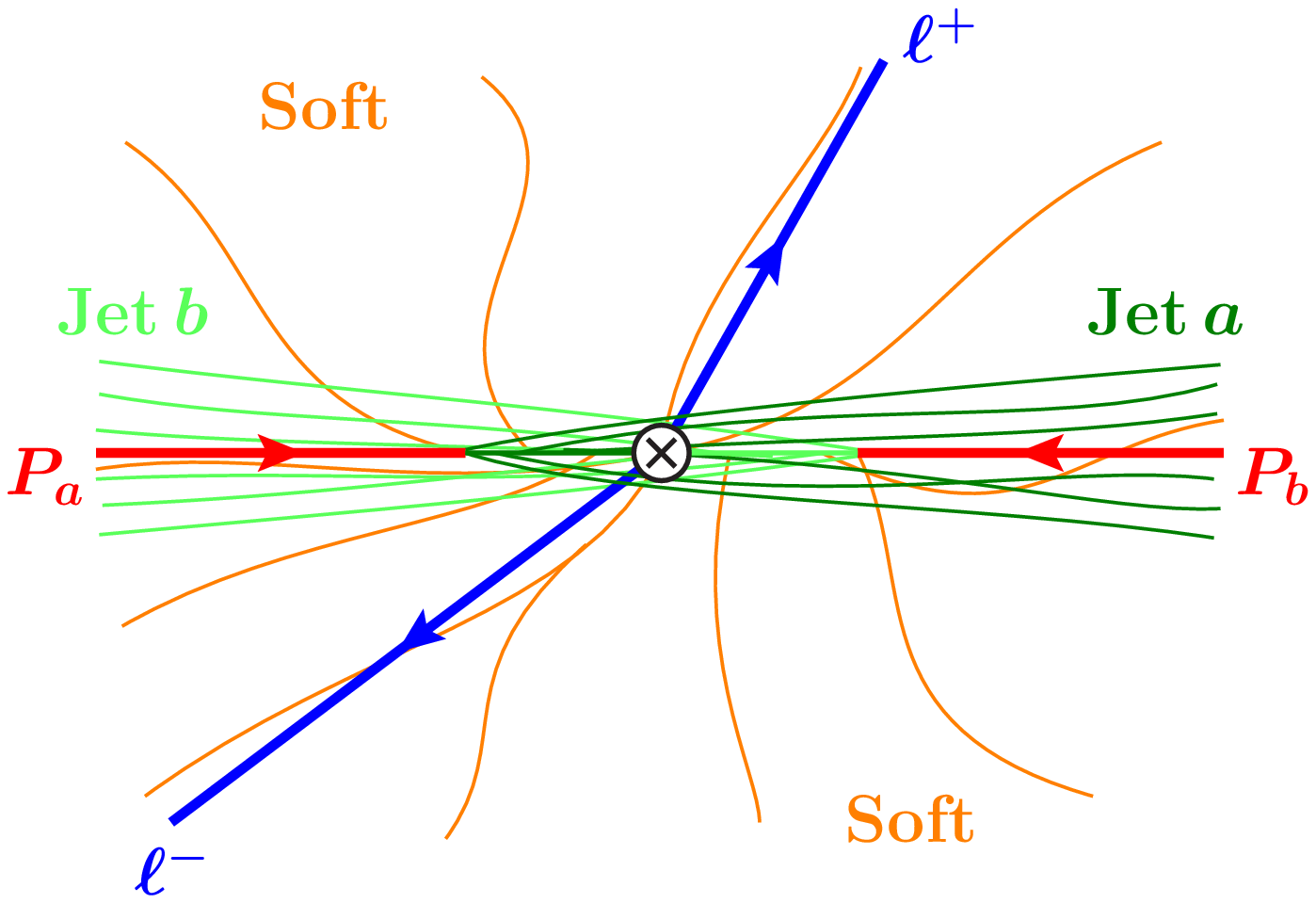}\label{fig:drellyan}}%
\caption{(a) Physics described by the beam function. Starting at a low hadronic
  scale $\mu_\Lambda$ the proton is described by a PDF $f$. At the scale
  $\mu_B$, the proton is probed by measuring radiation in the final state,
  identifying a parton $j$ described by $f_j(\xi,\mu_B)$. Above $\mu_B$, the
  initial state becomes an incoming jet described by $\cI_{ij}(t,x/\xi,\mu)$
  for an off-shell parton $i$ with spacelike virtuality $-t$, which enters the
  hard interaction at $\mu_H$.  (b) Schematic picture of
  the final state for isolated Drell-Yan.}
\end{figure}

The initial-state physics
described by beam functions is illustrated in \fig{beam}, and is characterized by
three distinct scales $\mu_\Lambda \ll \mu_B\ll \mu_H$.  At a low hadronic scale
$\mu_\Lambda$ the incoming proton contains partons of type $k$ whose distribution of
momentum is described by PDFs, $f_k(\xi',\mu_\Lambda)$. Here $\xi'$ is the
momentum fraction relative to the (massless) proton momentum.
Evolving $\mu$ to higher scales sums up single logarithms with the standard DGLAP
evolution~\cite{Gribov:1972ri, Georgi:1951sr, Gross:1974cs, Altarelli:1977zs, Dokshitzer:1977sg},
\begin{equation}
\mu \frac{\df}{\df\mu} f_{j}(\xi,\mu)
= \sum_k \int\!\frac{\df\xi'}{\xi'}\, \gamma^f_{jk}\Bigl(\frac{\xi}{\xi'}, \mu\Bigr) f_k(\xi',\mu)
\,.\end{equation}
This changes the type $k$ and momentum fraction $\xi'$ of the partons, but
constrains them to remain inside the proton.  At a scale $\mu_B$, the
measurement of radiation in the final state probes the proton, breaking it apart
as shown in \fig{beam} and identifying a parton $j$ with momentum fraction $\xi$
according to $f_j(\xi,\mu_B)$.  Measurements which have this effect at $\mu_B
\ll \mu_H$ are those that directly or indirectly constrain energetic radiation
in the forward direction, for example, by distinguishing hadrons in a central
jet from those in the forward directions.  The radiation emitted by the parton
$j$ builds up an incoming jet described by the function $\cI_{ij}$, and together
these two ingredients form the beam function,
\begin{equation} \label{eq:B_fact}
B_i(t', x, \mu_B)
= \sum_j\!\int_x^1 \frac{\df\xi}{\xi}\,
 \cI_{ij}\Bigl(t',\frac{x}{\xi},\mu_B \Bigr) f_j(\xi, \mu_B)
\,.\end{equation}
The sum indicates that the parton $i$ in the jet need not be the same as the
parton $j$ in the PDF. The emissions also change the momentum fraction from
$\xi$ to $x$ and push the parton $i$ off-shell with spacelike (transverse)
virtuality $-t' <0$.
The evolution for $\mu> \mu_B$ sums up the double-logarithmic
series associated with the $t$-channel singularity as $t'\to 0$. It
changes the virtuality $t'$ of the parton $i$, while leaving its identity and
momentum fraction unchanged,
\begin{equation} \label{eq:B_RGE_intro}
  \mu \frac{\df}{\df\mu} B_i(t,x,\mu) = \int\!\df t'\, \gamma^i_B(t-t',\mu)\, B_i(t',x,\mu)
\,.\end{equation}
This evolution stops at the hard scale $\mu_H$, where the off-shell parton $i$ enters
the hard partonic collision. For $\mu \geq \mu_B$ the initial state is
also sensitive to soft radiation as shown by the orange wider angle gluons in
\fig{beam}. For cases where the beam function description suffices this soft
radiation eikonalizes, and the corresponding soft Wilson line is one component
of the soft function $S$ that appears in the factorized cross section.

In general, a beam function combines the PDF with a description of
all energetic initial-state
radiation that is collinear to the incoming proton direction up to
$t \ll Q^2$. The parton's virtuality $t$ effectively measures the
transverse spread of the radiation around the beam axis.
The specific type of beam function may depend on
details of the measurements, much as how jet functions depend on the
algorithm used to identify radiation in the jet~\cite{Ellis:2009wj, Jouttenus:2009ns, Ellis:2010rwa}.
Our discussion here will focus
on the most inclusive beam function, which probes $t$ through the measurement of
hadrons in the entire forward hemisphere corresponding to the proton's
direction. The utility of beam functions is that for a class of cross sections they provide a universal
description of initial-state radiation that does not need to
be modeled or computed on a case by case basis.

An example of a factorization theorem that involves beam functions is the
``isolated Drell-Yan'' process, $pp\to X\ell^+\ell^-$. Here, as depicted in
\fig{drellyan}, $X$ is allowed to contain forward energetic radiation in jets
about the beam axis, but only soft wide-angle radiation with no central jets.
The presence of energetic forward
radiation is an unavoidable consequence for processes involving generic parton
momentum fractions $x$ that are away from the threshold limit $x\to 1$.
There are of course many ways one might enforce the events to have no central jets.
In Ref.~\cite{Stewart:2009yx}, a smooth central jet veto
is implemented by constructing a simple inclusive observable, called ``beam
thrust'', defined as
\begin{align} \label{eq:tauB}
\tau_B
= \frac{e^Y B_a^+(Y) + e^{-Y} B_b^+(Y)}{Q}
= \frac{x_a \Ecm B_a^+(Y) + x_b \Ecm B_b^+(Y)}{q^2}
\,.\end{align}
Here, $q^2$ and $Y$ are the total invariant mass and rapidity of the leptons, $Q = \sqrt{q^2}$, and
\begin{equation}
x_a = \frac{Q}{\Ecm}\,e^{Y}
\,,\qquad
x_b = \frac{Q}{\Ecm}\,e^{-Y}
\,\end{equation}
correspond to the partonic momentum fractions transferred to the leptons. The
hadronic momenta $B_a^\mu(Y)$ and $B_b^\mu(Y)$ measure the total momentum of all
hadrons in the final state at rapidities $y > Y$ and $y < Y$, respectively
(where the momenta are measured in the hadronic center-of-mass frame of the
collision and the rapidities are with respect to the beam axis). Their plus
components are defined as $B_a^+(Y) = n_a\cdot B_a(Y)$ and $B_b^+(Y) = n_b\cdot
B_b(Y)$ where $n_a^\mu = (1, 0, 0, 1)$ and $n_b^\mu = (1, 0, 0, -1)$ are
light-cone vectors corresponding to the directions of the incoming protons
(with the beam axis taken along the $z$ direction).  The interpretation of beam thrust
is analogous to thrust for $e^+e^-$ to jets, but with the thrust axis fixed to
be the beam axis. For $\tau_B \simeq 1$ the hadronic final state contains hard
radiation with momentum perpendicular to the beam axis of order $Q$, while
$\tau_B \ll 1$ corresponds to two-jet like events with hard radiation of order $Q$
only near the direction of the beams. The dependence of $B_{a,b}^+(Y)$ on $Y$
accounts for asymmetric collisions where the partonic center-of-mass frame is
boosted with respect to the hadronic center-of-mass frame. Requiring $\tau_B < \exp(-2y_B^\cut)$
essentially vetoes hard radiation in a rapidity interval of size $y_B^\cut -
1$ around $Y$, i.e. in the region $\abs{y-Y} < y_B^\cut - 1$, while radiation in
the larger interval $y_B^\cut + 1$ is essentially unconstrained, with a smooth transition in
between. Thus, interesting values for $y_B^\cut$ are around $1$ to $2$.

In Ref.~\cite{Stewart:2009yx}, a rigorous factorization theorem for the Drell-Yan
beam thrust cross section for small $\tau_B$ was derived,
\begin{align} \label{eq:DYbeam}
\frac{\df\sigma}{\df q^2\, \df Y\, \df \tau_B}
& = \sigma_0 \sum_{ij} H_{ij}(q^2, \mu)
\int\!\df t_a\,\df t_b\, B_i(t_a, x_a, \mu)\, B_j(t_b, x_b, \mu)
\nn\\ &\quad \times
Q S_B\Bigl(Q\,\tau_B - \frac{t_a + t_b}{Q}, \mu \Bigr)
 \biggl[1 + O\Bigl(\frac{\lqcd}{Q}, \tau_B \Bigr) \biggr]
\,,\end{align}
using the formalism of the soft-collinear effective theory
(SCET)~\cite{Bauer:2000ew,Bauer:2000yr, Bauer:2001ct, Bauer:2001yt},
supplemented with arguments to rule out the presence of Glauber gluons
partially based on Refs.~\cite{Collins:1988ig, Aybat:2008ct}.
The sum runs over partons $ij = \{u\bar u, \bar u u, d\bar d,
\ldots\}$. The hard function $H_{ij}(q^2, \mu)$ contains virtual radiation at
the hard scale $\mu_H \simeq Q$. It is given by the square of Wilson
coefficients from matching the relevant QCD currents onto SCET currents (and
hence it is identical to the hard function appearing in the threshold Drell-Yan
factorization theorem). The beam functions $B_i(t_a, x_a, \mu)$ and $B_j(t_b, x_b, \mu)$ describe
the formation of incoming jets prior to the hard collision due to collinear radiation
from the incoming partons, as described above. They are the initial-state analogs of the final-state
jet functions $J_i(t, \mu)$ (appearing for example in the analogous factorization theorem
for thrust in $e^+e^-\to 2$ jets), which describe the formation of a jet from
the outgoing partons produced in the hard interaction.
In contrast to the jet functions, the beam functions depend on the
parton's momentum fraction $x$ in addition to its virtuality.
For \eq{DYbeam}, the beam scale is set by $\mu_B \simeq \sqrt{\tau_B} Q$.
Finally, the soft function $S_B(k^+, \mu)$ describes the effect of
soft radiation from the incoming partons on the measurement of $\tau_B$, much
like the soft function for thrust encodes the effects of soft radiation from the
outgoing partons. It is defined in terms of incoming Wilson lines (instead of
outgoing ones) and is sensitive to the soft scale $\mu_S \simeq \tau_B Q$.

The cross section for $\tau_B$ contains double logarithms $\ln^2\tau_B$ which
become large for small $\tau_B \ll 1$. The factorization theorem in \eq{DYbeam}
allows us to systematically resum these to all orders in perturbation theory.
The logarithms of $\tau_B$ are split up
into logarithms of the three scale ratios $\mu/\mu_H$, $\mu/\mu_B$, $\mu/\mu_S$
that are resummed by evaluating all functions at their natural scale, i.e.\ $H_{ij}$
at $\mu_H$, $B_i$ and $B_j$ at $\mu_B$, $S_B$ at $\mu_S$, and then RG
evolving them to the (arbitrary) common scale $\mu$.

In this paper we give a detailed discussion of the beam function,
including a derivation of results that were quoted in
Ref.~\cite{Stewart:2009yx}. We start in \sec{general} with a discussion of
several formal aspects of the quark and gluon beam functions, including their
definition in terms of matrix elements of operators in SCET, their all-order
renormalization properties, their analytic structure, and the operator product
expansion in \eq{B_fact} relating the beam functions to PDFs.
In particular, we prove that the beam functions obey the RGE in \eq{B_RGE_intro}
with the same anomalous dimension as the jet function to all orders in perturbation theory.
(A part of the proof is relegated to \app{B_RGE}.) This result also implies that
the anomalous dimension of the hemisphere soft functions with incoming Wilson
lines is identical to the anomalous dimension of the hemisphere soft function
with outgoing Wilson lines appearing in $e^+e^-\to 2$ jets.

In \sec{oneloop}, we perform the one-loop matching of the quark beam function
onto quark and gluon PDFs. Using an offshellness regulator we first give
explicit details of the calculations for the quark beam function and PDFs. We
verify explicitly that the beam function contains the same IR singularities as
the PDFs at one loop, and extract results for $\cI_{qq}$ and $\cI_{qg}$ at
next-to-leading order (NLO).  In \app{dimreg} we repeat the matching
calculation for the quark beam function in pure dimensional regularization.

Our results show that beam functions must be defined with zero-bin subtractions~\cite{Manohar:2006nz},
but that in the OPE the subtractions are frozen out into the Wilson coefficients $\cI_{ij}$.
The subtractions are in fact necessary for the IR singularities
in the beam functions and PDFs to agree. We briefly discuss why PDFs formulated with SCET
collinear fields are identical with or without zero-bin subtractions.

In \sec{results}, we first give the full expression for the resummed beam thrust cross section
at small $\tau_B$ valid to any order in perturbation theory.
The necessary ingredients for its evaluation at next-to-next-to-leading logarithmic
(NNLL) order are collected in \app{pert}, which are
the three-loop QCD cusp anomalous dimension, the two-loop standard anomalous
dimensions, and the one-loop matching corrections for the various Wilson coefficients.
(We also comment on the still missing ingredients required at N$^3$LL order.)
We then show plots of the quark beam function both at NLO in
fixed-order perturbation theory and NNLL order in resummed perturbation theory.
We also discuss the relative size of the quark and gluon contributions as well
as the singular and nonsingular terms in the threshold limit $x\to 1$.
We conclude in \sec{conclusions}.

\section{Beam Functions}
\label{sec:general}

\subsection{Definition}
\label{subsec:definition}

In this subsection we discuss the definition of the quark and gluon beam functions in terms of matrix elements of operators in SCET, and compare them to the corresponding definition of the PDF. The operator language will be convenient to elucidate the renormalization structure and relation to jet functions in the following subsection.

We first discuss some SCET ingredients that are relevant later on. We introduce light-cone vectors $n^\mu$ and $\bn^\mu$ with $n^2 = \bn^2 = 0$ and $n\cdot\bn = 2$ that are used to decompose four-vectors into light-cone coordinates $p^\mu = (p^+, p^-, p^\mu_\perp)$, where $p^+ = n\cdot p$, $p^- = \bn\cdot p$ and $p_\perp^\mu$ contains the components perpendicular to $n^\mu$ and $\bn^\mu$.

In SCET, the momentum $p^\mu$ of energetic collinear particles moving close to the $n$ direction is separated into large and small parts
\begin{equation}
p^\mu = \lp^\mu + p_r^\mu = \bn\cdot\lp\, \frac{n^\mu}{2} + \lp_{n\perp}^\mu + p_r^\mu
\,.\end{equation}
The large part $\lp^\mu = (0, \lp^-, \lp_\perp)$ has components $\lp^- =
\bn\cdot\lp$ and $\lp_{n\perp} \sim \la \lp^-$, and the small residual piece
$p_r^\mu = (p_r^+, p_r^-, p_{r\perp}^\mu) \sim \lp^-(\la^2, \la^2, \la^2)$ with
$\la \ll 1$. The corresponding $n$-collinear quark and gluon fields are
multipole expanded (with expansion parameter $\la$). This means particles with
different large components are described by separate quantum fields,
$\xi_{n,\lp}(y)$ and $A_{n,\lp}(y)$, which are distinguished by explicit
momentum labels on the fields (in addition to the $n$ label specifying the
collinear direction). We use $y$ to denote the position of the fields in the
operators to reserve $x$ for the parton momentum fractions. Two different types
of collinear fields will be relevant for our discussion depending on whether or
not they contain perturbatively calculable components. For the beam functions
$\lambda \simeq \tau_B^{1/2}$, and the collinear modes have perturbative
components with $p^2\sim Q\tau_B \gg \Lambda_{\rm QCD}$.  Collinear fields such
as these are referred to as belonging to an \SCETa theory. For collinear modes
in the parton distribution functions $\lambda\simeq \Lambda_{\rm QCD}/Q$ and the
collinear modes are nonperturbative with $p^2\sim \Lambda_{\rm QCD}^2$. These
modes are a subset of the \SCETa collinear modes and we will refer to their
fields as belonging to \SCETb. For much of our discussion the distinction
between these two types of collinear modes is not important and we can just
generically talk about collinear fields. When it is important we will refer
explicitly to \SCETa and \SCETb.

Interactions between collinear fields cannot change the direction $n$ but change the momentum labels to satisfy label momentum conservation. Since the momentum labels are changed by interactions, it is convenient to use the short-hand notations
\begin{equation} \label{eq:xi}
\xi_n(y) = \sum_{\lp \neq 0} \xi_{n,\lp}(y)
\,,\qquad
A_n^\mu(y) = \sum_{\lp \neq 0} A^\mu_{n,\lp}(y)
\,.\end{equation}
The sum over label momenta explicitly excludes the case $\lp^\mu = 0$ to avoid double-counting the soft degrees of freedom (described by separate soft quark and gluon fields). In practice when calculating matrix elements, this is implemented using zero-bin subtractions~\cite{Manohar:2006nz} or alternatively by dividing out matrix elements of Wilson lines~\cite{Collins:1999dz, Lee:2006nr, Idilbi:2007ff}. The dependence on the label momentum is obtained using label momentum operators $\bnP_n$ or $\cP_{n\perp}^\mu$ which return the sum of the minus or perpendicular label components of all $n$-collinear fields on which they act.

The decomposition into label and residual momenta is not unique. Although the explicit dependence on the vectors $n^\mu$ and $\bn^\mu$ breaks Lorentz invariance, the theory must still be invariant under changes to $n^\mu$ and $\bn^\mu$ which preserve the power counting of the different momentum components and the defining relations $n^2 = \bn^2 = 0$, $n\cdot\bn = 2$. This reparametrization invariance (RPI)~\cite{Chay:2002vy, Manohar:2002fd} can be divided into three types. RPI-I and RPI-II transformations correspond to rotations of $n$ and $\bn$. We will mainly use RPI-III under which $n^\mu$ and $\bn^\mu$ transform as
\begin{equation} \label{eq:RPI}
n^\mu \to e^\alpha n^\mu
\,,\qquad
\bn^\mu \to e^{-\alpha} \bn^\mu
\,,\end{equation}
which implies that the vector components transform as $p^+ \to e^\alpha p^+$ and $p^- \to e^{-\alpha} p^-$. In this way, the vector $p^\mu$ stays invariant and Lorentz symmetry is restored within a cone about the direction of $n^\mu$. Since \eq{RPI} only acts in the $n$-collinear sector, it is not equivalent to a spacetime boost of the whole physical system.

We now define the following bare operators
\begin{align} \label{eq:tiop_def}
\tiop_q^\bare(y^-,\w)
&= e^{-\img\hp^+ y^-/2}\,
\bar{\chi}_n \Bigl(y^- \frac{n}{2}\Bigr) \frac{\bnslash}{2} \bigl[\delta(\w - \bnP_n)\chi_n(0)\bigr]
\,, \nn \\
\tiop_{\bar q}^\bare(y^-,\w)
&= e^{-\img\hp^+ y^-/2}\,
\tr \Bigl\{\frac{\bnslash}{2} \chi_n \Bigl(y^- \frac{n}{2}\Bigr) \bigl[\delta(\w - \bnP_n) \bar\chi_n(0)\bigr]\Bigr\}
\,, \nn \\
\tiop_g^\bare(y^-,\w)
& = -\w\, e^{-\img\hp^+ y^-/2}\,
 \cB_{n\perp\mu}^c \Bigl(y^- \frac{n}{2}\Bigr) \bigl[\delta(\w - \bnP_n) \cB_{n\perp}^{\mu c}(0) \bigr]
\,.\end{align}
Their renormalization will be discussed in the next subsection. The corresponding renormalized operators are denoted as $\tiop_i(y^-, \w, \mu)$ and are defined in \eq{op_ren_pos} below.
Here, $\hp^+$ is the momentum operator of the residual plus momentum and acts on everything to its right. The overall phase is included such that the Fourier-conjugate variable to $y^-$ corresponds to the plus momentum of the initial-state radiation, see \eq{oq_FT} below. The operator $\delta(\w - \bnP_n)$ only acts inside the square brackets and forces the total sum of the minus labels of all fields in $\chi_n(0)$ and $\cB_{n\perp}(0)$ to be equal to $\w$. The color indices of the quark fields are suppressed and summed over, $c$ is an adjoint color index that is summed over, and the trace in $\tiop_{\bar{q}}$ is over spin. The operators are RPI-III invariant, because the transformation of the $\delta(\w - \bnP_n)$ is compensated by that of the $\bnslash$ in $\tiop_{q,\bar{q}}$ and the overall $\w$ in $\tiop_g$.

The fields
\begin{equation} \label{eq:chiB}
\chi_n(y) = W_n^\dagger(y)\, \xi_n(y)
\,,\qquad
\cB_{n\perp}^\mu = \frac{1}{g} \bigl[W_n^\dagger(y)\, \img D_{n\perp}^\mu W_n(y) \bigr]
\,,\end{equation}
with $\img D_{n\perp}^\mu = \cP^\mu_{n\perp} + g A^\mu_{n\perp}$, are composite SCET fields of $n$-collinear quarks and gluons.
In \eq{tiop_def} they are at the positions $y^\mu = y^- n^\mu/2$ and $y^\mu = 0$. The Wilson lines
\begin{equation} \label{eq:Wn}
W_n(y) = \biggl[\sum_\text{perms} \exp\Bigl(-\frac{g}{\bnP_n}\,\bn\sdt A_n(y)\Bigr)\biggr]
\end{equation}
are required to make $\chi_n(y)$ and $\cB_{n\perp}^\mu(y)$ gauge invariant with
respect to collinear gauge transformations~\cite{Bauer:2000yr, Bauer:2001ct}.
They are Wilson lines in label momentum space consisting of $\bn\sdt A_n(y)$
collinear gluon fields. They sum up arbitrary emissions of $n$-collinear gluons
from an $n$-collinear quark or gluon, which are $\ord{1}$ in the SCET power
counting. Since $W_n(y)$ is localized with respect to the residual position $y$,
$\chi_n(y)$ and $\cB_{n\perp}^\mu(y)$ are local operators for soft interactions.
In \SCETa the fields in \eqs{tiop_def}{chiB} are those after the field
redefinition~\cite{Bauer:2001yt} decoupling soft gluons from collinear
particles. Thus at leading order in the power counting these collinear fields do
not interact with soft gluons through their Lagrangian and no longer transform
under soft gauge transformations. Hence, the operators in \eq{tiop_def} are
gauge invariant under both soft and collinear gauge transformations. The soft
interactions with collinear particles are factorized into a soft function, which
is a matrix element of soft Wilson lines.

Note that our collinear fields in \eq{tiop_def} have continuous labels and hence
are not the standard SCET fields with discrete labels. They only depend on the
minus coordinate, $y^-$, corresponding to the residual plus momentum, $p_r^+$,
and not a full four-vector $y^\mu$. As discussed in detail in the derivation of
the factorization theorem in Ref.~\cite{Stewart:2009yx}, it is convenient to
absorb the residual minus and perpendicular components into the label momenta
which then become continuous variables. For example, for the minus momentum
(suppressing the perpendicular dependence)
\begin{align}
\sum_{\lp^-}\, e^{-\img \lp^- y^+/2} \chi_{n,\lp^-}(y^-, y^+)
&= \sum_{\lp^-} \int\!\df p_r^-\,e^{-\img (\lp^- + p_r^-) y^+/2} \chi_{n,p^-}(y^-)
\nn\\
&= \int\!\df p^-\,e^{-\img p^- y^+/2} \chi_{n, p^-}(y^-)
\,.\end{align}
In this case, $W_n(y^-n/2)$ can also be written in position space where all gluon fields sit at the same residual minus coordinate, $y^-$, and are path ordered in the plus coordinate (corresponding to the label minus momentum) from $y^+$ to infinity.

Next, we introduce the Fourier-transformed operators
\begin{equation} \label{eq:op_FT}
\op_i^\bare(\abs{\w} b^+, \w)
= \frac{1}{2\pi} \int \! \frac{\df y^-}{2\abs{\w}}\, e^{\img b^+ y^-/2}\, \tiop_i^\bare(y^-,\w)
\,,\end{equation}
and the corresponding renormalized operators $\op_i(\abs{\w} b^+, \w, \mu)$ [see \eq{op_ren} below].
For example, for the quark operator
\begin{align} \label{eq:oq_FT}
\op_q^\bare(\abs{\w} b^+, \w)
&= \frac{1}{2\pi} \int \! \frac{\df y^-}{2\abs{\w}}\, e^{\img (b^+ - \hp^+) y^-/2}
\Bigl(e^{\img\hp^+ y^-/2} \bar \chi_n(0) e^{-\img\hp^+ y^-/2}\Bigr) \frac{\bnslash}{2}
\bigl[\delta(\w - \bnP_n) \chi_n(0)\bigr]
\nn\\
&= \bar \chi_n(0)\, \delta(\w b^+ - \w\hp^+)\, \frac{\bnslash}{2} \bigl[\delta(\w - \bnP_n) \chi_n(0)\bigr]
\,.\end{align}
In the first step we used residual momentum conservation to shift the position of the field. Here we see that the overall phase in \eq{tiop_def} allows us to write the $b^+$ dependence in terms of $\delta(\w b^+ - \w\hp^+)$, which means that $b^+$ measures the plus momentum of any intermediate state that is inserted between the fields.

We divide by $\abs{\w}$ in \eq{op_FT} to make the integration measure of the
Fourier transform RPI-III invariant. Using the absolute value $\abs{\w}$ ensures
that the definition of the Fourier transform does not depend on the sign of $\w$
and that the first argument of $\op_q$, $t = \abs{\w} b^+$, always has the same
sign as $b^+$. The Fourier-transformed operators are still RPI-III invariant and
only depend on $b^+$ through the RPI-III invariant combination $t$. The beam
functions are defined as the proton matrix elements of the renormalized
operators $\op_i(t, \w,\mu)$ in \SCETa,
\begin{equation} \label{eq:B_def}
B_i(t, x = \w/P^-,\mu) = \Mae{p_n(P^-)}{\theta(\w) \op_i(t,\w,\mu)}{p_n(P^-)}
\,.\end{equation}
The matrix elements are always averaged over proton spins, which we suppress in our notation. Note that part of the definition in \eq{B_def} is the choice of the direction $n$ such that the proton states have no perpendicular momentum, $P^\mu = P^- n^\mu/2$, which is why we denote them as $\ket{p_n(P^-)}$. By RPI-III invariance, the beam functions can then only depend on the RPI-III invariant variables $t = \w b^+$ and $x = \w/P^-$. The restriction $\theta(\w)$ on the right-hand side of \eq{B_def} is included to enforce that the $\chi_n(0)$, $\bar\chi_n(0)$, or $\cB_{n\perp}(0)$ fields annihilate a quark, antiquark, or gluon out of the proton, as we discuss further at the beginning of \subsec{beamT}.

The definition of the beam functions can be compared with that of the quark and gluon PDFs. In SCET, the PDFs are defined~\cite{Bauer:2002nz} in terms of the RPI-III invariant operators
\begin{align} \label{eq:oq_def}
\oq^\bare_q(\w')
&= \theta(\w')\, \bar{\chi}_n(0) \frac{\bnslash}{2} \bigl[\delta(\w' - \bnP_n) \chi_n(0)\bigr]
\,, \nn \\
\oq^\bare_{\bar q}(\w')
&= \theta(\w')\, \tr \Bigl\{\frac{\bnslash}{2} \chi_n(0) \bigl[\delta(\w' - \bnP_n) \bar\chi_n(0)\bigr] \Bigr\}
\,, \nn \\
\oq^\bare_g(\w')
& = -\w'\theta(\w')\, \cB_{n\perp\mu}^c(0) \bigl[\delta(\w' - \bnP_n) \cB_{n\perp}^{\mu c}(0) \bigr]
\,,\end{align}
as the proton matrix elements in \SCETb of the corresponding renormalized
operators $\oq_i(\w', \mu)$ defined in \eq{oq_ren} below,
\begin{equation} \label{eq:f_def_SCET}
f_i(\w'/P^-,\mu) = \Mae{p_n(P^-)}{\oq_i(\w',\mu)}{p_n(P^-)}
\,.\end{equation}
By RPI-III invariance, the PDFs can only depend on the momentum fraction $\xi =
\w'/P^-$. Beyond tree level $\xi$ or $\w'$ are not the same as $x$ or $\w$,
which is why we denote them differently.  Without the additional $\theta(\w')$
in the operators in \eq{oq_def} the quark and anti-quark PDFs would combine into
one function, with the quark PDF corresponding to $\w>0$ and the antiquark PDF to
$\w<0$. We explicitly separate these pieces to keep analogous definitions for
the PDFs and beam functions.

It is important to note that the \SCETb collinear fields in \eq{oq_def} do not
require zero-bin subtractions, because as is well-known, the soft region does
not contribute to the PDFs. If one makes the field redefinitions $\xi_n\to Y
\xi_n$ and $A_n \to Y A_n Y^\dagger$ to decouple soft gluons, then the soft
Wilson lines $Y$ cancel in \eq{oq_def}. Equivalently, if the fields in
\eq{oq_def} include zero-bin subtractions then the subtractions will cancel in
the sum of all diagrams, just like the soft gluons.  (This is easy to see by
formulating the zero-bin subtraction as a field redefinition~\cite{Lee:2006nr}
analogous to the soft one but with Wilson lines in a different light-cone
direction.)  In contrast, the \SCETa collinear fields in the beam function
operator in \eq{tiop_def} must include zero-bin subtractions. We will see this
explicitly at one loop in our PDF and beam function calculations in
\sec{oneloop}.

The SCET definitions of the PDFs are equivalent to the standard definition in terms of full QCD quark fields $\psi$ in position space. For example, the quark PDF in QCD is defined as~\cite{Collins:1981uw}
\begin{equation} \label{eq:f_def_QCD}
f_q(\w'/P^-, \mu) = \theta(\w') \int\! \frac{\df y^+}{4\pi}\,
  e^{-\img \w' y^+/2}
  \MAe{p_n(P^-)}{\Bigl[\bar\psi \Bigl(y^+\frac{\bn}{2}\Bigr)
   W\Bigl(y^+\frac{\bn}{2},0\Bigr) \frac{\bnslash}{2} \psi(0) \Bigr]_\mu}{p_n(P^-)}
\,.\end{equation}
The square brackets denote the renormalized operator.  Here, the fields are
separated along the $\bn$ direction and the lightlike Wilson line
$W(y^+\bn/2,0)$ is required to render the product of the fields gauge invariant.
The relation to the SCET definition is that the \SCETb fields in \eq{oq_def}
(without zero-bin subtractions) involve a Fourier transform of $\psi$ in $y^+$
to give the conjugate variable $\w'$. The corresponding Wilson lines in
\eq{f_def_SCET} are precisely the $W_n$ contained in the definitions of $\chi_n$
and $\cB_{n\perp}^\mu$. Hence, the QCD and SCET definitions of the PDF are
equivalent (provided of course that one uses the same renormalization scheme,
which we do).

Comparing \eq{oq_def} to \eq{tiop_def}, the difference between the beam
functions and PDFs is that for the beam functions the fields are additionally
separated along the $n$ light-cone, with a large separation $y^- \gg y^+$
corresponding to the small momentum $b^+ \ll \w$. Thus, formulating equivalent
definitions of the beam functions directly in QCD would be more challenging, as
it would require QCD fields that are simultaneously separated in the $n$ and
$\bn$ directions.  For this case, it is not clear a priori how to obtain an
unambiguous gauge-invariant definition, because Wilson lines connecting the
fields along different paths are not equivalent.  This ambiguity is resolved in
\SCETa, where the multipole expansion distinguishes the different scales and
divides the possible gauge transformations into global, collinear, and soft
transformations, allowing one to treat the separations along the two orthogonal
light-cones independently. The large $y^-$ separation corresponds to soft Wilson
lines and soft gauge transformations that are independent from collinear gauge
transformations corresponding to the small $y^+$ dependence. As already
mentioned, the operators in \eq{tiop_def} are separately gauge invariant under
both types of gauge transformations.

\subsection{Renormalization and RGE}
\label{subsec:B_RGE}

The beam functions and PDFs are defined as the matrix elements of renormalized operators. The renormalization of the operators immediately yields that of the functions defined by their matrix elements. In this subsection we derive the RG equations and show that the anomalous dimensions of the beam and jet functions are the same to all orders in perturbation theory.

We start by considering the known renormalization of the PDF, but in the SCET operator language. The renormalized PDF operators are given in terms of the bare operators in \eq{oq_def} as
\begin{equation} \label{eq:oq_ren}
\oq_i^\bare(\w) = \sum_j \int\! \frac{\df \w'}{\w'}\, Z^f_{ij}\Big(\frac{\w}{\w'},\mu\Big) \oq_j(\w',\mu)
\,.\end{equation}
In general, operators with different $i$ and $\w$ can (and will) mix into each
other, so the renormalization constant $Z^f_{ij}(\w/\w',\mu)$ is a matrix in
$i,j$ and $\w,\w'$. RPI-III invariance then restricts the integration measure to
be $\df\w'/\w'$ and $Z^f_{ij}(\w/\w',\mu)$ to only depend on the ratio $\w/\w'$.
Hence, the form of \eq{oq_ren} is completely specified by the SCET symmetries.
The $\mu$ independence of the bare operators $\oq_i^\bare(\w)$ yields an RGE for
the renormalized operators in $\overline{\mathrm{MS}}$
\begin{align} \label{eq:oq_RGE}
\mu\frac{\df}{\df\mu} \oq_i(\w,\mu)
&= \sum_j \int\! \frac{\df \w'}{\w'}\, \gamma^f_{ij}\Bigl(\frac{\w}{\w'},\mu\Bigr)\, \oq_j(\w',\mu)
\,, \nn \\
\gamma^f_{ij}(z,\mu) &= -\sum_k \int\! \frac{\df z'}{z'}\, (Z^f)_{ik}^{-1}\Bigl(\frac{z}{z'},\mu\Bigr)\, \mu\frac{\df}{\df\mu} Z^f_{kj}(z',\mu)
\,,\end{align}
where the inverse $(Z^f)_{ik}^{-1}(z, \mu)$ is defined as
\begin{equation}
\sum_k  \int\! \frac{\df z'}{z'}\, (Z^f)_{ik}^{-1}\Bigl(\frac{z}{z'},\mu\Bigr) Z^f_{kj}(z',\mu)
= \delta_{ij}\, \delta(1-z)
\,.\end{equation}
Taking the proton matrix element of \eq{oq_RGE} yields the RGE for the PDFs
\begin{equation} \label{eq:PDF_RGE}
\mu\frac{\df}{\df\mu} f_i(\xi,\mu)
= \sum_j\int\! \frac{\df\xi'}{\xi'}\, \gamma^f_{ij}\Bigl(\frac{\xi}{\xi'},\mu\Bigr)\, f_j(\xi',\mu)
\,.\end{equation}
The solution of this RGE can be written in terms of an evolution function $U^f$ which acts on the initial PDF $f_j(\xi',\mu_0)$ and takes it to $f_i(\xi,\mu)$,
\begin{equation} \label{eq:Uf_def}
f_i(\xi,\mu) = \int\! \frac{\df\xi'}{\xi'}\, U^f_{ij}\Bigl(\frac{\xi}{\xi'},\mu,\mu_0\Bigr) f_j(\xi',\mu_0)
\,.\end{equation}

From \eq{PDF_RGE} we can identify the anomalous dimensions $\gamma_{ij}^f(z)$ in terms of the
QCD splitting functions. For example, in dimensional regularization in the $\overline{\mathrm{MS}}$ scheme, the one-loop anomalous dimensions for the quark PDF are the standard ones
\begin{equation} \label{eq:gammaf}
\gamma_{qq}^f(z, \mu) = \frac{\alpha_s(\mu)C_F}{\pi}\, \theta(z) P_{qq}(z)
\,,\qquad
\gamma_{qg}^f(z, \mu) = \frac{\alpha_s(\mu)T_F}{\pi}\, \theta(z) P_{qg}(z)
\,,\end{equation}
with the $q\to qg$ and $g\to q\bar{q}$ splitting functions
\begin{align} \label{eq:Pqq_def}
P_{qq}(z)
&= \cL_0(1-z)(1+z^2) + \frac{3}{2}\,\delta(1-z)
= \biggl[\theta(1-z)\frac{1+z^2}{1-z} \biggr]_+
\,,\nn\\
P_{qg}(z) &= \theta(1-z)\bigl[(1-z)^2+ z^2\bigr]
 \,.\end{align}
The plus distribution $\cL_0(x) = [\theta(x)/x]_+$ is defined in the standard
way, see \eq{plusdef}. For later convenience we do not include the overall color
factors in the definitions in \eq{Pqq_def}.

We now go through an analogous discussion for the beam functions. The renormalized operators $\tiop_i(y^-, \w, \mu)$ are given in terms of the bare operators in \eq{tiop_def} by
\begin{equation} \label{eq:op_ren_pos}
\tiop_i^\bare(y^-, \w) = \tZ_B^i\Bigl(\frac{y^-}{2\abs{\w}}, \mu\Bigr) \tiop_i(y^-,\w,\mu)
\,,\end{equation}
where $\tZ_B^i(y^-/2\abs{\w}, \mu)$ is the position-space renormalization constant. In \app{B_RGE}, we give an explicit proof that the beam function renormalization is multiplicative in this way to all orders in perturbation theory.%
\footnote{
With our definitions of $b^+$ and $t = \abs{\w} b^+$, they are always positive
irrespective of the sign of $\w$ (i.e.\ for both beam and jet functions). Since
$y^-$ is the Fourier conjugate variable to $b^+$, the Fourier-conjugate variable
to $t$ is $u = y^-/2\abs{\w}$. The proof in the appendix, which is for $\w > 0$,
shows that $\tga_B$ only depends on $u$ through $\ln[\img(u - \img 0)]$. Its
most general RPI-III invariant form is $\tga_B(u, \w/\abs{\w})$. From
\eq{tiop_def} $\Vev{\tiop_i^{\bare\,\dagger}(y^-, \w)} =
\Vev{\tiop_i^\bare(-y^-, -\w)}$ for any forward matrix element. Since the
renormalization does not change the analytic structure, the same is true for the
renormalized matrix elements so $\tga_B^*(u, \w/\abs{\w}) = \tga_B(-u,
-\w/\abs{\w})$, and also $\tga_B$ can only be a real function of
$\w/\abs{\w}$. Because of its simple $u$ dependence, we can conclude that
$\tga_B \equiv \tga_B(u)$ only. With
the tree-level boundary condition this then implies $\tZ_B \equiv \tZ_B(u)$.}
The underlying reason is that the renormalization of the theory should preserve locality, so renormalizing the nonlocal beam function operator should not affect the $y^-$ separation between the fields. For example, mixing between operators with different $y^-$ would destroy locality at distance scales within the validity range of the effective theory. RPI-III invariance then implies that $\tZ_B^i$ can only depend on the ratio $y^-/2\abs{\w}$ (the factor of $1/2$ is for convenience). In principle, one might think there could also be mixing between operators with different $i$ or $\w$ in \eq{op_ren_pos} [as was the case for the PDFs in \eq{oq_ren}]. Our derivation in \app{B_RGE} shows that this is not the case.

Taking the Fourier transform of \eq{op_ren_pos} according to \eq{op_FT}, we find
\begin{align} \label{eq:op_ren}
\op_i^\bare(t, \w) &= \int\! \df t'\, Z_B^i(t - t', \mu)\, \op_i(t', \w, \mu)
\,, \nn \\
Z_B^i(t, \mu) &= \frac{1}{2\pi} \int\! \frac{\df y^-}{2\abs{\w}}\, e^{\img t y^-/2\abs{\w}}\, \tZ_B^i \Bigl(\frac{y^-}{2\abs{\w}}, \mu\Bigr)
\,.\end{align}
Since the bare operator is $\mu$ independent, taking the derivative with respect to $\mu$, we find the RGE for the renormalized operator
\begin{align} \label{eq:op_RGE}
\mu \frac{\df}{\df \mu} \op_i(t, \w, \mu) &= \int\! \df t' \gamma_B^i(t - t', \mu)\, \op_i(t', \w, \mu)
\,,\nn\\
\gamma_B^i(t,\mu) &= - \int\! \df t'\, (Z_B^i)^{-1}(t-t', \mu)\, \mu\frac{\df}{\df\mu} Z_B^i (t', \mu)
\,,\end{align}
where the inverse of $Z_B^i(t, \mu)$ is defined as usual,
\begin{equation}
\int\! \df t'\, (Z_B^i)^{-1}(t-t',\mu)\, Z_B^i(t',\mu) = \delta(t)
\,.\end{equation}

Taking the proton matrix element of \eq{op_RGE} we obtain the corresponding RGE for the beam function,
\begin{equation} \label{eq:B_RGE}
\mu \frac{\df}{\df \mu} B_i(t, x, \mu) = \int\! \df t'\, \gamma_B^i(t-t',\mu)\, B_i(t', x, \mu)
\,.\end{equation}
As discussed in \app{B_RGE}, to all orders in perturbation theory the anomalous dimension has the form
\begin{equation} \label{eq:gaB_gen}
\gamma_B^i(t, \mu)
= -2 \Gamma^i_\cusp(\alpha_s)\,\frac{1}{\mu^2}\cL_0\Bigl(\frac{t}{\mu^2}\Bigr)
+ \gamma_B^i(\alpha_s)\,\delta(t)
\,,\end{equation}
where $\cL_0(x) = [\theta(x)/x]_+$ is defined in \eq{plusdef}, $\Gamma_\cusp^i(\alpha_s)$ is the cusp anomalous dimension for quarks/antiquarks ($i = q$) or gluons ($i = g$), and $\gamma_B^i(\alpha_s)$ denotes the non-cusp part. Since there is no mixing between operators $\op_i(t, \w, \mu)$ with different $i$ or $\w$, the beam function RGE only changes the virtuality $t$ but not the momentum fraction $x$ and does not mix quark and gluon beam functions. By rescaling the plus distribution,
\begin{equation}
\frac{1}{\mu^2}\cL_0\Bigl(\frac{t}{\mu^2}\Bigr)
= \frac{1}{\mu_0^2}\cL_0\Bigl(\frac{t}{\mu_0^2}\Bigr) - 2\ln\frac{\mu}{\mu_0}\, \delta(t)
\end{equation}
we can see that $\gamma_B^i(t, \mu)$ has logarithmic $\mu$-dependence, which means that the RGE sums Sudakov double logarithms.

The solution of the RGE in \eq{B_RGE} with the form of the anomalous dimension in \eq{gaB_gen} is known~\cite{Balzereit:1998yf, Neubert:2004dd, Fleming:2007xt}. It takes the form
\begin{equation} \label{eq:Brun}
B_i(t,x,\mu) =  \int\! \df t'\, B_i(t - t',x,\mu_0)\, U_B^i(t',\mu_0,\mu)
\,,\end{equation}
where the evolution kernel can be written as~\cite{Ligeti:2008ac}
\begin{equation} \label{eq:UB}
U_B^i(t, \mu_0, \mu) = \frac{e^{K_B^i -\gamma_E\, \eta_B^i}}{\Gamma(1 + \eta_B^i)}\,
\biggl[\frac{\eta_B^i}{\mu_0^2} \cL^{\eta_B^i} \Bigl( \frac{t}{\mu_0^2} \Bigr) + \delta(t) \biggr]
\,.\end{equation}
The distribution $\cL^\eta(x)$ is defined in \eq{plusdef}, and the RGE functions $K_B^i \equiv K_B^i(\mu_0, \mu)$ and $\eta_B^i \equiv \eta_B^i(\mu_0, \mu)$ are given in \eq{Brun_full}.

The SCET quark, antiquark, and gluon jet functions are given by~\cite{Bauer:2001yt, Fleming:2003gt}
\begin{align} \label{eq:J_def}
&J_q(\w p^+\! + \w_\perp^2, \mu)
\nn\\ & \qquad
= \frac{(2\pi)^2}{N_c} \!\int\! \frac{\df y^-}{2\abs{\w}}\, e^{\img p^+ y^-/2}\,
   \tr\MAe{0}{ \Bigl[ \frac{\bnslash}{2} \chi_n \Bigl(y^- \frac{n}{2}\Bigr)
   \bigl[\delta(\w + \bnP_n)\delta^2(\w_\perp + \cP_{n\perp}) \bar \chi_n(0) \bigr] \Bigr]_\mu}{0}
\,,\nn\\
&J_{\bar q}(\w p^+\! + \w_\perp^2, \mu)
\nn\\ & \qquad
= \frac{(2\pi)^2}{N_c} \!\int\! \frac{\df y^-}{2\abs{\w}}\, e^{\img p^+ y^-/2}\,
   \MAe{0}{\Bigl[ \bar \chi_n \Bigl(y^- \frac{n}{2}\Bigr)\frac{\bnslash}{2}
   \bigl[\delta(\w + \bnP_n) \delta^2(\w_\perp + \cP_{n\perp}) \chi_n(0) \bigr] \Bigr]_\mu}{0}
\,,\\\nn
&J_g(\w p^+\! + \w_\perp^2, \mu)
\nn\\\nn & \qquad
= -\frac{(2\pi)^2}{N_c^2 - 1}\, \w \!\int\!\frac{\df y^-}{2\abs{\w}}\, e^{\img p^+ y^-/2}
   \MAe{0}{ \Bigl[ \cB_{n\perp\mu}^c \Bigl(y^- \frac{n}{2}\Bigr)
   \bigl[ \delta(\w + \bnP_n) \delta^2(\w_\perp + \cP_{n\perp}) \cB_{n\perp}^{\mu c}(0) \bigr] \Bigr]_\mu}{0}
\,,\end{align}
where the notation $[\ldots]_\mu$ again denotes the renormalized operators.
Here, we used the same conventions as for the beam functions where the large label momenta $\w$ and $\w_\perp$ are continuous, so the only position dependence of the fields is in the minus component. RPI invariance requires that the jet function only depends on the total invariant mass of the jet, $p^2 = \w p^+ + \w_\perp^2$. When the jet function appears in a factorization theorem, the direction of the jet is either measured (e.g. by measuring the thrust axis in $e^+e^-\to 2$ jets) or fixed by kinematics (e.g. in $B\to X_s\gamma$ the jet direction is fixed by the direction of the photon) and $n$ is chosen along the jet direction, so one typically has $\w_\perp = 0$. Taking the vacuum matrix element of $\op_q^\bare(t,\w)$, we get
\begin{align}\label{eq:qjet}
&\frac{(2\pi)^2}{N_c}\Mae{0}{\op_q^\bare(t,-\w)}{0}
\nn\\ & \qquad
= \frac{(2\pi)^2}{N_c}
\int\! \df^2 \w_\perp \frac{1}{2\pi} \int \! \frac{\df y^-}{2\abs{\w}}\, e^{\img t y^-/(2\abs{\w})}\,
  \MAe{0}{\bar \chi_n \Bigl(y^- \frac{n}{2}\Bigr) \frac{\bnslash}{2}\, \delta(\w + \bnP_n) \delta^2(\w_\perp - \cP_{n\perp}) \chi_n(0)}{0}
\nn \\ & \qquad
= \int\! \df^2 \vec\w_\perp J^\bare_{\bar q}(t - \vec\w_\perp^2)
\equiv \hJ^\bare_{\bar q}(t) = \hJ^\bare_q(t)
\,.\end{align}
In the last step we used that the quark and antiquark jet functions are the same. The analogous relation holds for the antiquark operator, $\op_{\bar q}(t,\w,\mu)$. The $\vec\w_\perp$ integral is bounded and does not lead to new UV divergences, because the jet function only has support for nonnegative argument, $0 < \vec\w_\perp^2 < t$, and $t$ is fixed. Similarly, for the gluon operator we have
\begin{equation}\label{eq:gjet}
\frac{(2\pi)^2}{N_c^2 - 1}\Mae{0}{\op_g^\bare(t, -\w)}{0}
= \int\! \df^2 \vec\w_\perp J_g^\bare(t - \vec\w_\perp^2, \mu) \equiv \hJ_g^\bare(t)
\,.\end{equation}
The renormalization of $J_i^\bare(t)$ does not depend on the choice of $\w_\perp$ in \eq{J_def}. Since $\hJ_i^\bare(t)$ is simply an average over different choices for $\w_\perp$ it has the same renormalization. Hence $J_i(t, \mu)$ and $\hJ_i(t, \mu)$ have the same anomalous dimension,
\begin{align}
\mu\frac{\df}{\df\mu} \hJ_i(t, \mu)
&= \int\! \df^2 \vec\w_\perp\, \df s\, \gamma_J^i(t - \vec\w_\perp^2 - s, \mu)\, J_i(s,\mu)
= \int\! \df t' \gamma_J^i(t - t', \mu) \int\! \df^2 \vec\w_\perp\, J_i(t' - \vec\w_\perp^2,\mu)
\nn \\
&= \int\!\df t'\,\gamma_J^i(t - t', \mu)\, \hJ_i(t', \mu)
\,.\end{align}
On the other hand, taking the vacuum matrix element of \eq{op_RGE} we get
\begin{equation}
\mu\frac{\df}{\df\mu} \hJ_i(t, \mu) = \int\!\df t'\,\gamma_B^i(t - t', \mu)\, \hJ_i(t', \mu)
\,.\end{equation}
We thus conclude that the beam and jet function anomalous dimensions are identical to all orders in perturbation theory,
\begin{equation} \label{eq:gaJgaB}
\gamma_B^i(t, \mu) = \gamma_J^i(t, \mu)
\,.\end{equation}
For the cusp part this result already follows from our explicit one-loop calculation, since $\Gamma^i_\cusp$ is universal and its coefficients are the same at one loop. Our one-loop result provides a cross check for the identity of the one-loop non-cusp part of the anomalous dimension, which agree. Furthermore, $\gamma_J^q(\alpha_s)$ and hence $\gamma_B^q(\alpha_s)$ can be obtained to three loops from Refs.~\cite{Moch:2004pa, Moch:2005id}, and for completeness the result is given in \app{pert}.

\subsection{Operator Product Expansion}
\label{subsec:OPE}

The difference between the beam function operators in \eq{tiop_def} and the PDF operators in \eq{oq_def} is the additional separation in the $y^-$ coordinate between the fields. Hence, by performing an operator product expansion about the limit $y^-\!\to 0$ we can expand the renormalized operators $\tiop_i(y^-, \w, \mu)$ in terms of a sum over $\oq_i(\w', \mu)$,
\begin{equation} \label{eq:tiop_RGE}
\tiop_i(y^-, \w, \mu)
= \widetilde{J}_i\Bigl(\frac{y^-}{2\abs{\w}},\mu\Bigr) 1
 + \sum_j \int\! \frac{\df \w'}{\w'}\,
  \widetilde{\cI}_{ij}\Bigl(\frac{y^-}{2\abs{\w}},\frac{\w}{\w'},\mu \Bigr) \oq_j(\w',\mu)
+ \ORd{\frac{y^-}{\w}}
\,.\end{equation}
For completeness we included the identity operator on the right-hand side. The
form of the matching coefficients $\widetilde{\cI}_{ij}$ and $\widetilde{J}_i$
is again constrained by RPI-III invariance so the structure of the OPE is
completely determined by the SCET symmetries. Equation~(\ref{eq:tiop_RGE})
encodes a matching computation between the operator $\tiop_i$ in \SCETa, and the
operators $1$ and $\oq_j$ in \SCETb, where $\widetilde J_i$ and $\widetilde
{\cI}_{ij}$ are the corresponding Wilson coefficients.

Fourier transforming both sides of \eq{tiop_RGE} with respect to $y^-$ we get
\begin{equation} \label{eq:op_OPE}
\op_i(t, \w, \mu) = \hJ_i(t,\mu) 1 + \sum_j \int\! \frac{\df \w'}{\w'}\,
\cI_{ij}\Bigl(t,\frac{\w}{\w'},\mu \Bigr) \oq_j(\w',\mu) + \ORd{\frac{y^-}{\w}}
\,.\end{equation}
Taking the vacuum matrix element of both sides, and using $\mae{0}{\oq_j}{0} =
0$, we just get the coefficient of the identity operator on the right-hand side,
which from \eqs{qjet}{gjet} is thus given by $\hJ_i(t,\mu)$. Taking the proton
matrix element of \eq{op_OPE} with $\w > 0$ according to \eq{B_def}, this first
term drops out, because the jet functions only have support for $-\w>0$ (or
alternatively because the corresponding diagrams are disconnected), and we
obtain the OPE for the beam function
\begin{equation} \label{eq:beam_fact}
B_i(t,x,\mu)
= \sum_j \int \! \frac{\df \xi}{\xi}\, \cI_{ij}\Bigl(t,\frac{x}{\xi},\mu \Bigr) f_j(\xi,\mu)
  \biggl[1 + \ORd{\frac{\lqcd^2}{t}}\biggr]
\,.\end{equation}
For $B_g$ this equation was first derived in Ref.~\cite{Fleming:2006cd} using a
moment-space OPE for the matrix element (modulo missing the mixing contribution
from the quark PDF). The higher-order power corrections in \eq{beam_fact} must
scale like $1/t$ and are therefore of $\ord{\lqcd^2/t}$ where $\lqcd^2$ is the
typical invariant mass of the partons in the proton.  Equivalently, one can
think of the scaling as $(\lqcd^2/\w)/b^+$ where $\lqcd^2/\w$ is the typical
plus momentum of the parton in the proton. These power corrections are given in
terms of higher-twist proton structure functions. Since \eq{op_OPE} is valid for
$t \gg \lqcd^2$, this also means that we can calculate the matching coefficients
in perturbation theory at the beam scale $\mu_B^2 \simeq t$. This \SCETa to
\SCETb matching calculation is carried out in the usual way by computing
convenient matrix elements of the operators on both sides of \eq{op_OPE} and
extracting the Wilson coefficients from the difference. This is carried out at
tree level in the next subsection, while the full one-loop matching calculation
for the quark beam function is given in \sec{oneloop}. On the other hand, for
$t\sim\lqcd^2$ the beam functions are nonperturbative and the OPE would require
an infinite set of higher-twist proton structure functions. In this case, the
beam functions essentially become nonperturbative $b^+$-dependent PDFs.

The physical interpretation of the beam function OPE in \eq{beam_fact} leads exactly to the physical picture shown in \fig{beam} and discussed in the introduction. At the beam scale $\mu_B\simeq t$, the PDFs are evaluated and a parton $j$ with momentum fraction $\xi$ is taken out of the proton. It then undergoes further collinear interactions, which are described by the perturbative Wilson coefficients $\cI_{ij}(t, z, \mu)$. By emitting collinear radiation it looses some of its momentum, and the final momentum fraction going into the hard interaction is $x < \xi$. In addition, the sum on $j$ indicates that there is a mixing effect from terms without large logarithms, e.g. the quark beam function gets contributions from the quark, gluon, and antiquark PDFs. For example, when an incoming gluon from the proton pair-produces, with the quark participating in the hard interaction and the antiquark going into the beam remnant, then this is a mixing of the gluon PDF into the quark beam function. These are the physical effects that would usually be described by the PDF evolution. The difference is that once we are above the beam scale these effects only cause non-logarithmic perturbative corrections, which means the parton mixing and $x$-reshuffling now appears in the matching, while the RG evolution of the beam function only changes $t$, as we saw above. In \sec{results}, we will see that these matching corrections are still important numerically and must be taken into account. For example, since the gluon PDF at small $\xi$ is very large compared to the quark and antiquark PDFs, it still gives an important contribution to the quark and antiquark beam functions.

The consistency of the RGE requires that the $\mu$ dependence of the Wilson coefficients $\cI_{ij}(t, z, \mu)$ turns the RG running of the PDFs into the proper RG running of the beam functions. Taking the $\mu$ derivative of \eq{beam_fact} we find the evolution equation for the Wilson coefficients
\begin{align}
\mu\frac{\df}{\df\mu}\cI_{ij}(t,z,\mu)
= \sum_k \int\!\df t'\, \frac{\df z'}{z'}\, \cI_{ik}\Bigl(t - t', \frac{z}{z'},\mu\Bigr)\Bigl[\gamma_B^i(t', \mu)\, \delta_{kj} \delta(1 - z')
- \delta(t') \gamma^f_{kj}(z',\mu) \Bigr]
\,.\end{align}
The solution to this RGE can be easily obtained in terms of the evolution factors for the PDF and beam function in \eqs{Uf_def}{UB},
\begin{align}
\cI_{ij}(t,z,\mu) &= \sum_k\int\! \df t'\, \frac{\df z'}{z'}\,
  \cI_{ik}\Bigl(t - t',\frac{z}{z'},\mu_0\Bigr)\, U_B^i(t',\mu_0,\mu)\, U^f_{kj}(z',\mu_0,\mu)
\,.\end{align}
Hence as expected, the RGE running of $\cI_{ij}(t, z, \mu)$ cancels the running of the PDFs and adds in the running of the beam function.

\subsection{Tree-level Matching  onto PDFs}
\label{subsec:tree_match}

\begin{figure}[t]
\hfill%
\subfigure[]{\includegraphics[scale=0.75]{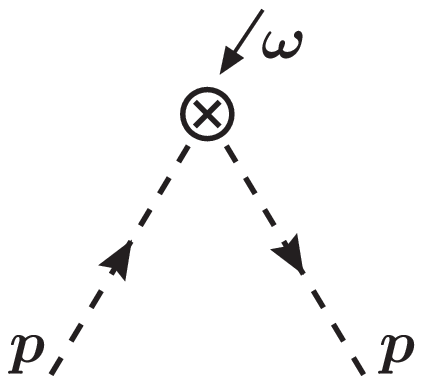}\label{fig:ftree}}%
\hfill%
\subfigure[]{\includegraphics[scale=0.75]{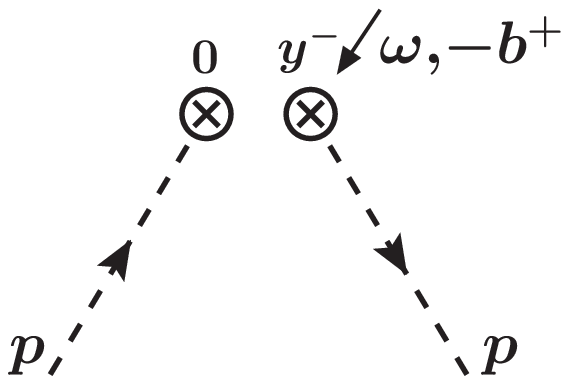}\label{fig:Btree}}%
\hspace*{\fill}%
\caption{Tree-level diagram for the quark PDF (a) and the quark beam function (b).
For the latter, the $y^-$ coordinate separation in the operator is indicated
by drawing separated vertices for each field.}
\end{figure}

To illustrate the application of the OPE, we will calculate the Wilson
coefficients $\cI_{ij}$ at tree level, starting with $\cI_{qq}$. We can use any
external states for the computation of the Wilson coefficient as long as they
have nonzero overlap with our operator. Thus, we pick the simplest choice,
$n$-collinear quark and gluon states, $\ket{q_n(p)}$ and $\ket{g_n(p)}$, with
momentum $p^\mu = (p^+,p^-,0)$ where $p^->0$ is the large momentum. In the
following section we will use a small $p^+<0$ as an IR regulator, but otherwise
$p^+$ is set to zero. The tree-level diagrams with an external quark for the
quark PDF and beam function are shown in \figs{ftree}{Btree}. They give
\begin{align} \label{eq:op_tree}
\Mae{q_n(p)}{\oq_q(\w',\mu)}{q_n(p)}^\zero
&= \theta(\w')\,\bar u_n(p)\delta(\w'-p^-)\, \frac{\bnslash}{2} u_n(p)
= \theta(\w')\,\delta(1-\w'/p^-)
\,, \nn \\
\Mae{q_n(p)}{\op_q(t,\w,\mu)}{q_n(p)}^\zero
&=  \bar u_n(p)\,\delta(t)\,\delta(\w-p^-)\, \frac{\bnslash}{2} u_n(p)
= \delta(t) \, \delta(1 - \w/p^-)
\,.\end{align}
Here and in the following the superscript $(i)$ indicates the $\ord{\alpha_s^i}$ contribution. Note that the results in \eq{op_tree} are the same whether we use a state with fixed spin and color or whether we average over spin and color. Taking the matrix element of both sides of \eq{op_OPE} and using \eq{op_tree}, we can read off the tree-level matching coefficient
\begin{equation}
\cI_{qq}^\zero(t,z,\mu) = \cI_{\bar{q}\bar{q}}^\zero(t,z,\mu) = \delta(t)\, \delta(1-z)
\,.\end{equation}

Similarly, the tree-level results for the gluon PDF and beam function are
\begin{align} \label{eq:opg_tree}
\Mae{g_n(p)}{\oq_g(\w',\mu)}{g_n(p)}^\zero
&= \theta(\w')\,\delta(1-\w'/p^-)
\,, \nn \\
\Mae{g_n(p)}{\op_g(t,\w,\mu)}{g_n(p)}^\zero
&= -\w\, \varepsilon^*\sdt \varepsilon \, \delta(t)\, \delta(\w-p^-)
= \delta(t)\, \delta(1-\w/p^-)
\,,\end{align}
leading to
\begin{equation}
\cI_{gg}^\zero(t,z,\mu) = \delta(t)\, \delta(1-z)
\,.\end{equation}
Finally, since at tree level the quark (gluon) matrix elements of the gluon
(quark) operators vanish,
\begin{align} \label{eq:qg_tree}
\Mae{g_n(p)}{\oq_q(\w',\mu)}{g_n(p)}^\zero  &= \Mae{q_n(p)}{\oq_g(\w',\mu)}{q_n(p)}^\zero = 0
\,,\nn\\
\Mae{g_n(p)}{\op_q(t,\w,\mu)}{g_n(p)}^\zero &= \Mae{q_n(p)}{\op_g(t,\w,\mu)}{q_n(p)}^\zero = 0
\,,\end{align}
we obtain
\begin{equation}
\cI_{qg}^\zero(t,z,\mu) = \cI_{gq}^\zero(t,z,\mu) = 0
\,.\end{equation}
To summarize, the complete tree-level results are
\begin{equation}
\cI_{ij}^\zero(t,z,\mu) = \delta_{ij} \delta(t)\, \delta(1-z)
\,,\qquad
B_i^\zero (t, x, \mu) = \delta(t) f_i(x, \mu)
\,.\end{equation}
The interpretation is simply that at tree level the parton taken out of the
proton goes straight into the hard interaction. However, even at tree level the
OPE already provides nontrivial information. From our general discussion we know
that the matching should be performed at the beam scale $\mu_B^2 \simeq t$ to avoid
large logarithms in the $\ord{\alpha_s}$ terms, and this determines the scale at
which the PDFs must be evaluated to be $\mu = \mu_B$.

\subsection{Analytic Structure and Time-Ordered Products}
\label{subsec:beamT}

In this subsection we discuss the analytic structure of the beam functions. For the OPE matching calculation we want to calculate partonic matrix elements of $\op_q(t, \w, \mu)$. For this purpose it is convenient to relate the matrix elements of the products of fields in $\op_q(t, \w, \mu)$ to discontinuities of matrix elements of time-ordered products of fields, since the latter are easily evaluated using standard Feynman rules. For notational simplicity we only consider the quark operator $\op_q(t, \w)$ and suppress the spin indices and $\mu$ dependence. The discussion for the antiquark and gluon operators are analogous.

We are interested in the forward matrix element of $\op_q(t, \w)$ between some $n$-collinear state $\ket{p_n} \equiv \ket{p_n(p^+, p^-)}$ with large momentum $p^-$ and small residual momentum $p^+$. Inserting a complete set of states $\sum_X \ket{X}\bra{X}$, we get
\begin{align} \label{eq:Bkin}
\Mae{p_n}{\op_q(t, \w)}{p_n}
&= \sum_X \MAe{p_n}{\bar\chi_n(0) \frac{\bnslash}{2}\,\delta(t - \abs{\w}\hp^+)}{X}
\Mae{X}{\bigl[\delta(\w - \bnP_n) \chi_n(0)\bigr]}{p_n}
\\\nn
&= \sum_X \delta(t - \abs{\w} p_X^+)\, \delta(\w - p^- + p_X^-)
   \MAe{p_n}{\bar\chi_n(0) \frac{\bnslash}{2}}{X} \Mae{X}{\chi_n(0)}{p_n}
\,.\end{align}
The $\delta(\w-\bnP_n)$ by definition only acts on the field inside the square bracket, returning its minus momentum, which by momentum conservation must be equal to the difference of the minus momenta of the external states. Since $\w = p^- - p_X^-$, requiring $\w > 0$ implies $p_X^- < p^-$. This means that the action of the field reduces the momentum of the initial state so it effectively annihilates a parton in the initial state $\ket{p_n}$. Similarly, for $\w < 0$ we would have $p_X^- > p^-$ and the field would effectively create an antiquark in $\bra{X}$. Also, since $\ket{X}$ are physical states, we have $p_X^\pm \geq 0$ so $\w \leq p^-$ and $t = \abs{\w} p_X^+ \geq 0$.

Hence, for the beam function, where $\ket{p_n}\equiv\ket{p_n(P^-)}$ is the proton state, the restriction to $\w > 0$ in its definition, \eq{B_def}, enforces that we indeed take a quark out of the proton. (Note that $\w<0$ does not correspond to the anti-quark beam function.) Taking the states $\ket{X}$ to be a complete set of physical intermediate states, the beam function has the physical support
\begin{equation}
0 < x < 1 - \frac{p_{X\mathrm{min}}^-}{P^-} < 1
\,,\qquad
t > \w\, p_{X\mathrm{min}}^+ > 0
\,,\end{equation}
where $p_{X\min}^\pm > 0$ are the smallest possible momenta (which are strictly
positive because with an incoming proton $\ket{X}$ can neither be massless nor
the vacuum state).  For the jet function the external state is the vacuum
$\ket{p_n} = \ket{0}$ yielding $\delta(\w + p_X^-)$, so the matrix element in
\eq{Bkin} vanishes for $\w>0$.

Next, consider the following time-ordered analog of $\Mae{p_n}{\theta(\w)\op_q(t, \w)}{p_n}$,
\begin{align}
\Mae{p_n}{T_q(\w b^+, \w)}{p_n}
&\equiv \frac{\theta(\w)}{2\pi} \int\! \frac{\df y^-}{2\w}\, e^{\img (b^+ - p^+) y^-/2}
   \MAe{p_n}{T\Bigl\{\bar\chi_n\Bigl(y^-\frac{n}{2}\Bigr) \frac{\bnslash}{2} \bigl[\delta(\w-\bnP_n) \chi_n(0)\bigr]\Bigr\}}{p_n}
\,.\end{align}
Writing out the time-ordering,
\begin{align}
&T\Bigl\{\bar\chi_n\Bigl(y^-\frac{n}{2}\Bigr) \frac{\bnslash}{2} \bigl[\delta(\w-\bnP_n) \chi_n(0)\bigr]\Bigr\}
\nn\\ & \qquad
= \theta(y^-) \bar\chi_n\Bigl(y^-\frac{n}{2}\Bigr) \frac{\bnslash}{2} \bigl[\delta(\w-\bnP_n) \chi_n(0)\bigr]
 -\theta(-y^-) \bigl[\delta(\w-\bnP_n) \chi_n(0)\bigr] \bar\chi_n\Bigl(y^-\frac{n}{2}\Bigr) \frac{\bnslash}{2}
\,,\end{align}
using
\begin{equation}
\theta(\pm y^-) = \frac{\img}{2\pi} \int\!\df\kappa\, \frac{e^{\mp \img\kappa y^-}}{\kappa + \img 0}
\,,\end{equation}
inserting a complete set of states, and translating the fields to spacetime position zero, we arrive at
\begin{align} \label{eq:Tresult}
&\Mae{p_n}{T_q(\w b^+,\w)}{p_n}
\nn\\ & \quad
=\frac{\img \theta(\w)}{(2\pi)^2} \int\!\frac{\df y^-}{2\w}\, \frac{\df \kappa}{\kappa+\img 0}
\sum_X \biggl[ e^{\img(b^+ - p_X^+ -\kappa)y^-/2}\, \delta(\w - p^- + p_X^-)\,
\MAe{p_n}{\bar\chi_n(0)\frac{\bnslash}{2}}{X} \Mae{X}{\chi_n(0)}{p_n}
\nn\\ &\qquad
  + e^{\img(b^+ + p_X^+ + \kappa)y^-/2}\, \delta(\w + p^- - p_X^-)\,
  \Mae{p_n}{\chi_n(0)}{X}\MAe{X}{\bar \chi_n(0)\frac{\bnslash}{2}}{p_n} \biggr]
\nn\\ & \quad
= \frac{\img \theta(\w)}{2\pi \w} \sum_X \biggl[
  \frac{\delta(\w - p^- + p_X^-)}{b^+ - p_X^+ + \img 0}
  \MAe{p_n}{\bar\chi_n(0)\frac{\bnslash}{2}}{X} \Mae{X}{\chi_n(0)}{p_n}
\nn\\ & \qquad\qquad\qquad\quad
- \frac{\delta(\w + p^- - p_X^-)}{b^+ + p_X^+ - \img 0}
  \Mae{p_n}{\chi_n(0)}{X}\MAe{X}{\bar \chi_n(0)\frac{\bnslash}{2}}{p_n} \biggr]
\,.\end{align}
The first term creates a cut in the complex $b^+$ plane for $b^+ \geq p_{X\mathrm{min}}^+$. This cut is shown as the dark red line in \fig{Bcuts}. The second term produces a cut at $b^+ \leq - p_{X\mathrm{min}}^+$, shown as the light blue line in \fig{Bcuts}.

\FIGURE[t]{%
\parbox{\columnwidth}{\centering
\includegraphics[scale=0.85]{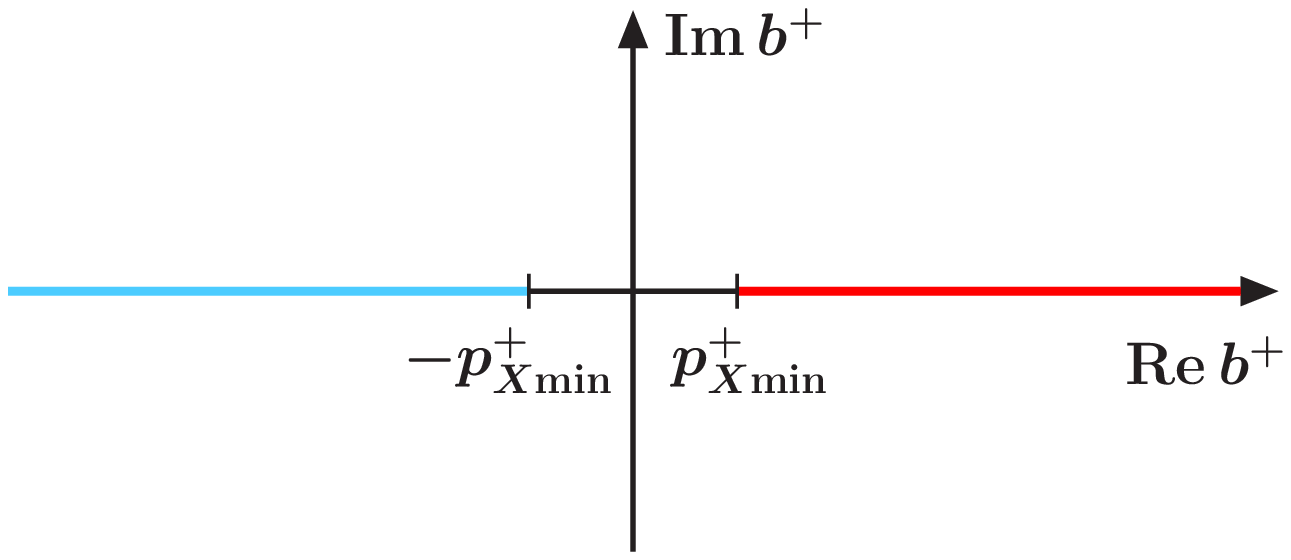}}
\caption{Cuts in the complex $b^+$ plane for the time-ordered product in \eq{Tresult}.}
\label{fig:Bcuts}}

The beam function matrix element in \eq{Bkin} can be identified as precisely the discontinuity of the first term in \eq{Tresult} with respect to $b^+$. Thus, for the beam function we have
\begin{equation}
B_q(\w b^+, \w) = \Disc_{b^+>0}\, \Mae{p_n(P^-)}{T_q(\w b^+, \w)}{p_n(P^-)}
\,.\end{equation}
Taking the discontinuity only for $b^+>0$ ensures that we only pick out the cut due to the first term in \eq{Tresult}. Here, the discontinuity of a function $g(x)$ for $x>x_0$ is defined as
\begin{align} \label{eq:Disc_def}
\Disc_{x>x_0}\, g(x) = \lim_{\beta\to 0} \theta(x - x_0) \bigl[ g(x + \img\beta) - g(x - \img\beta) \bigr]
\,,\end{align}
and we used \eq{disc_os} to take the discontinuity of $1/(b^+ - p_X^+)$,
\begin{equation}
\Disc_{b^+>0}\,\frac{\img}{2\pi\abs{\w}}\,\frac{1}{b^+ - p_X^+} = \frac{1}{\abs{\w}}\,\delta(b^+ - p_X^+) = \delta(\w b^+ - \w p_X^+)
\,.\end{equation}
Since we explicitly specify how to take the discontinuity, we can drop the $\img 0$ prescription in the denominators. (Alternatively, we could multiply by $\img$ and take the imaginary part using the $\img 0$ prescription.) Since $b^+$ and $t = \abs{\w}b^+$ always have the same sign we can also take the discontinuity for $t > 0$, so
\begin{equation} \label{eq:DiscTB}
\Mae{p_n}{\theta(\w)\op_q(t, \w)}{p_n} = \Disc_{t > 0}\, \Mae{p_n}{T_q(t, \w)}{p_n}
\,.\end{equation}

For the matching calculation $\ket{p_n}$ is a partonic quark or gluon state. For any contributions with real radiation in the intermediate state, i.e.\ diagrams where the two $\chi_n$ or $\cB_{n\perp}$ fields in the operator $\op_i$ are joined by a series of propagators and vertices, we can use the standard Feynman rules to evaluate the time-ordered matrix element of $T_q(t, \w)$. However, with partonic external states, we can also have the vacuum state as an intermediate state, because the fields in the operator are spacetime separated. For such purely virtual contributions it is simpler to directly start from $\op_q(t, \w)$, insert the vacuum state between the fields, and then use standard Feynman rules to separately compute the two pieces $\Mae{p_n}{\bar\chi_n(0)\bnslash/2}{0}$ and $\Mae{0}{\chi_n(0)}{p_n}$. In fact, this is exactly what we already did in our tree-level calculation in \subsec{tree_match}, and we will see another example in \sec{oneloop}. Thus, we will obtain the total partonic matrix element as
\begin{align} \label{eq:DiscTB_full}
\Mae{p_n}{\theta(\w)\op_q(t, \w)}{p_n}
&= \Mae{p_n}{\theta(\w)\op_q(t, \w)}{p_n}_\mathrm{virtual} + \Mae{p_n}{\theta(\w)\op_q(t, \w)}{p_n}_\mathrm{radiation}
\nn\\
&= \delta(t)\,\delta(\w - p^-)\, \MAe{p_n}{\bar\chi_n(0)\frac{\bnslash}{2}}{0}_\mathrm{connected} \Mae{0}{\chi_n(0)}{p_n}_\mathrm{connected}
\nn\\ & \quad
+ \Disc_{t > 0}\, \Mae{p_n}{T_q(t, \w)}{p_n}_\mathrm{connected}
\,.\end{align}
The virtual contribution must be kept, since it only looks superficially disconnected because the operator itself is spacetime separated. As always, we still disregard genuinely disconnected diagrams, e.g.\ diagrams involving vacuum bubbles, when calculating the matrix elements in the second line.

\section{NLO Calculation of the Quark Beam Function}
\label{sec:oneloop}

In this section, we compute the matching coefficients $\cI_{qq}(t, z, \mu)$ and $\cI_{qg}(t, z, \mu)$ in the OPE for the quark beam function in \eq{beam_fact} to next-to-leading order in $\alpha_s(\mu)$. As explained in \subsec{OPE} and \subsec{tree_match}, this can be done by computing the partonic matrix elements of both sides of \eq{op_OPE} to NLO. We use the same $n$-collinear quark and gluon states, $\ket{q_n}\equiv \ket{q_n(p)}$ and $\ket{g_n}\equiv\ket{g_n(p)}$, as in the tree-level matching in \subsec{tree_match}, with momentum $p^\mu = (p^+,p^-,0)$. Since only $\cI_{qq}(t, z, \mu)$ is nonzero at leading order, we will only need the NLO matrix elements of the quark operators, $\op_q(t, \w, \mu)$ and $\oq_q(\w, \mu)$. We write the results for all matrix elements in terms of the RPI-III invariant variables (in this section we will always have $\w > 0$)
\begin{equation} \label{eq:defA}
t = \w b^+
\,,\qquad
t' = -\w p^+ = - zp^+p^-
\,,\qquad
z=\frac{\w}{p^-}
\,.\end{equation}
Here, $z$ is the partonic momentum fraction of the quark annihilated by the operator relative to the momentum of the incoming quark or gluon, and will coincide with the argument of $\cI_{ij}(t, z, \mu)$.

To regulate the UV we use dimensional regularization with $d=4-2\eps$ dimensions and renormalize using the $\overline{\text{MS}}$ scheme. Since the matching coefficients in the OPE must be IR finite, the matrix elements of $\op_q$ and $\oq_q$ must have the same IR divergences, i.e., the beam function must contain the same IR divergences as the PDF. To explicitly check that this is the case, we separate the UV and IR divergences by regulating the IR with a small $p^+ < 0$. This forces the external states to have a small offshellness $p^+ p^- < 0$, and since $p^+p^- = -t'/z$ the IR divergences will appear as $\ln t'$. This also allows us to directly obtain the one-loop renormalization constants and anomalous dimensions for $\op_q$ and $\oq_q$ from their one-loop matrix elements.

We first compute the renormalized one-loop matrix elements of the quark PDF operator $\oq_q$ in \subsec{NLO_PDF}. This calculation of the PDF for general $x$ using the SCET operator definition and with an offshellness IR regulator is quite instructive, both by itself and in comparison to the beam function calculation, which is why we give it in some detail. In \subsec{NLO_B}, we compute the renormalized one-loop matrix elements of the quark beam function operator $\op_q$. Finally in \subsec{NLO_matching}, we use these results to extract expressions for $\cI_{qq}(t, z, \mu)$ and $\cI_{qg}(t, z, \mu)$ valid to NLO.

Assuming that the IR divergences in the beam function and PDF will cancel, the matching calculation can be performed more easily using dimensional regularization for both UV and IR. We do this as an illustrative exercise in \app{dimreg}, which, as it should, yields the same result for the matching coefficients.

\subsection{PDF with Offshellness Infrared Regulator}
\label{subsec:NLO_PDF}

\begin{figure}
\subfigure[]{\includegraphics[scale=0.75]{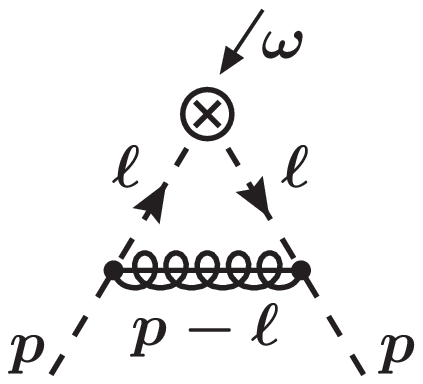}\label{fig:fone_a}}%
\hfill%
\subfigure[]{\includegraphics[scale=0.75]{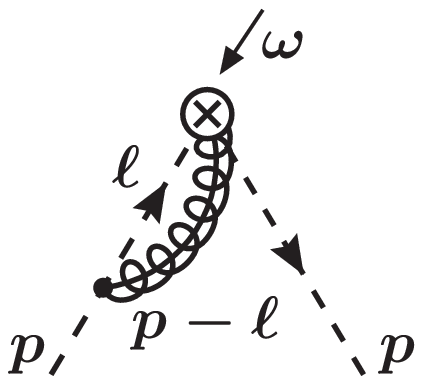}\label{fig:fone_b}}%
\hfill%
\subfigure[]{\includegraphics[scale=0.75]{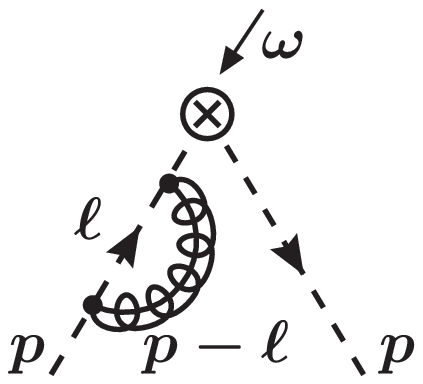}\label{fig:fone_c}}%
\hfill%
\subfigure[]{\includegraphics[scale=0.75]{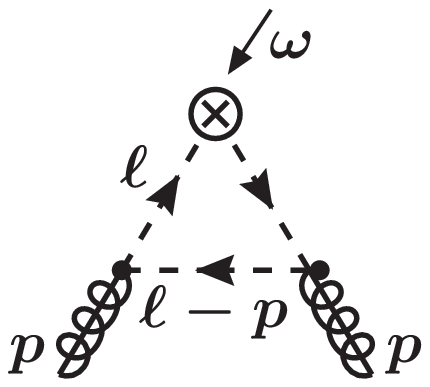}\label{fig:fone_d}}%
\caption{Nonzero one-loop diagrams for the quark PDF. The minus momentum $\w$ enters the vertex through its outgoing fermion line and leaves through its incoming fermion line. Diagram (c) represents the inclusion of the wave-function renormalization constant for the renormalized fields together with the corresponding
residue  factor in the LSZ formula for the $S$-matrix. Diagrams (b) and (c) have symmetric counterparts which are included in their computation.}
\label{fig:fone}
\end{figure}

We start by calculating the bare $S$-matrix elements
\begin{equation}
\Mae{q_n(p)}{\oq_q^\bare(\w)}{q_n(p)}
\,,\qquad
\Mae{g_n(p)}{\oq_q^\bare(\w)}{g_n(p)}
\,,\end{equation}
using Feynman gauge to compute the gauge-invariant sum of all diagrams. The relevant one-loop diagrams are shown in \fig{fone}. Since $\oq_q$ is a local SCET operator, we can use the usual time-ordered Feynman rules in SCET (without any of the complications discussed in \subsec{beamT} for $\op_q$). The collinear $q_n q_n g_n$ vertex factor is
\begin{equation}
\img g\, T^a V_n^\mu(p, \ell)\,\frac{\bnslash}{2}
\qquad\text{with}\qquad
V_n^\mu(p, \ell)
= n^\mu + \frac{\pslash_\perp\gamma_\perp^\mu}{p^-} + \frac{\gamma_\perp^\mu \ellslash_\perp}{\ell^-}
- \frac{\pslash_\perp\ellslash_\perp}{p^-\ell^-} \bn^\mu
\,,\end{equation}
where $p^\mu$ and $\ell^\mu$ are the label momenta of the outgoing and incoming quark lines.
(Because we have a single collinear direction the computation can also be done with QCD Feynman rules, still accounting for zero-bin subtractions, with the only difference being the Dirac algebra in the numerator of the loop integral. We checked that the final results for each diagram are indeed the same either way.)

The diagram in \fig{fone_a} is
\begin{equation}
 \Mae{q_n}{\oq_q^\bare(\w)}{q_n}^{(a)}
\!= -\img \Bigl(\frac{e^{\gamma_E} \mu^2}{4\pi}\Bigr)^{\eps} g^2 C_F\! \int\! \frac{\df^d\ell}{(2\pi)^d}
 \frac{
\bar u_n(p) V_n^\mu(p, \ell) V_{n\mu}(\ell, p) \frac{\bnslash}{2} u_n(p) (\ell^-)^2
}{(\ell^2+\img 0)^2 [(\ell-p)^2+\img 0] } \,
 \delta(\ell^-\! -\w)
\,,\end{equation}
where $g \equiv g(\mu)$ is the renormalized $\overline{\textrm{MS}}$ coupling.
The Dirac algebra for the numerator gives
\begin{equation}
\bar u_n(p) V_n^\mu(p, \ell) V_{n\mu}(\ell, p) \frac{\bnslash}{2} u_n(p) (\ell^-)^2
= \bar u_n(p) \gamma_\perp^\mu \ellslash_\perp \ellslash_\perp \gamma_{\perp\mu} \frac{\bnslash}{2} u_n(p)
= p^- (d-2) \ell_\perp^2
\,.\end{equation}
To compute the loop integral we write $\df^d\ell = \df\ell^+ \df\ell^- \df^{d-2} \vec\ell_\perp/2$, where $\vec\ell_\perp$ is Euclidean, so $\ell_\perp^2 = -\vec\ell_\perp^2$. The $\ell^+$ integral is done by contour integration as follows. For $\ell^-<0$ all poles are above the axis and for $\ell^- > p^-$ all poles are below the axis, so both cases give zero. Hence, the $\ell^-$ integration range is restricted to $0 < \ell^- < p^-$, where there is a double pole below the axis from the $1/(\ell^2 + \img 0)^2$ and a single pole above the axis from the $1/[(\ell - p)^2 + \img 0]$. Taking the single pole above amounts to replacing the second denominator by $2\pi\img/(\ell^-\! - p^-)$ and setting
\begin{align} \label{eq:ellpole}
\ell^+ = p^+ - \frac{\vec \ell_\perp^2}{p^-\! -\ell^-}
\end{align}
everywhere else. After performing the contour integral the $\img 0$ have served their purpose and can be set to zero everywhere. The $\ell^-$ integral is trivial using the $\delta(\ell^-\! - \w)$ and turns the $\ell^-$ limits into an overall $\theta(\w)\theta(p^-\! - \w)$. The remaining $\vec\ell_\perp$ integration is done in $d - 2 = 2(1 - \eps)$ Euclidean dimensions as usual. Putting everything together, we obtain
\begin{align} \label{eq:PDFa1}
&\Mae{q_n}{\oq_q^\bare(\w)}{q_n}^{(a)}
\nn\\ & \qquad
= \Bigl(\frac{e^{\gamma_E} \mu^2}{4\pi}\Bigr)^{\eps} g^2 C_F\, \theta(\w) \theta(p^-\! - \w)\, \frac{(d - 2) (p^-\! - \w)}{4\pi\, p^-} \!\int\! \frac{\df^{d-2}\vec\ell_\perp}{(2\pi)^{d-2}}\,
\frac{\vec\ell_\perp^2\, }{[\vec\ell_\perp^{2} + (1-z)t']^2 }
\nn\\ &\qquad
= \frac{\alpha_s(\mu)C_F}{2\pi}\,
  \theta(z) \theta(1-z)\, \Gamma(\eps) \Bigl(\frac{e^{\gamma_E}\mu^2}{t'}\Bigr)^{\eps} (1-z)^{1-\eps} (1-\eps)^2
\nn \\ &\qquad
= \frac{\alpha_s(\mu) C_F}{2\pi} \,\theta(z) \theta(1-z)\,(1-z) \biggl\{ \frac{1}{\eps} - \ln\frac{t'}{\mu^2} - \ln(1-z) -2 \biggr\}
\,,\end{align}
where in the last line we expanded in $\eps$.

In the diagram in \fig{fone_b}, the gluon is annihilated by the Wilson line inside one of the $\chi_n$ fields. The contraction with the one in $\bar{\chi}_n$ is $\propto\delta(\ell^-\! - \w)$ and the contraction with the one in $\chi_n$ is $\propto\delta(p^-\! - \w)$. The $1/\bnP_n$ in the Wilson lines [see \eq{Wn}] contributes a factor $1/(\ell^-\!-p^-)$ with a relative minus sign between the two contractions. (There is also a diagram where the gluon connects both Wilson lines which vanishes because the Wilson lines only contain $\bn \sdt A$ gluons and we use Feynman gauge.) Adding \fig{fone_b} and its mirror graph, which gives an identical contribution, we get
\begin{align} \label{eq:PDFb1}
& \Mae{q_n}{\oq_q^\bare(\w)}{q_n}^{(b)}\,
  \nn \\ & \quad
  = 2\img \Bigl(\frac{e^{\ga_E} \mu^2}{4\pi}\Bigr)^{\eps} g^2 C_F \int\! \frac{\df^d\ell}{(2\pi)^d}\,
  \frac{
\bn_\mu \bar u_n(p) V_n^\mu \frac{\bnslash}{2} u_n(p) \ell^-
}{(\ell^-\! - p^-)(\ell^2 + \img 0) [(\ell-p)^2 + \img 0] } \,
  \bigl[\delta(\ell^-\! -\w) - \delta(p^- \! -\w) \bigr]
\nn\\ & \quad
 = \frac{\alpha_s(\mu) C_F }{\pi}\,\Gamma(\eps)\biggl(\frac{e^{\gamma_E} \mu^2}{-p^+p^-} \biggr)^\eps
\! \int\! \df\ell^-\,\theta(\ell^-)\,\theta(p^-\!-\ell^-) \frac{(\ell^-/p^-)^{1-\eps}}{(1 -\ell^-/p^- )^{1+\eps} }
  \bigl[\delta(\ell^-\! -\w) - \delta(p^- \! -\w) \bigr]
\nn\\ & \quad
 =\frac{\alpha_s(\mu) C_F}{\pi}\,
 \Gamma(\eps) \Bigl(\frac{e^{\gamma_E}\mu^2}{t'}\Bigr)^\eps
\biggl\{ \frac{\theta(z)\theta(1-z) z}{(1-z)^{1+\eps} } - \delta(1 -z ) \, \frac{\Gamma(2-\eps)\Gamma(-\eps)}{\Gamma(2-2\eps)}
  \biggr\}
\,.\end{align}
In the first step we used $\bn_\mu V_n^\mu = 2$ and $\bar u_n(p) \bnslash u_n(p) = 2p^-$, performed the $\ell^+$ integral by contours and did the $\vec\ell_\perp$ integral as usual. The $\ell^+$ integral has the same pole structure as in \fig{fone_a} (except that the double pole at $\ell^+ = 0$ is now a single pole), which restricts the $\ell^-$ integral to the finite range $0 < \ell^- < p^-$. Expanding \eq{PDFb1} in $\eps$, using the distribution identity in \eq{distr_id}, we get
\begin{align} \label{eq:PDFb2}
\Mae{q_n}{\oq_q^\bare(\w)}{q_n}^{(b)}
&=\frac{\alpha_s(\mu) C_F}{\pi}\, \Gamma(\eps)\Bigl(\frac{e^{\gamma_E}\mu^2}{t'}\Bigr)^\eps
\biggl\{ \theta(z)
 \biggl[-\frac{1}{\eps}\,\delta(1-z) + \cL_0(1-z)z -\eps \cL_1(1-z)z\biggr]
\nn\\ &\quad\hspace{25ex}
 +\delta(1 -z ) \biggl[\frac{1}{\eps} + 1 + \eps\Bigl(2-\frac{\pi^2}{6}\Bigr) \biggr]
  \biggr\} \,
\nn\\
&=\frac{\alpha_s(\mu) C_F}{\pi}\, \theta(z) \biggl\{
  \Bigl(\frac{1}{\eps} - \ln\frac{t'}{\mu^2} \Bigr) \bigl[\cL_0(1-z)z + \delta(1-z)  \bigr] - \cL_1(1-z)z
\nn\\* & \quad\hspace{15ex}
+ \delta(1 -z) \Bigl(2-\frac{\pi^2}{6} \Bigr)
  \biggr\}
\,,\end{align}
where $\cL_n(x) = [\theta(x)(\ln^n x)/x]_+$ are the usual plus distributions defined in \eq{plusdef}.

In the last step in \eq{PDFb1}, the $\ell^-$ integral produces an additional $1/\eps$ pole in each of the two terms corresponding to real and virtual radiation from the two different Wilson line contractions. It comes from the singularity at $\ell^- = p^-$, where the gluon in the loop becomes soft. (This soft IR divergence appears as a pole in $\eps$ because the offshellness only regulates the collinear IR divergence here.) The soft IR divergences cancel in the sum of the virtual and real contributions, as can be seen explicitly in the first line of \eq{PDFb2} where the $1/\eps$ poles in curly brackets cancel between the two terms. One can already see this in the $\ell^-$ integral in \eq{PDFb1}, because for $\ell^- = p^-$ the two $\delta$ functions cancel so there is no soft divergence in the total integral. Thus, in agreement with our discussion in \subsec{definition}, we explicitly see that contributions from the soft region drop out in the PDF. As a consequence, the PDF only contains a single $1/\eps$ pole and correspondingly its RGE will sum single logarithms associated with this purely collinear IR divergence.

Since the gluon in the loop is supposed to be collinear, the soft gluon region must be explicitly removed from the collinear loop integral, which is the condition $\lp \neq 0$ in \eq{xi}. For continuous loop momenta this is achieved by a zero-bin subtraction. However, since the soft region does not contribute to the PDF, it also does not require zero-bin subtractions in SCET. (If we were to include separate zero-bin subtractions for the virtual and real contributions, they would simply cancel each other.) We will see shortly that the situation for the beam function is quite different.

The last diagram with external quarks, \fig{fone_c}, is
\begin{align}  \label{eq:PDFc1}
\Mae{q_n}{\oq_q^\bare(\w)}{q_n}^{(c)} &= \delta(1-z) (Z_\xi - 1)
  = -\frac{\alpha_s(\mu) C_F}{4\pi}\, \delta(1-z) \biggl\{\frac{1}{\eps} - \ln\frac{t'}{\mu^2} + 1 \biggr\}\,
\,.\end{align}
Here we used the result for the one-loop on-shell wave-function renormalization with an offshellness IR regulator, which is the same in SCET and QCD.

Adding up the results in Eqs.~\eqref{eq:PDFa1}, \eqref{eq:PDFb2}, and \eqref{eq:PDFc1} we obtain for the bare one-loop quark matrix element
\begin{align} \label{eq:fqq_bare}
\Mae{q_n}{\oq_q^\bare(\w)}{q_n}^\one
&= \frac{\alpha_s(\mu) C_F}{2\pi}\,  \theta(z) \biggl\{
\Bigl(\frac{1}{\eps} - \ln\frac{t'}{\mu^2} \Bigr) P_{qq}(z) - \cL_1(1-z)(1 + z^2)
\nn\\ &\quad\hspace{15ex}
+ \delta(1- z ) \Bigl(\frac72 -\frac{\pi^2}{3} \Bigr) - \theta(1-z) 2 (1-z) \biggr\}
\,,\end{align}
where
\begin{equation}
P_{qq}(z) = \cL_0(1-z)(1 + z^2) + \frac{3}{2}\, \delta(1-z) = \biggl[\theta(1-z)\frac{1+z^2}{1-z} \biggr]_+
\end{equation}
is the $q\to qg$ splitting function, see \eq{Pqq_def}.

Next, we consider the matrix element of $\oq_q$ between gluon states $\ket{g_n} \equiv \ket{g_n(p)}$. The only relevant diagram is shown in
\fig{fone_d},
\begin{equation}
\Mae{g_n}{\oq_q^\bare(\w)}{g_n}^{(d)}
\!= \img\Bigl(\frac{ e^{\gamma_E} \mu^2}{4\pi}\Bigr)^\eps g^2 T_F\!
 \int\! \frac{\df^d \ell}{(2\pi)^d} \,
 \frac{(-\varepsilon_\mu^* \varepsilon_\nu)
\tr\bigl[V_n^\mu V_n^{\nu} \frac{\bnslash\nslash}{4}\bigr]
 (\ell^-)^2(\ell^-\! - p^-)}{(\ell^2+\img 0)^2 [(\ell-p)^2+\img 0] } \, \delta(\ell^-\! -\w)
\,.\end{equation}
Here $\varepsilon \equiv \varepsilon(p)$, $V_n^\mu \equiv V_n^\mu(\ell-p, \ell)$  and $V_n^\nu\equiv V_n^\nu(\ell, \ell-p)$. Since the physical polarization vector is perpendicular, $n\cdot\varepsilon(p) = \bn\cdot\varepsilon(p) = 0$, we only need the perpendicular parts of the collinear vertices. The numerator then becomes
\begin{align}
\tr\Bigl[V_n^\mu V_n^{\nu} \frac{\bnslash\nslash}{4} \Bigr]
 (\ell^-)^2(\ell^-\! - p^-)
&= \frac{1}{2}\tr\biggl[
\biggl( \frac{\ellslash_\perp \gamma_\perp^\mu}{\ell^- - p^-} + \frac{\gamma_\perp^\mu \ellslash_\perp}{\ell^-} \biggr)
\biggl( \frac{\ellslash_\perp \gamma_\perp^\nu}{\ell^-} + \frac{\gamma_\perp^\nu \ellslash_\perp}{\ell^- - p^-} \biggr)
\biggr] (\ell^-)^2(\ell^-\! - p^-)
\nn\\
&= 2\frac{(p^-)^2}{\ell^- - p^-}\,\ell_\perp^2 g_\perp^{\mu\nu} + 8 \ell^- \ell_\perp^\mu \ell_\perp^\nu
= 2 g_\perp^{\mu\nu} p^-\Bigl(\frac{1}{1-z} - \frac{4z}{d - 2}\Bigr) \vec\ell_\perp^2
\,.\end{align}
In the last step we used that under the integral we can replace $\ell^- = \w = z p^-$ and $\ell_\perp^\mu \ell_\perp^\nu = \ell_\perp^2 g_\perp^{\mu\nu}/(d-2)$. The remaining loop integral is exactly the same as in \fig{fone_a}, so the bare one-loop gluon matrix element becomes
\begin{align} \label{eq:fqg_bare}
\Mae{g_n}{\oq_q^\bare(\w)}{g_n}^\one
&= \frac{\alpha_s(\mu) T_F}{2\pi}\, \theta(z)\,\theta(1 - z)
\Gamma(\eps) \Bigl(\frac{e^{\gamma_E}\mu^2}{t'}\Bigr)^\eps (1-z)^{-\eps}(1- 2z + 2z^2 - \eps)
\nn\\
&= \frac{\alpha_s(\mu) T_F}{2\pi}\,\theta(z)
\biggl\{ \Bigl[\frac{1}{\eps} - \ln\frac{t'}{\mu^2} - \ln(1-z) \Bigr] P_{qg}(z) - \theta(1 - z) \biggr\}
\,.\end{align}
Here
\begin{equation}
P_{qg}(z) = \theta(1-z)\,(1 - 2z + 2z^2)
\end{equation}
is the $g\to q\bar{q}$ splitting function from \eq{Pqq_def}.

Note that the diagram analogous to \fig{fone_d} with the two gluons crossed can be obtained from \fig{fone_d} by taking $p^\mu \to -p^\mu$, which takes $z\to -z$. The limits resulting from the $\ell^+$ integral are then $-1\leq z\le 0$ or $-p^- < \w < 0$, and since we require $\w > 0$ for $\oq_q$, this diagram does not contribute. The diagram involving the SCET vertex with two collinear gluons vanishes because here the $\ell^+$ integral does not have poles on both sides of the axis.

From the bare matrix elements in \eqs{fqq_bare}{fqg_bare} we can obtain the renormalization of $\oq_q$. Taking the quark and gluon matrix elements of \eq{oq_ren} and expanding to NLO,
\begin{align} \label{eq:fbare2}
& \Mae{q_n}{\oq_q^\bare(\w)}{q_n}^\one
\nn \\ & \quad
= \sum_j \int\! \frac{\df \w'}{\w'}\,
\biggl[Z^{f\one}_{qj}\Bigl(\frac{\w}{\w'}, \mu \Bigr) \Mae{q_n}{\oq_j(\w',\mu)}{q_n}^\zero +
Z^{f\zero}_{qj}\Bigl(\frac{\w}{\w'}, \mu\Bigr) \Mae{q_n}{\oq_j(\w',\mu)}{q_n}^\one \biggr]
\nn \\ & \quad
= Z_{qq}^{f\one}(z, \mu) + \Mae{q_n}{\oq_q(\w,\mu)}{q_n}^\one
\,, \nn \\
&\Mae{g_n}{\oq_q^\bare(\w)}{g_n}^\one
\nn \\ & \quad
= \sum_j \int\! \frac{\df \w'}{\w'}\,
\biggl[Z^{f\one}_{qj}\Bigl(\frac{\w}{\w'}, \mu\Bigr) \Mae{g_n}{\oq_j(\w',\mu)}{g_n}^\zero +
Z^{f\zero}_{qj}\Bigl(\frac{\w}{\w'}, \mu\Bigr) \Mae{g_n}{\oq_j(\w',\mu)}{g_n}^\one \biggr]
\nn \\ & \quad
= Z_{qg}^{f\one}(z, \mu) + \Mae{g_n}{\oq_q(\w,\mu)}{g_n}^\one
\,,\end{align}
where we used the tree-level matrix elements in \eqs{op_tree}{qg_tree} and $Z^{f\zero}_{ij}(z, \mu) = \delta_{ij}\,\delta(1 - z)$. The $\overline{\mathrm{MS}}$ counter terms required to cancel the $1/\eps$ poles in the bare PDF matrix elements are then
\begin{equation} \label{eq:Zpdf}
  Z_{qq}^f(z) = \delta(1-z) + \frac{1}{\eps}\, \frac{\alpha_s(\mu) C_F}{2\pi}\,
  \theta(z) P_{qq}(z)
  \,,\qquad
  Z_{qg}^f(z) = \frac{1}{\eps}\, \frac{\alpha_s(\mu) T_F}{2\pi}\,
  \theta(z) P_{qg}(z)
 \,.\end{equation}
Expanding \eq{oq_RGE} to NLO, the one-loop anomalous dimensions are obtained by
\begin{equation}
\gamma_{ij}^f(z, \mu) = -\mu \frac{\df}{\df\mu} Z^{f\one}_{ij}(z, \mu)
\,,\qquad
\mu \frac{\df}{\df\mu} \alpha_s(\mu) = -2\eps\, \alpha_s(\mu) + \beta[\alpha_s(\mu)]
\,,\end{equation}
which with \eq{Zpdf} yields the anomalous dimension for the quark PDF in \eq{gammaf},
\begin{equation}
  \gamma_{qq}^f(z, \mu) = \frac{\alpha_s(\mu) C_F}{\pi}\, \theta(z) P_{qq}(z)
\,,\qquad
 \gamma_{qg}^f(z, \mu) = \frac{\alpha_s(\mu)T_F}{\pi}\,\theta(z) P_{qg}(z)
\,.\end{equation}
Finally, the renormalized NLO PDF matrix elements, which we will need for the matching computation in \subsec{NLO_matching} below, are
\begin{align} \label{eq:fren}
\Mae{q_n}{\oq_q(\w, \mu)}{q_n}^\one
&= -\frac{\alpha_s(\mu) C_F}{2\pi}\,  \theta(z) \biggl\{
P_{qq}(z) \ln\frac{t'}{\mu^2} + \cL_1(1-z)(1 + z^2)
\nn\\ &\quad\hspace{17ex}
- \delta(1- z ) \Bigl(\frac72 -\frac{\pi^2}{3} \Bigr) + \theta(1-z) 2 (1-z) \biggr\}
\,, \nn \\
\Mae{g_n}{\oq_q(\w, \mu)}{g_n}^\one
&= -\frac{\alpha_s(\mu) T_F}{2\pi}\,\theta(z)
\biggl\{P_{qg}(z) \Bigl[\ln\frac{t'}{\mu^2} + \ln(1-z)\Bigr] + \theta(1 - z) \biggr\}
\,.\end{align}

\subsection{Quark Beam Function with Offshellness Infrared Regulator}
\label{subsec:NLO_B}

\begin{figure}
\subfigure[]{\includegraphics[scale=0.75]{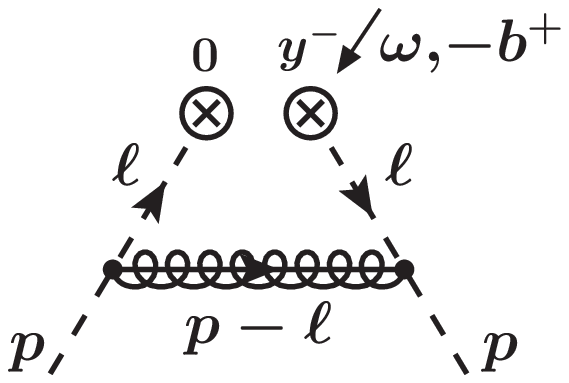}\label{fig:Bone_a}}%
\hfill%
\subfigure[]{\includegraphics[scale=0.75]{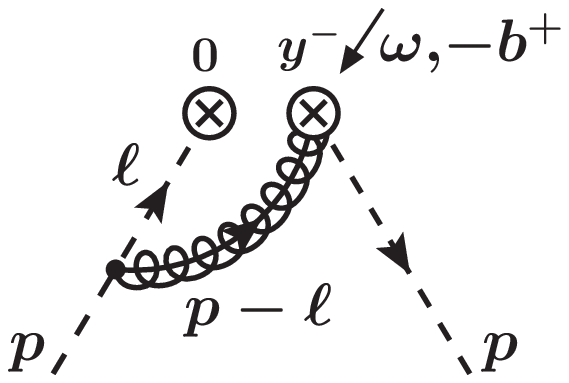}\label{fig:Bone_b}}%
\hfill%
\subfigure[]{\includegraphics[scale=0.75]{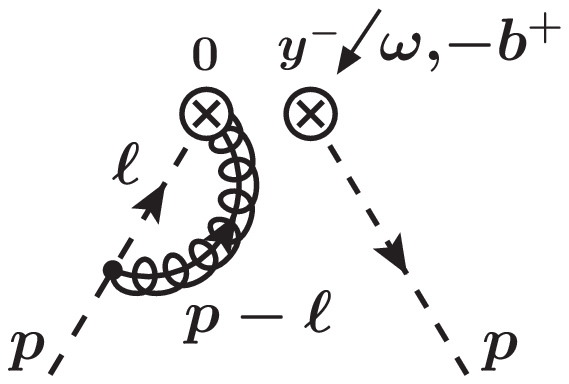}\label{fig:Bone_c}}%
\\
\subfigure[]{\includegraphics[scale=0.75]{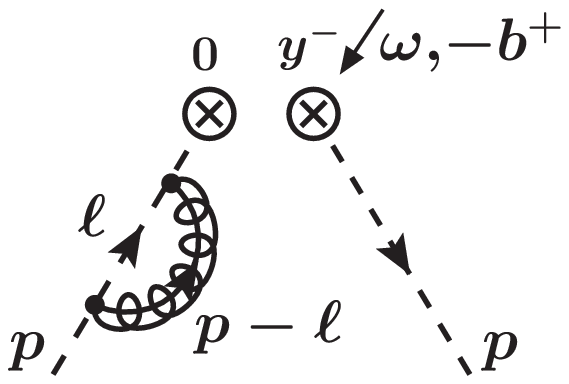}\label{fig:Bone_d}}%
\hfill%
\subfigure[]{\includegraphics[scale=0.75]{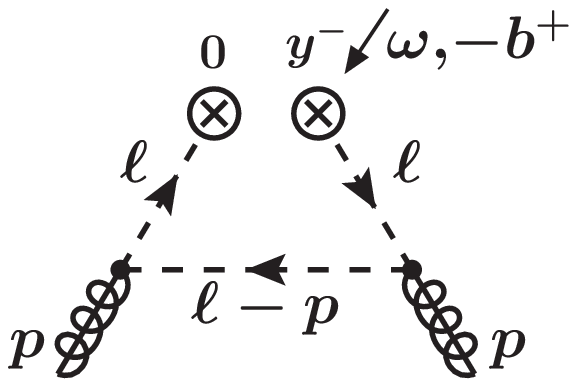}\label{fig:Bone_e}}%
\hfill%
\subfigure[]{\includegraphics[scale=0.75]{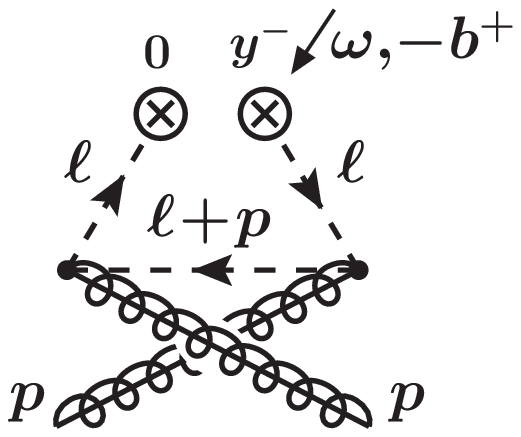}\label{fig:Bone_f}}%
\caption{One-loop diagrams for the quark beam function. The minus momentum $\w$ is incoming at the vertex and the $b^+$ momentum is outgoing. Diagram (d) denotes the wave-function contribution. Diagrams (b), (c), and (d) have symmetric counterparts which are equal to the ones shown and included in the computation. Diagram (f) and the diagram with the gluon connecting both vertices vanish.}
\label{fig:Bone}
\end{figure}

Next, we calculate the bare beam function $S$-matrix elements,
\begin{equation}
\Mae{q_n(p)}{\theta(\w) \op_q^\bare(t, \w)}{q_n(p)}
\,,\qquad
\Mae{g_n(p)}{\theta(\w) \op_q^\bare(t, \w)}{g_n(p)}
\,,\end{equation}
to NLO. The corresponding one-loop diagrams are shown in \fig{Bone}. The matrix elements are calculated as explained in \subsec{beamT} in \eq{DiscTB_full}: For the virtual diagrams with vacuum intermediate state we explicitly insert the vacuum state, while for the real-emission diagrams we use \eq{DiscTB}. In the latter case, we first take the $\Disc$, then expand in $\eps$ to extract the UV divergences, and at last take the $t' \to 0$ limit to isolate the IR divergences into $\ln t'$ terms. Some helpful formulas for calculating the discontinuity and taking the limit $t'\to 0$ are given in \app{disc}.

For the beam function calculation the $p^+ < 0$ actually plays a dual role: For the UV divergent piece we can treat the calculation as in \SCETa, and so $p^+ \sim b^+ \sim \la^2 p^-$, which allows us to explicitly check the structure of the convolution in \eq{op_ren}. The renormalized result contributes to the matching onto PDFs, matching from \SCETa onto \SCETb. In the matching, $-p^+ \ll b^+$ plays the role of the IR regulator, since we are required to use the same states as in the PDF calculation. We will see that the IR divergences $\ln t'$ match up with those present in the PDF calculation, and hence drop out in the matching coefficients $\cI_{ij}$.

The diagrams in \fig{Bone} have the same Dirac and propagator structure and overall factors as the corresponding PDF diagrams in \fig{fone}, so we can reuse those parts from the previous subsection. The difference compared to the PDF calculation is that for the real-emission diagrams, instead of doing the $\ell^+$ integral by contours, $\ell^+$ is fixed by the additional $\delta$ function in $b^+$, and since we use time-ordered perturbation theory we must now take the discontinuity. This also alters the structure of the remaining $\vec\ell_\perp$ integral, for which we now use Feynman parameters to combine the denominators. After carrying out the $\vec\ell_\perp$ integration, we will need the following two Feynman parameter integrals
\begin{align} \label{eq:I12}
I_1(A, B, \eps)
&= \int_0^1\! \df \alpha\, [(1 - \alpha)A - \alpha B]^{-1-\eps}
= \frac{(-B)^{-\eps} - A^{-\eps} }{\eps(A+B)}
\,, \nn \\
I_2(A, B, \eps)
&= \int_0^1\! \df \alpha\, (1 - \alpha) [(1 - \alpha)A - \alpha B]^{-1-\eps}
= -\frac{(-B)^{1-\eps} - A^{1-\eps}}{\eps(1-\eps)(A+B)^2} - \frac{A^{-\eps} }{\eps(A+B)}
\,.\end{align}

The first diagram, \fig{Bone_a}, has real radiation in the final state, so we use \eq{DiscTB} for the computation,
\begin{align} \label{eq:Ba}
 & \Mae{q_n}{\theta(\w)\op^\bare_q(t,\w)}{q_n}^{(a)}
 \nn \\ & \quad
\!= -\img \Bigl(\frac{e^{\gamma_E} \mu^2}{4\pi}\Bigr)^{\eps}g^2 C_F \frac{\theta(\w)}{\w}\, \Disc_{t>0}\! \int\! \frac{\df^d\ell}{(2\pi)^d}
 \frac{p^- (d-2)\ell_\perp^2
}{(\ell^2+\img 0)^2 [(\ell-p)^2+\img 0] } \,
 \delta(\ell^-\! -\w) \delta(\ell^+\! + b^+\! - p^+)
\nn \\ & \quad
= -\img \Bigl(\frac{e^{\gamma_E} \mu^2}{4\pi}\Bigr)^{\eps}g^2 C_F\,\frac{\theta(z)(d-2)}{(2\pi)^2 z}\, \Disc_{t>0}
   \int_0^1 \!\df\alpha \int\! \frac{\df^{d-2}\vec\ell_\perp}{(2\pi)^{d-2}}\,
    \frac{(1-\alpha)\, \vec\ell_\perp^2 }{[\vec \ell_\perp^2  +  (1 - \alpha)A - \alpha B]^3 }
\nn \\ & \quad
= \frac{\alpha_s(\mu)C_F}{2\pi}\,\frac{\theta(z)}{z}\, \Gamma(1+\eps) (e^{\gamma_E} \mu^2)^\eps (1-\eps)^2
  \Bigl[-\frac{\img}{2\pi}\Disc_{t>0}\, I_2(A, B, \eps) \Bigr]
\,,\end{align}
where we abbreviated
\begin{align}
A= t + t'
\,,\qquad
B= \frac{1 - z}{z}\,t
\,,\qquad
A+B = \frac{t}{z} + t'
\,.\end{align}
Since $t' > 0$ and $z > 0$, the only discontinuity in $I_2(A, B, \eps)$ for $t > 0$ arises from $(-B)$. Using \eq{disc_os} to take the $\Disc$, we obtain
\begin{align}
-\frac{\img}{2\pi}\Disc_{t>0}\,I_2(A, B, \eps)
&= \frac{\img}{2\pi}\Disc_{t>0} \frac{(-B)^{1-\eps}}{\eps(1-\eps)(A+B)^2}
= \theta(t)\frac{\sin\pi\eps}{\pi\eps(1-\eps)}\, \frac{\theta(B) B^{1-\eps}}{(A+B)^2}
\,,\nn\\
&=  \theta\Bigl(\frac{1-z}{z}\Bigr)\theta(t)\frac{\sin\pi\eps}{\pi\eps(1-\eps)}\, \frac{[(1-z)t]^{1-\eps}z^{1+\eps}}{(t + zt')^2}
\,.\end{align}
Note that there is only a discontinuity for $B > 0$, so taking the discontinuity for $t > 0$ requires $(1-z)/z > 0$, and since $z > 0$ we obtain the expected limit $z < 1$. Since there are no UV divergences, we can let $\eps \to 0$, and \eq{Ba} becomes
\begin{equation} \label{eq:DiscBa}
\Mae{q_n}{\theta(\w)\op^\bare_q(t,\w)}{q_n}^{(a)}
= \frac{\alpha_s(\mu)C_F}{2\pi}\,\theta(z)\theta(1-z)(1-z)\frac{\theta(t)\,t}{(t + zt')^2}
\,.\end{equation}
The above result has a collinear IR singularity for $t\to 0$ which is regulated by the nonzero $t'$. We can isolate the IR singularity using \eqs{plusdef}{limits} by letting $\beta\equiv z t'/\mu^2 \to 0$ while holding $\tilde t = t + zt'$ fixed%
\footnote{We keep the dependence on $\tilde t$ in our calculation as it will be useful for checking the
structure of the renormalization in the following subsection.},
\begin{equation} \label{eq:Ba_limit}
\lim_{t'\to 0} \frac{\theta(t)\,t}{(t + zt')^2}
= \lim_{z t'/\mu^2 \to 0}\biggl[
  \frac{\theta(\tilde t - z t')}{\tilde t} - \frac{\theta(\tilde t - z t') zt'}{\tilde t^2} \biggr]
= \frac{1}{\mu^2}\cL_0\Bigl(\frac{\tilde t}{\mu^2}\Bigr)
- \delta(\tilde t) \Bigl(\ln \frac{z t'}{\mu^2} + 1\Bigr)
\,.\end{equation}
The final result for \fig{Bone_a} is thus
\begin{equation} \label{eq:Ba_final}
\Mae{q_n}{\theta(\w)\op^\bare_q(t,\w)}{q_n}^{(a)}
= \frac{\alpha_s(\mu)C_F}{2\pi}\,\theta(z)\theta(1-z)(1-z)
\biggl\{\frac{1}{\mu^2}\cL_0\Bigl(\frac{\tilde t}{\mu^2}\Bigr) - \delta(\tilde t) \Bigl(\ln \frac{z t'}{\mu^2} + 1\Bigr)\biggr\}
\,.\end{equation}

Next, we consider the real-emission diagram in \fig{Bone_b}. It corresponds to the $\delta(\ell^- - \w)$ term in \eq{PDFb1}. Together with its mirror graph, giving an identical contribution, we obtain
\begin{align} \label{eq:Bb1}
 & \Mae{q_n}{\theta(\w)\op_q^\bare(t,\w)}{q_n}^{(b)}
 \nn \\ &\qquad
= 2\img \Bigl(\frac{e^{\gamma_E} \mu^2}{4\pi}\Bigr)^{\eps}g^2 C_F \frac{\theta(\w)}{\w}\, \Disc_{t>0}\!
   \int\!\frac{\df^d\ell}{(2\pi)^d}\,\frac{2p^-\ell^-\,  \delta(\ell^-\! -\w)\, \delta(\ell^+\! + b^+\! - p^+)}
   {(\ell^-\! - p^-)(\ell^2+\img 0) [(\ell-p)^2+\img 0] } \,
\nn \\ &\qquad
= \frac{\alpha_s(\mu)C_F}{\pi}\, \frac{\theta(z)}{1 -z}\, \Gamma(1+\eps)(e^{\gamma_E} \mu^2)^\eps
  \Bigl[-\frac{\img}{2\pi}\Disc_{t>0}\, I_1(A, B, \eps) \Bigr]
\nn \\ &\qquad
= \frac{\alpha_s(\mu)C_F}{\pi}\,\theta(z) \Gamma(1+\eps)\Bigl(\frac{e^{\gamma_E} \mu^2}{t}\Bigr)^\eps
\frac{\sin\pi\eps}{\pi\eps} \frac{\theta(t) }{t + zt'}\, \frac{\theta(1-z)z^{1+\eps}}{(1-z)^{1+\eps}}
\,,\end{align}
where in the second step we performed the loop integral as before, and in the last step we used \eq{disc_os} to take the discontinuity. As for \fig{Bone_a}, the loop integral produces no UV divergence. However, as in the PDF calculation for \fig{fone_b}, there is a soft gluon IR divergence at $z\to 1$ or $\ell^-\to p^-$ producing a $\delta(1-z)/\eps$ IR pole when expanding the last factor using \eq{distr_id}. In contrast to the PDF calculation, the soft gluon region must now be explicitly excluded from the collinear loop integral. In dimensional regularization with an offshellness IR regulator the relevant zero-bin integral is scaleless and vanishes. Thus, including the zero-bin subtraction removes the $1/\eps$ IR divergence and replaces it by an equal $1/\eps$ UV divergence such that all $1/\eps$ poles in the final result are UV divergences.
Expanding in $\eps$, we have
\begin{equation} \label{eq:Bb2}
\Mae{q_n}{\theta(\w)\op_q^\bare(t,\w)}{q_n}^{(b)}\,
 =  \frac{\alpha_s(\mu) C_F }{\pi} \, \theta(z)\,\frac{\theta(t)}{t + zt'}
  \Bigl\{ \delta(1-z) \Bigl(-\frac{1}{\eps}  + \ln\frac{t}{\mu^2}\Bigr) + \cL_0(1-z)z \Bigr\}
\,,\end{equation}
and taking the same limit as in \eq{Ba_limit} to isolate the IR divergences,
\begin{align}
\lim_{t'\to 0} \frac{\theta(t)}{t + zt'}
&= \lim_{z t'/\mu^2\to 0} \frac{\theta(\tilde t - z t')}{\tilde t}
= \frac{1}{\mu^2}\cL_0\Bigl(\frac{\tilde t}{\mu^2}\Bigr) - \delta(\tilde t) \ln \frac{z t'}{\mu^2}
\,,\\\nn
\lim_{t'\to 0} \frac{\theta(t)}{t + zt'}\ln\frac{t}{\mu^2}
&= \lim_{z t'/\mu^2\to 0} \frac{\theta(\tilde t - z t')}{\tilde t}\ln\frac{\tilde t - zt'}{\mu^2}
= \frac{1}{\mu^2}\cL_1\Bigl(\frac{\tilde t}{\mu^2}\Bigr) - \delta(\tilde t) \Bigl(\frac{1}{2}\ln^2\frac{z t'}{\mu^2} + \frac{\pi^2}{6}\Bigr)
\,,\end{align}
the final result for \fig{Bone_b} is
\begin{align} \label{eq:Bb_final}
&\Mae{q_n}{\theta(\w)\op_q^\bare(t,\w)}{q_n}^{(b)}
\nn\\ & \qquad
= \frac{\alpha_s(\mu) C_F}{\pi}\, \theta(z) \biggl\{
 \biggl[
 \frac{1}{\mu^2} \cL_0 \Bigl(\frac{\tilde t}{\mu^2}\Bigr) -\delta(\tilde t) \ln \frac{zt'}{\mu^2} \biggr]
\Bigl[-\frac{1}{\eps}\, \delta(1-z) + \cL_0(1 - z)z \Bigr]
\nn\\ & \qquad\hspace{18ex}
 + \biggl[\frac{1}{\mu^2} \cL_1\Bigl(\frac{\tilde t}{\mu^2}\Bigr) -\delta(\tilde t) \Bigl(\frac{1}{2}\ln^2\frac{t'}{\mu^2} + \frac{\pi^2}{6}\Bigr)
   \biggr] \delta(1 - z)
 \biggr\}
\,.\end{align}

For the diagram in \fig{Bone_c} (and its mirror diagram) we insert the vacuum intermediate state between the fields in $\op_q$ as in \eq{DiscTB_full}, resulting in a one-loop virtual diagram involving a single field. The calculation is exactly the same as for the $\delta(p^-\! -\w)$ term in \eq{PDFb1} times an overall $\delta(t)$,
\begin{align} \label{eq:Bc_final}
  &\Mae{q_n}{\theta(\w) \op_q^\bare(t,\w)}{q_n}^{(c)}\,
  \nn \\ & \qquad
  = - 2\img \Bigl(\frac{e^{\gamma_E} \mu^2}{4\pi}\Bigr)^\eps g^2 C_F\,
  \delta(t) \delta(p^-\! -\w) \int\! \frac{\df^d\ell}{(2\pi)^d}\,
  \frac{2p^-\ell^-}{(\ell^-\! - p^-)(\ell^2+\img 0) [(\ell-p)^2+\img 0] }
\nn \\ & \qquad
 = -\frac{\alpha_s(\mu) C_F}{\pi}\,
 \Gamma(\eps) \Bigl(\frac{e^{\gamma_E}\mu^2}{t'}\Bigr)^\eps
  \delta(t) \delta(1 -z )\, \frac{\Gamma(2-\eps)\Gamma(-\eps)}{\Gamma(2-2\eps)}
\nn \\ & \qquad
 = \frac{\alpha_s(\mu) C_F}{\pi}\, \delta(\tilde t)\delta(1 -z ) \biggl\{
 \frac{1}{\eps^2} + \frac{1}{\eps} \Bigl(1 - \ln\frac{t'}{\mu^2}\Bigr)
  + \frac12 \ln^2\frac{t'}{\mu^2} - \ln\frac{t'}{\mu^2} + 2 -\frac{\pi^2}{12} \biggr\}
\,.\end{align}
In the last step we expanded in $\eps$ and took the IR limit. To be consistent we have to use the same IR limit in the virtual diagrams as in the real-emission diagrams above, which simply turns the overall $\delta(t)$ into a $\delta(\tilde t)$,
\begin{equation} \label{eq:Bc_limit}
\lim_{t'\to 0} \delta(t) = \lim_{zt'/\mu^2\to 0} \delta(\tilde t - zt') = \delta(\tilde t)
\,.\end{equation}
As in the PDF calculation, the UV divergence in the loop produces a $\Gamma(\eps)$ and the soft IR divergence a $\Gamma(-\eps)$. The latter is converted by the zero-bin subtraction into a UV divergence, producing the $1/\eps^2$ pole. The $1/\eps^2$ poles do not cancel anymore between \figs{Bone_b}{Bone_c} as they did for the PDF in \fig{fone_b}, because the phase space of the real emission in \fig{Bone_b} is now restricted by the measurement of $b^+$ via the $\delta(\ell^+ + b^+ - p^+)$. For the same reason \fig{Bone_a} has no UV divergence anymore, while \fig{fone_a} did. The $(1/\eps)\ln t'$ terms in \eqs{Bb_final}{Bc_final}, which are a product of UV and collinear IR divergences, still cancel between the real and virtual diagrams, ensuring that the UV renormalization is independent of the IR, as should be the case.

The final one-loop contribution to the quark matrix element, \fig{Bone_d} and its mirror diagram, comes from wave-function renormalization,
\begin{equation} \label{eq:Bd_final}
\Mae{q_n}{\theta(\w)\op_q^\bare(t,\w)}{q_n}^{(d)}
= \delta(t) \delta(1-z) (Z_\xi -1)
= -\frac{\alpha_s(\mu) C_F}{4\pi}\, \delta(\tilde t)\delta(1-z)
  \biggl\{\frac{1}{\eps} - \ln\frac{t'}{\mu^2} + 1 \biggr\}
\,.\end{equation}
Adding up the results in Eqs.~\eqref{eq:Ba_final}, \eqref{eq:Bb_final}, \eqref{eq:Bc_final}, and \eqref{eq:Bd_final}, we obtain the bare beam function quark matrix element at one loop,
\begin{align} \label{eq:Bqbare}
   \Mae{q_n}{\theta(\w)\op_q^\bare(t,\w)}{q_n}^\one
  &= \frac{\alpha_s(\mu) C_F}{2\pi}\,\theta(z) \biggl\{
   \biggl[ \delta(\tilde t) \Bigl(\frac{2}{\eps^2} +
    \frac{3}{2\eps} \Bigr) -
    \frac{2}{\eps}\, \frac{1}{\mu^2} \cL_0\Bigl(\frac{\tilde t}{\mu^2}\Bigr)
  \biggr]\delta(1-z)
  \nn \\ & \quad
  +\frac{2}{\mu^2} \cL_1\Bigl(\frac{\tilde t}{\mu^2}\Bigr)\delta(1-z) +
  \frac{1}{\mu^2} \cL_0\Bigl(\frac{\tilde t}{\mu^2}\Bigr)\cL_0(1-z)(1 + z^2)
  \\\nn & \quad
  -\delta(\tilde t) \biggl[ P_{qq}(z)\ln\frac{zt'}{\mu^2} - \delta(1-z) \Bigl(\frac{7}{2} - \frac{\pi^2}{2}\Bigr) + \theta(1-z)(1-z) \biggr]
  \biggr\}
\,.\end{align}

We now consider the beam function matrix element with external gluons. The corresponding diagrams are shown in \figs{Bone_e}{Bone_f}. For \fig{Bone_e}, which is analogous to \fig{fone_d}, we find
\begin{align}  \label{eq:Be}
 & \Mae{g_n}{\theta(\w)\op_q^\bare(t,\w)}{g_n}^{(e)}
\nn \\ & \quad
 = \img\Bigl(\frac{ e^{\gamma_E} \mu^2}{4\pi}\Bigr)^\eps g^2 T_F \frac{\theta(\w)}{\w}\,
2p^- \Bigl(\frac{1}{1-z} - \frac{4z}{d - 2}\Bigr)
\Disc_{t>0}\! \int\! \frac{\df^d\ell}{(2\pi)^d}\,
 \frac{\vec\ell_\perp^2\, \delta(\ell^-\! -\w) \delta(\ell^+\! + b^+\! - p^+)}{(\ell^2+\img 0)^2 [(\ell-p)^2+\img 0]}
\nn\\ & \quad
= \frac{\alpha_s(\mu)T_F}{2\pi}\,\frac{\theta(z)}{z}\,
 \Gamma(1+\eps) (e^{\gamma_E} \mu^2)^\eps \Bigl(\frac{1-\eps}{1-z} - 2z\Bigr)
  \Bigl[-\frac{\img}{2\pi}\Disc_{t>0}\, I_2(A, B, \eps) \Bigr]
\nn\\ & \quad
= \frac{\alpha_s(\mu)T_F}{2\pi}\,\theta(z) P_{qg}(z)\,\frac{\theta(t)t}{(t + zt')^2}
\,.\end{align}
The loop integral and discontinuity are exactly the same as for \fig{Bone_a}. The diagram in \fig{Bone_f} does not contribute to the quark beam function. It can be obtained from \eq{Be} by replacing $p^\mu \to -p^\mu$, which takes $t'\to -t'$ and $z\to -z$. Doing so, the only contribution to the discontinuity is still from $B = -(1 + z)t/z$ for $B > 0$, which for $t>0$ requires $-1 < z < 0$. Hence, \fig{Bone_f} does not contribute. Using \eq{Ba_limit} to take $t'\to 0$ in \eq{Be}, we get the final result for the bare one-loop gluon matrix element
\begin{equation} \label{eq:Bgbare}
\Mae{g_n}{\theta(\w) \op_q^\bare(t,\w)}{g_n}^\one
 = \frac{\alpha_s(\mu)  T_F}{ 2\pi }\, \theta(z) P_{qg}(z)
\biggl\{
  \frac{1}{\mu^2}\cL_0\Bigl(\frac{\tilde t}{\mu^2}\Bigr) - \delta(\tilde t) \Bigl(\ln \frac{z t'}{\mu^2} + 1 \Bigr)
\biggr\}
\,.\end{equation}
As for \fig{Bone_a}, it has no UV divergences because of the measurement of $b^+$, which means that the renormalization does not mix $\op_q$ and $\op_g$.

\subsection{Renormalization and Matching}
\label{subsec:NLO_matching}

Using the bare matrix elements calculated in the previous section, we can extract the renormalization of $\op_q$. We first take $\tilde t = t + zt'\to t$ in the bare matrix elements. Then, expanding the quark matrix element of \eq{op_ren} to one-loop order,
\begin{align} \label{eq:Bbare_ren}
&\Mae{q_n}{\op_q^\bare(t,\w)}{q_n}^\one
\nn \\ &\qquad
= \int\! \df t'\,
\biggl[Z^{q\one}_B(t-t', \mu) \Mae{q_n}{\op_q(t',\w,\mu)}{q_n}^\zero + Z_B^{q\zero}(t-t', \mu) \Mae{q_n}{\op_q(t',\w,\mu)}{q_n}^\one \biggr]
\nn \\ &\qquad
= Z_B^{q\one}(t, \mu)\, \delta(1-z) + \Mae{q_n}{\op_q(t,\w,\mu)}{q_n}^\one
\,,\end{align}
we can then read off the $\overline{\mathrm{MS}}$ renormalization constant from \eq{Bqbare}
\begin{equation} \label{eq:ZB}
 Z_B^q(t,\mu) = \delta(t) +
  \frac{\alpha_s(\mu) C_F}{2\pi}
   \biggl[ \delta(t) \Bigl(\frac{2}{\eps^2} +
    \frac{3}{2\eps} \Bigr) -
    \frac{2}{\eps} \frac{1}{\mu^2} \cL_0\Bigl(\frac{t}{\mu^2}\Bigr)
  \biggr]
\,.\end{equation}
The fact that the gluon matrix element is UV finite and the UV divergences in the quark matrix element are proportional to $\delta(1-z)$ confirms at one loop our general result that the renormalization of the beam function does not mix quarks and gluons or change the momentum fraction.

In \eqs{Bbare_ren}{ZB} we used that we already know the structure of the renormalization from our general arguments in \subsec{B_RGE}, i.e.\ that $Z_B^q$ only depends on the difference $t - t'$. Alternatively, we can also use the dependence on $z$ and the finite dependence on $t'$ via $\tilde t$ to explicitly check the structure of the renormalization. In this case, we must use the same IR limit also for the tree-level result in \eq{op_tree}, which using \eq{Bc_limit} becomes
\begin{equation} \label{eq:Btree}
\Mae{q_n}{\theta(\w)\op_q(t,\w, \mu)}{q_n}^\zero
=\lim_{t'\to 0}\delta(t)\, \delta(1-z) = \delta(\tilde t)\,\delta(1-z)
\,.\end{equation}
Taking $Z_B^q(t, t', \w/\w', \mu)$ to be a general function of $t$, $t'$ and $\w/\w'$, we now get for \eq{Bbare_ren}
\begin{align}
&\Mae{q_n}{\op_q^\bare(t,\w)}{q_n}^\one
\nn\\ & \qquad
= \int\!\df t''\frac{\df\w'}{\w'}\, Z_B^{q\one}\Bigl(t, t'', \frac{\w}{\w'}, \mu\Bigr)\, \delta(t'' + z' t')\, \delta\Bigl(1- \frac{\w'}{p^-}\Bigr) + \Mae{q_n}{\op_q(t,\w,\mu)}{q_n}^\one
\nn\\ & \qquad
= Z_B^{q\one}(t, -t', z) + \Mae{q_n}{\op_q(t,\w,\mu)}{q_n}^\one
\,.\end{align}
In the first step we used \eq{Btree} and $Z_B^{q\zero}(t, t', z) = \delta(t - t')\delta(1 - z)$. From \eq{Bqbare} we now find
\begin{equation}
 Z_B^q(t, t', z, \mu)
 = \biggl\{\delta(t - t') + \frac{\alpha_s(\mu) C_F}{2\pi}
   \biggl[ \delta(t - t') \Bigl(\frac{2}{\eps^2} +
    \frac{3}{2\eps} \Bigr) - \frac{2}{\eps} \frac{1}{\mu^2} \cL_0\Bigl(\frac{t - t'}{\mu^2}\Bigr)
  \biggr] \biggr\}\, \delta(1-z)
\,,\end{equation}
thus explicitly confirming at one loop that $Z_B^q(t, t', z, \mu) \equiv Z_B^q(t - t', \mu)\, \delta(1-z)$.

The one-loop anomalous dimension for the quark beam function follows from \eq{ZB},
\begin{align} \label{eq:gaB1}
\gamma_B^q(t,\mu) &=
-\mu\, \frac{\df}{\df\mu} Z_B^{q\one}(t,\mu)
= \frac{\alpha_s(\mu) C_F}{\pi}
  \biggl[ - \frac{2}{\mu^2} \cL_0\Bigl(\frac{t}{\mu^2}\Bigr) + \frac{3}{2}\, \delta(t)
  \biggr]
\,.\end{align}
It is identical to the one-loop anomalous dimension of the quark jet function. The coefficient of $\cL_0(t/\mu^2)/\mu^2$ can be identified as the one-loop expression for $-2\Gamma_\cusp^q$. Thus, \eq{gaB1} explicitly confirms the general results in \eqs{gaB_gen}{gaJgaB} at one loop.

Taking the bare matrix elements in \eqs{Bqbare}{Bgbare} and subtracting the UV divergences using \eqs{Bbare_ren}{ZB} gives the renormalized one-loop beam function matrix elements,
\begin{align} \label{eq:Bren}
\Mae{q_n}{\theta(\w) \op_q(t,\w, \mu)}{q_n}^\one
  &= \frac{\alpha_s(\mu) C_F}{2\pi}\,
 \theta(z) \biggl\{
  \frac{2}{\mu^2} \cL_1\Bigl(\frac{t}{\mu^2}\Bigr) \delta(1\!-\!z) +
  \frac{1}{\mu^2} \cL_0\Bigl(\frac{t}{\mu^2}\Bigr) \cL_0(1\!-\!z)(1 \!+\! z^2)
  \nn\\ & \quad
  -\delta(t) \biggl[ P_{qq}(z)\ln\frac{zt'}{\mu^2}
  -\delta(1-z) \Bigl(\frac{7}{2} - \frac{\pi^2}{2}\Bigr) + \theta(1-z)(1-z) \biggr]
   \biggr\}
  \,, \nn \\
\Mae{g_n}{\theta(\w) \op_q(t,\w, \mu)}{g_n}^\one
 &= \frac{\alpha_s(\mu) T_F}{ 2\pi }\, \theta(z) P_{qg}(z)
\biggl\{
  \frac{1}{\mu^2}\cL_0\Bigl(\frac{t}{\mu^2}\Bigr) - \delta(t) \Bigl(\ln \frac{z t'}{\mu^2} + 1 \Bigr)
\biggr\}
\,.\end{align}
For the matching onto the PDFs, we must take $t' \to 0$ and have therefore set $\tilde t = t$ everywhere, only keeping $t'$ in the IR divergent $\ln t'$ terms.

Expanding the OPE for the quark beam function, \eq{beam_fact}, to one loop, we have
\begin{align}
&\Mae{q_n}{\theta(\w) \op_q(t,\w,\mu)}{q_n}^\one
\nn \\&\qquad
  = \sum_j \int\! \frac{\df \w'}{\w'}\,
  \bigg[\cI_{qj}^\one\Big( t,\frac{\w}{\w'},\mu \Big) \Mae{q_n}{\oq_j(\w',\mu)}{q_n}^\zero +
  \cI_{qj}^\zero\Big( t,\frac{\w}{\w'},\mu \Big) \Mae{q_n}{\oq_j(\w',\mu)}{q_n}^\one \bigg]
\nn \\ & \qquad
  = \cI_{qq}^\one (t,z,\mu) + \delta(t) \Mae{q_n}{\oq_q(\w,\mu)}{q_n}^\one
  \,, \nn \\
&\Mae{g_n}{\theta(\w) \op_q(t,\w,\mu)}{g_n}^\one
\nn\\ & \qquad
= \sum_j \int\! \frac{\df \w'}{\w'}\,
  \bigg[\cI_{qj}^\one\Big( t,\frac{\w}{\w'},\mu \Big) \Mae{g_n}{\oq_j(\w',\mu)}{g_n}^\zero +
  \cI_{qj}^\zero\Big( t,\frac{\w}{\w'},\mu \Big) \Mae{g_n}{\oq_j(\w',\mu)}{g_n}^\one \bigg]
\nn \\ & \qquad
= \cI_{qg}^\one (t,z,\mu) + \de(t) \Mae{g_n}{\oq_q(\w,\mu)}{g_n}^\one
\,.\end{align}
Thus, the one-loop matching coefficients, $\cI^\one_{qi}(t,z,\mu)$, are obtained by subtracting the renormalized PDF matrix elements in \eq{fren} from those in \eq{Bren}. Doing so, we see that the $\ln t'$ IR divergences in \eqs{fren}{Bren} precisely cancel, as they must, such that the matching coefficients are independent of the IR regulator and only involve large logarithms that are minimized at the scale $\mu^2\simeq t$. The final result for the NLO matching coefficients is
\begin{align} \label{eq:Iresult}
\cI_{qq}(t,z,\mu)
&= \delta(t)\, \delta(1 - z)
\\ & \quad
  + \frac{\alpha_s(\mu)C_F}{2\pi}\, \theta(z) \biggl\{
  \frac{2}{\mu^2} \cL_1\Bigl(\frac{t}{\mu^2}\Bigr) \delta(1 - z) +
  \frac{1}{\mu^2} \cL_0\Bigl(\frac{t}{\mu^2}\Bigr) \Bigl[P_{qq}(z) - \frac{3}{2}\,\delta(1-z)\Bigr]
  \nn \\ & \quad
  + \delta(t) \biggl[
  \cL_1(1 - z)(1 + z^2)
  -\frac{\pi^2}{6}\, \delta(1 - z)
  + \theta(1 - z)\Bigl(1 - z - \frac{1 + z^2}{1 - z}\ln z \Bigr) \biggr]
  \biggr\}
  \,, \nn\\\nn
\cI_{qg}(t,z,\mu)
 &= \frac{ \alpha_s(\mu) T_F }{2\pi}\, \theta(z) \biggl\{
\frac{1}{\mu^2} \cL_0\Bigl(\frac{t}{\mu^2}\Bigr) P_{qg}(z)
 + \delta(t) \biggl[P_{qg}(z)\Bigl(\ln\frac{1-z}{z} - 1\Bigr) +  \theta(1-z) \biggr]
\biggr\}
\,.\end{align}

\section{Numerical Results and Plots}
\label{sec:results}

Including the RGE running in \eq{DYbeam}, the full result for the resummed cross
section for isolated Drell-Yan differential in $q^2$, $Y$, and $\tau_B$ is
\begin{align} \label{eq:DYbeamrun}
\frac{\df\sigma}{\df q^2 \df Y \df \tau_B}
&= \frac{4\pi\alem^2}{3N_c\Ecm^2 q^2}
\sum_{ij} H_{ij}(q^2, \mu_H)\, U_H(q^2, \mu_H, \mu) \int\!\df t_a\,\df t_b
\nn\\* &\quad \times
\int\!\df t_a' B_i(t_a - t_a', x_a, \mu_B)\, U^i_B(t_a', \mu_B, \mu)
\int\!\df t_b' B_j(t_b - t_b', x_b, \mu_B)\, U^j_B(t_b', \mu_B, \mu)
\nn\\* &\quad \times
\int\!\df k^+ Q\, S_B\Bigl(Q\,\tau_B - \frac{t_a + t_b}{Q} - k^+, \mu_S \Bigr)\, U_S(k^+, \mu_S, \mu)
\,,\end{align}
where the sum runs over quark flavors $ij = \{u\bar u, \bar u u, d\bar d, \ldots\}$ and the additional contributions from the leptonic matrix element are contained in the hard function. Equation~\eqref{eq:DYbeamrun} is valid to all orders in perturbation theory. The all-order solutions for the evolution factors
in terms of the respective anomalous dimensions are given in \app{pert}. The hard, beam, and soft functions are each evaluated at their natural scales, $\mu_H \simeq Q$, $\mu_B \simeq \sqrt{\tau_B} Q$, $\mu_S \simeq \tau_B Q$, and are then evolved to the common arbitrary scale $\mu$ by the evolution kernels $U_H$, $U_B^{i,j}$, and $U_S$, respectively. With the one-loop results for the beam function presented above, \eq{DYbeamrun} can be evaluated at NNLL order in resummed perturbation theory, which requires the one-loop matching corrections, the two-loop standard anomalous dimensions, and the three-loop cusp anomalous dimension (see Table~\ref{tab:counting}). All the necessary ingredients are given in \app{pert}.
If we let $v_B-\img 0$ be the Fourier conjugate variable to $\tau_B$, then the NNLL cross section resums the following terms
\begin{equation}
\ln\frac{\df\sigma}{\df q^2 \df Y \df v_B} \sim \ln v_B (\alpha_s \ln v_B)^k + (\alpha_s \ln v_B)^k + \alpha_s (\alpha_s \ln v_B)^k
\end{equation}
for all integers $k>0$.

\begin{table}
  \centering
  \begin{tabular}{l | c c c c}
  \hline \hline
  & matching & $\gamma_x$ & $\Gamma_\cusp$ & $\beta$  \\ \hline
  LO & $0$-loop & - & - & - \\
  NLO & $1$-loop & - & - & - \\
  NLL & $0$-loop & $1$-loop & $2$-loop & $2$-loop\\
  NNLL & $1$-loop & $2$-loop & $3$-loop & $3$-loop\\
  \hline\hline
  \end{tabular}
\caption{Order counting in fixed-order and resummed perturbation theory.}
\label{tab:counting}
\end{table}

In the remainder we will focus on the beam functions, which are the topic of this paper. Below we compare results for the quark beam function at LO and NLO in fixed-order perturbation theory as well as at NLL and NNLL in resummed perturbation theory. Our conventions for the $\alpha_s$ loop counting are given in Table~\ref{tab:counting}. To evaluate the required convolutions of plus distributions at NNLL we use the identities from App.~B of Ref.~\cite{Ligeti:2008ac}. We always use the MSTW2008~\cite{Martin:2009bu} parton distributions at NLO for $\alpha_s(m_Z) = 0.117$ and with two-loop, five-flavor running for $\alpha_s(\mu)$. The uncertainty bands in the plots show the perturbative uncertainties, which are estimated by varying the appropriate scales as explained in each case. They do not include the additional uncertainties from the PDFs and $\alpha_s(m_Z)$.

The order of the running of $\alpha_s(\mu)$ deserves some comment. Working consistently to NLO in the matching corrections requires us to use NLO PDFs, for which the two-loop running of $\alpha_s$ was used in Ref.~\cite{Martin:2009bu}. On the other hand, the double-logarithmic running of the hard function and beam functions at NNLL requires the three-loop running of $\alpha_s$, which poses a slight dilemma. Ideally, we would need NLO PDFs using three-loop running for $\alpha_s(\mu)$, which as far as we know is not available. The numerical difference between $\alpha_s$ run at two and three loops is very small, at most $2\%$. Hence, we use the following compromise. To be consistent with our PDF set, we use the above $\alpha_s(m_Z)$ and two-loop, five-flavor running to obtain the numerical value of $\alpha_s$ at some required scale, and to be consistent with the RGE, we use the two- and three-loop expression for the QCD $\beta$ function in the RGE solutions at NLL and NNLL. (For simplicity we use the same NLO PDFs and $\alpha_s$ also at NLL.)

To illustrate the importance of the various contributions to the quark beam function, we also consider the beam function in the threshold limit and without the gluon mixing contribution. In the threshold limit we only keep the terms in \eq{Iresult} which are singular as $z \to 1$,
\begin{align} \label{eq:Ithres}
  \cI_{qq}^\text{thresh}(t,z,\mu) &= \delta(t) \delta(1-z)
  + \frac{\alpha_s(\mu)C_F}{2\pi}\, \theta(z) \biggl\{
  \frac{2}{\mu^2} \cL_1\Bigl(\frac{t}{\mu^2}\Bigr) \delta(1-z) +
  \frac{2}{\mu^2} \cL_0\Bigl(\frac{t}{\mu^2}\Bigr) \cL_0(1-z)
  \nn \\ & \hspace{32ex}
  + \delta(t) \Bigl[  2 \cL_1(1-z) -\frac{\pi^2}{6} \delta(1-z) \Bigr]
  \biggr\}
\,,\nn\\
\cI_{qg}^\text{thresh}(t,z,\mu) &= 0
\,.\end{align}
The gluon mixing term $\cI_{qg}$ contains no threshold term (which reflects the fact that in threshold Drell-Yan the gluon PDF does not contribute). For the result without the gluon mixing contribution we keep the full $\cI_{qq}$ but set $\cI_{qg}$ to zero, which corresponds to adding the remaining non-threshold terms in $\cI_{qq}$ to the threshold result. In the plots below, the results in the threshold limit are shown by a dotted line and are labeled ``$x\to 1$'', and the results without the gluon contribution are shown by a dashed line and are labeled ``no $g$''. The full result, including both $\cI_{qq}$ and $\cI_{qg}$, is shown by a solid line. Hence, the size of the non-threshold terms in $\cI_{qq}$, and therefore the applicability of the threshold limit, is seen by the shift from the dotted to the dashed line, and the effect of the gluon mixing is given by the shift from the dashed to the solid line.

\FIGURE[ht]{%
\includegraphics[width=0.49\textwidth]{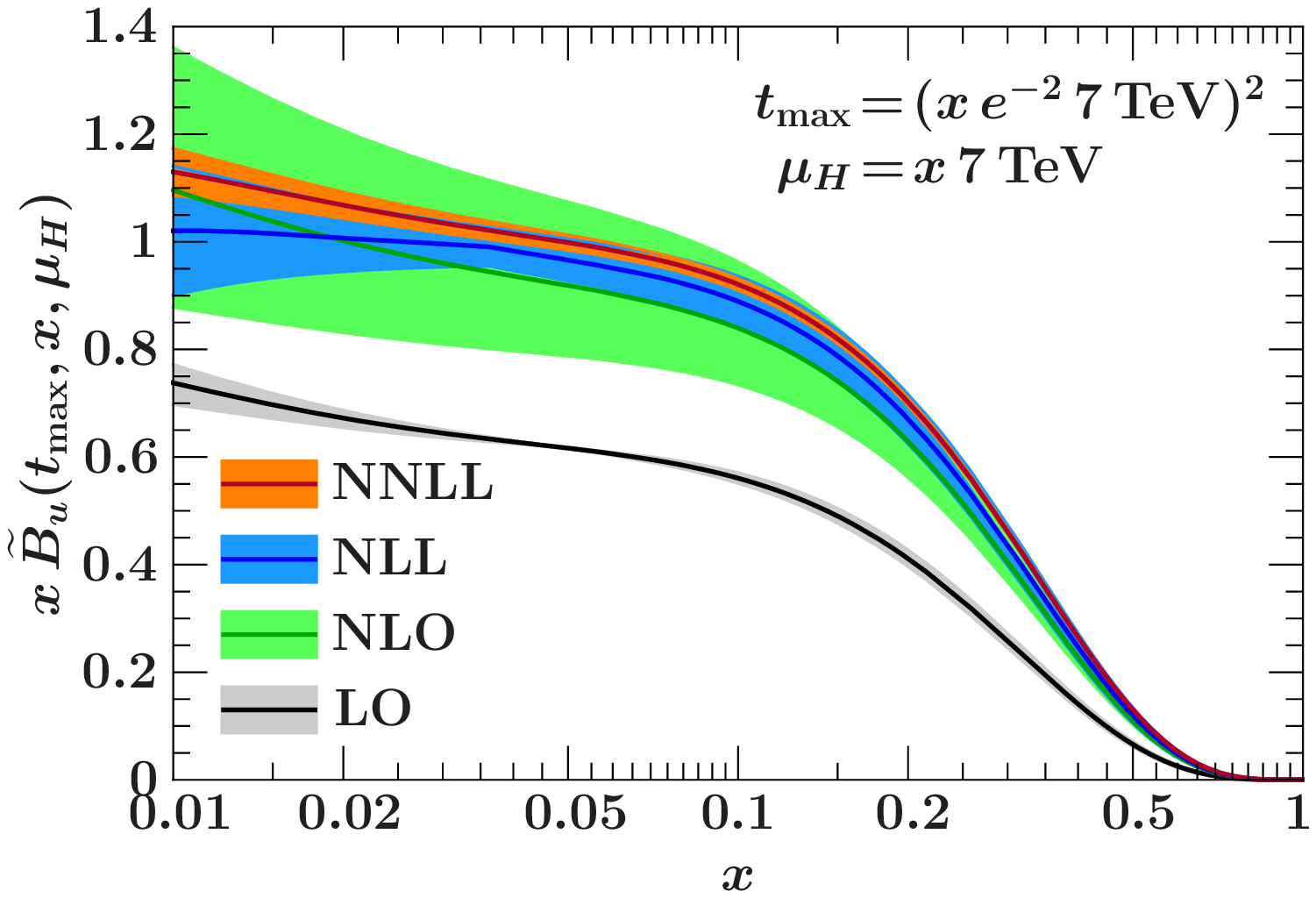}%
\hfill%
\includegraphics[width=0.49\textwidth]{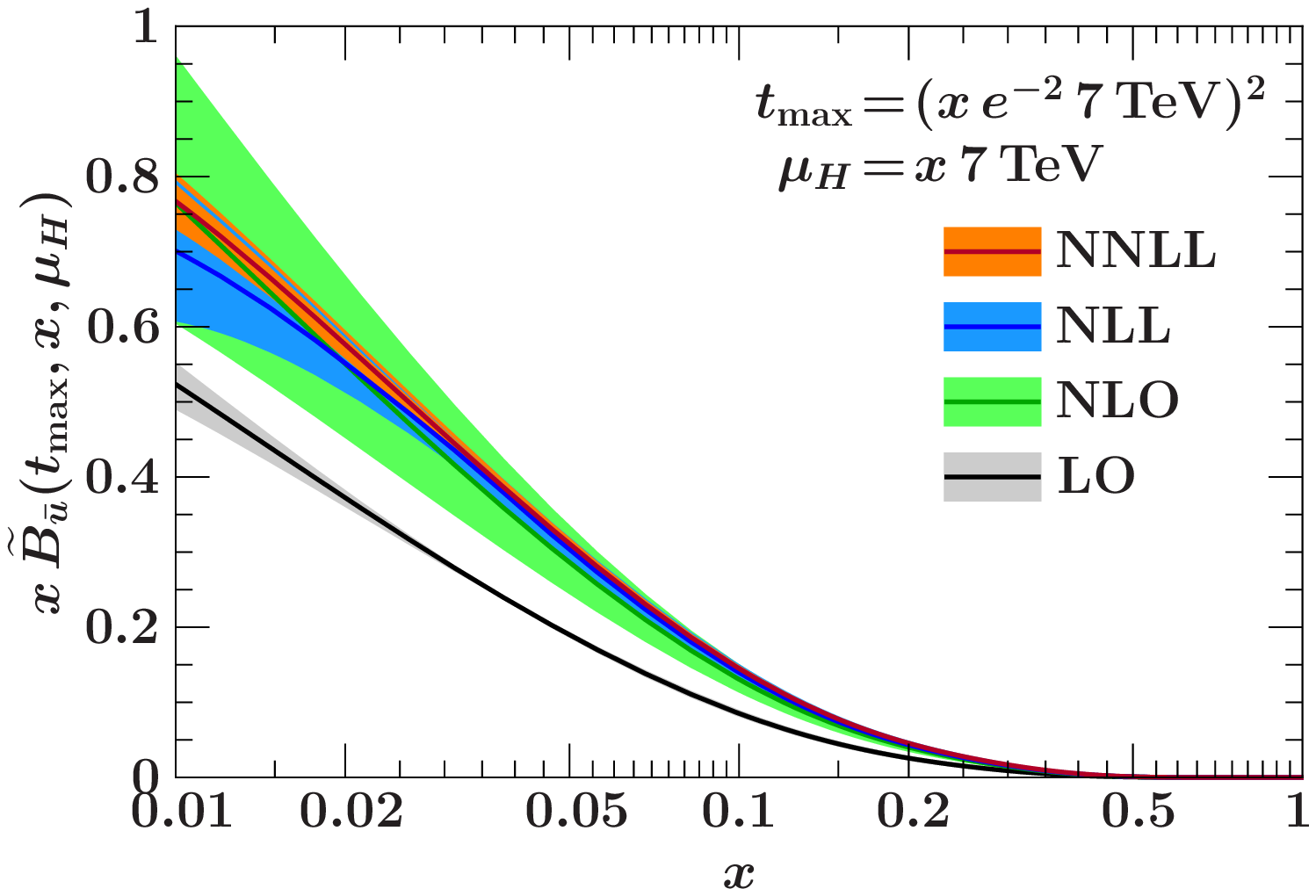}%
\caption{The $u$ (left) and $\bar u$ (right) beam functions at the hard scale $\mu_H = x\, 7 \TeV$ at LO, NLO, NLL, and NNLL, integrated up to $t_\textrm{max} = (x\, e^{-2}\, 7 \TeV)^2$. The bands show the perturbative uncertainties estimated by varying  $\mu_H$ for the fixed-order results and the matching scale $\mu_B^2 \simeq t_\mathrm{max}$ for the resummed results, as explained in the text.}
\label{fig:Bu_muH}}

\FIGURE[ht]{%
\includegraphics[width=0.49\textwidth]{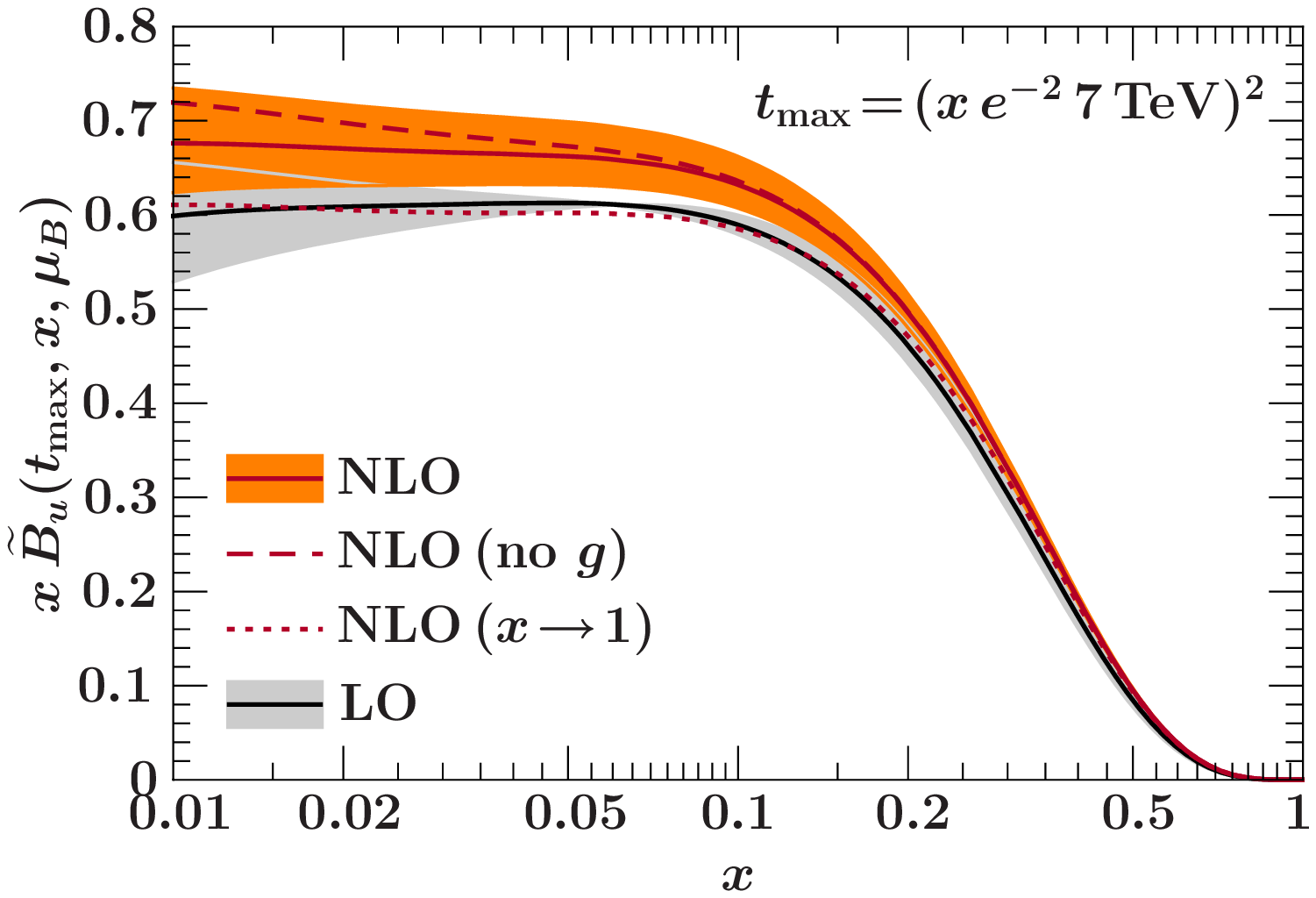}%
\hfill%
\includegraphics[width=0.49\textwidth]{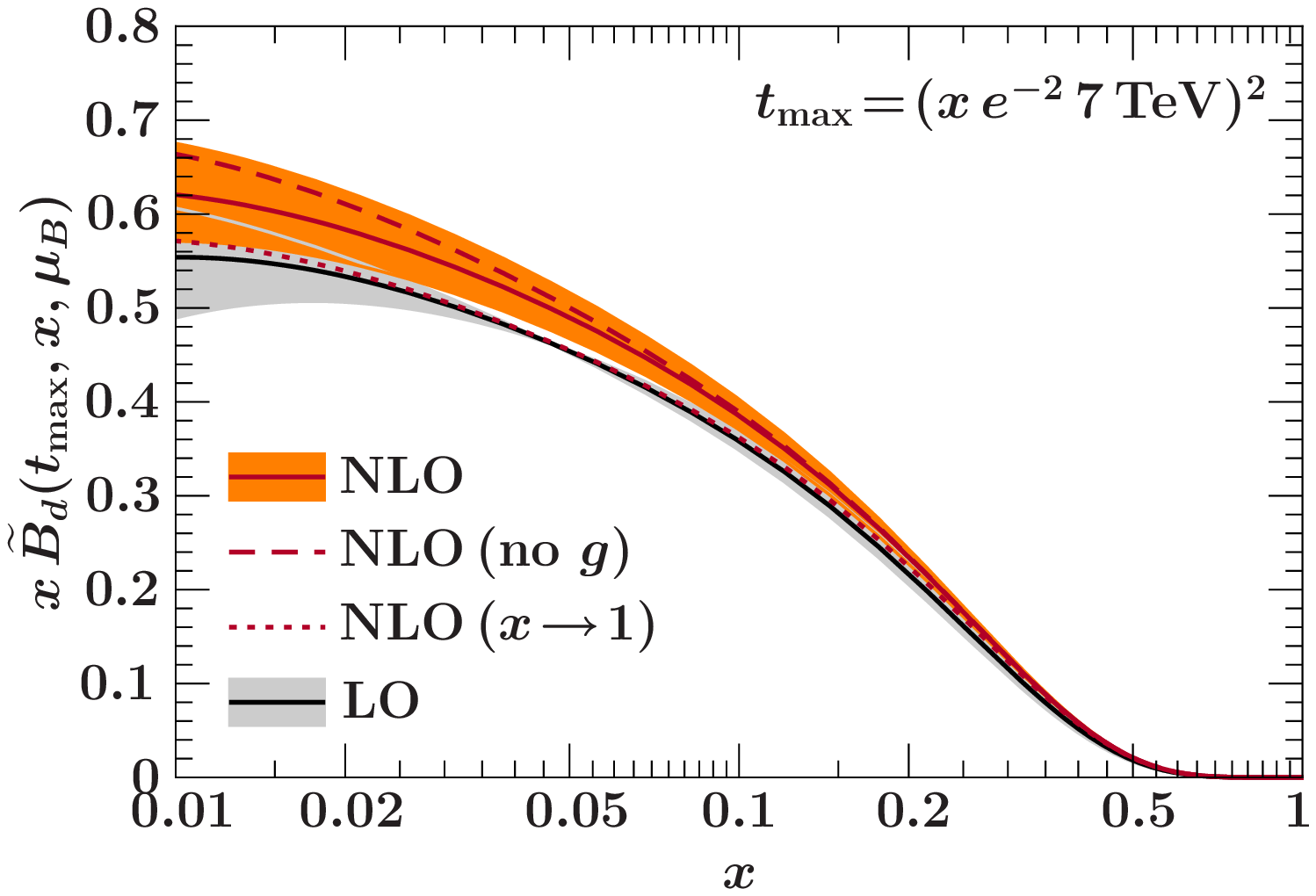}%
\\
\includegraphics[width=0.49\textwidth]{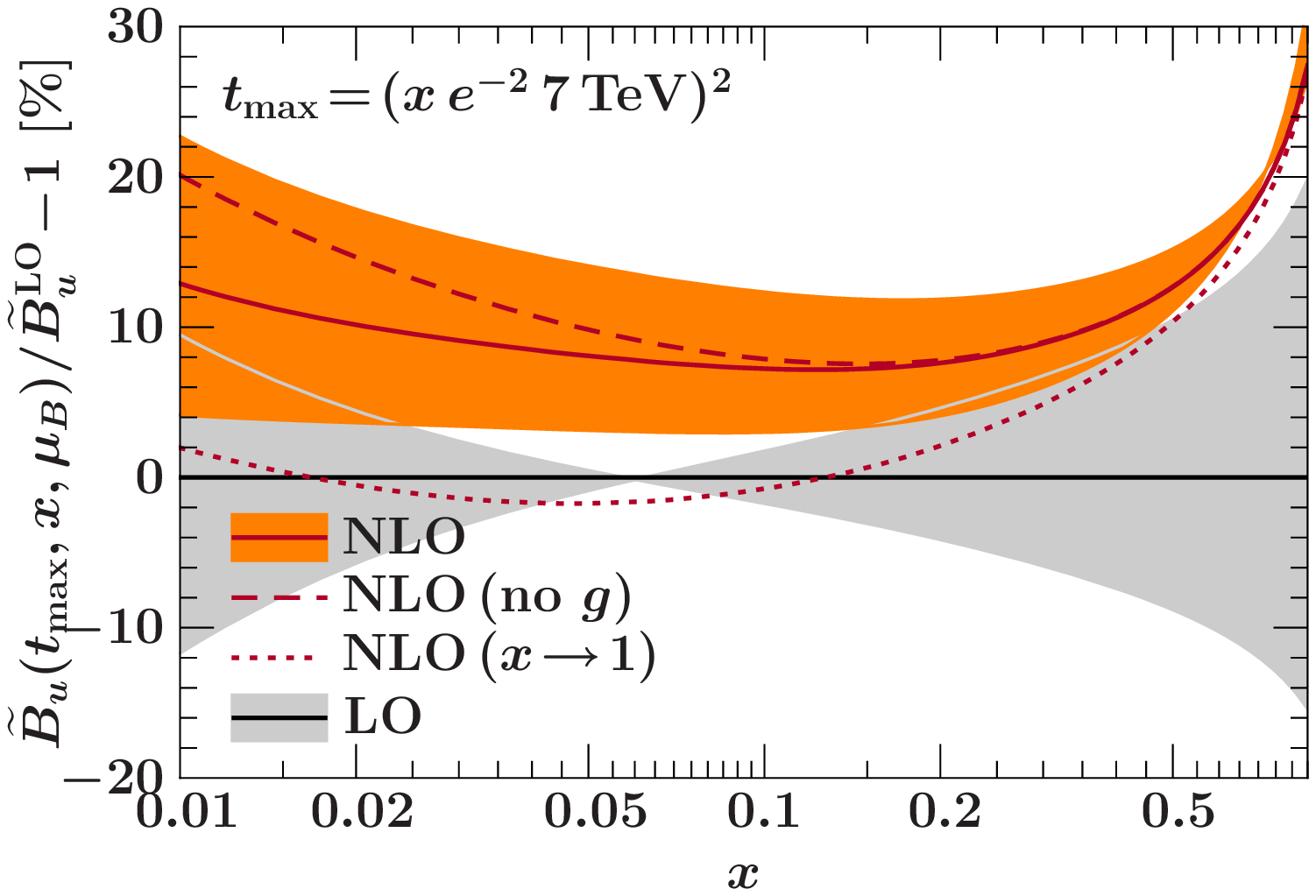}%
\hfill%
\includegraphics[width=0.49\textwidth]{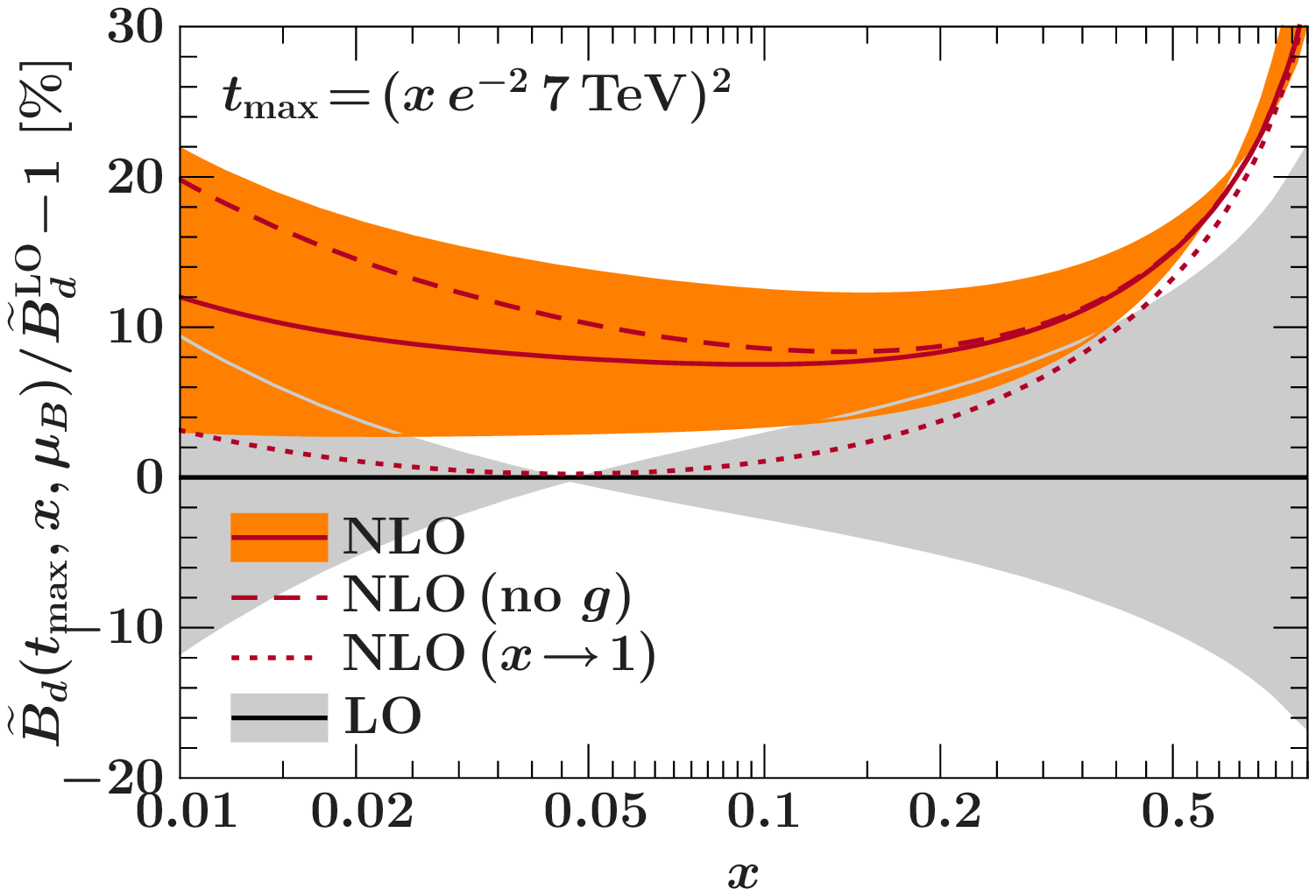}%
\caption{The $u$ (left column) and $d$ (right column) beam functions at the beam scale $\mu_B^2 \simeq t_\mathrm{max}$ at LO and NLO, integrated up to $t_\mathrm{max} = (x e^{-2} 7 \TeV)^2$. The top row shows the functions times $x$. The bottom row shows the relative differences compared to the LO result. Also shown are the NLO beam functions in the threshold limit (dotted) and without the gluon contribution (dashed). The bands show the perturbative scale uncertainties as explained in the text.}
\label{fig:Bq_muB}}

\FIGURE[t]{%
\includegraphics[width=0.49\textwidth]{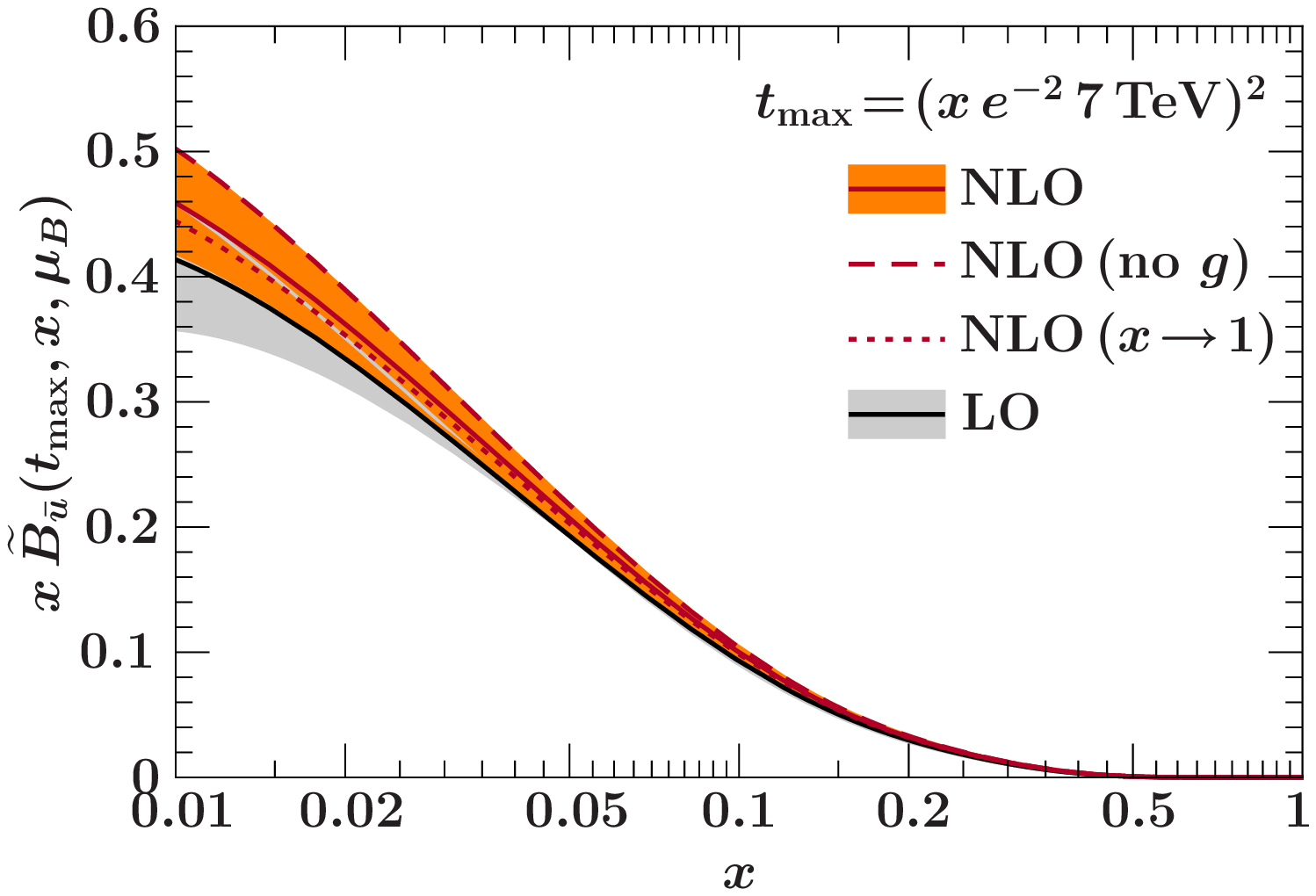}%
\hfill%
\includegraphics[width=0.49\textwidth]{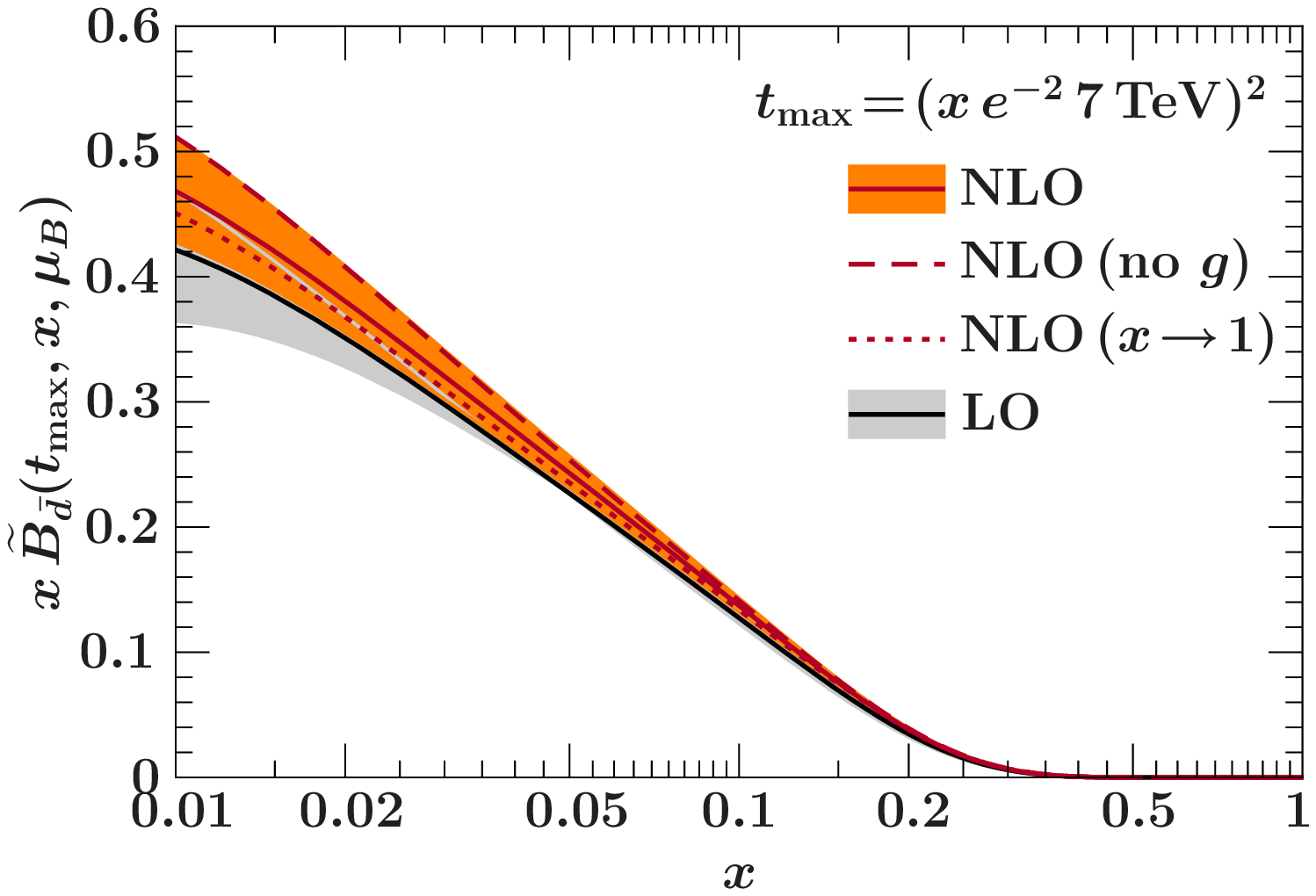}%
\\
\includegraphics[width=0.49\textwidth]{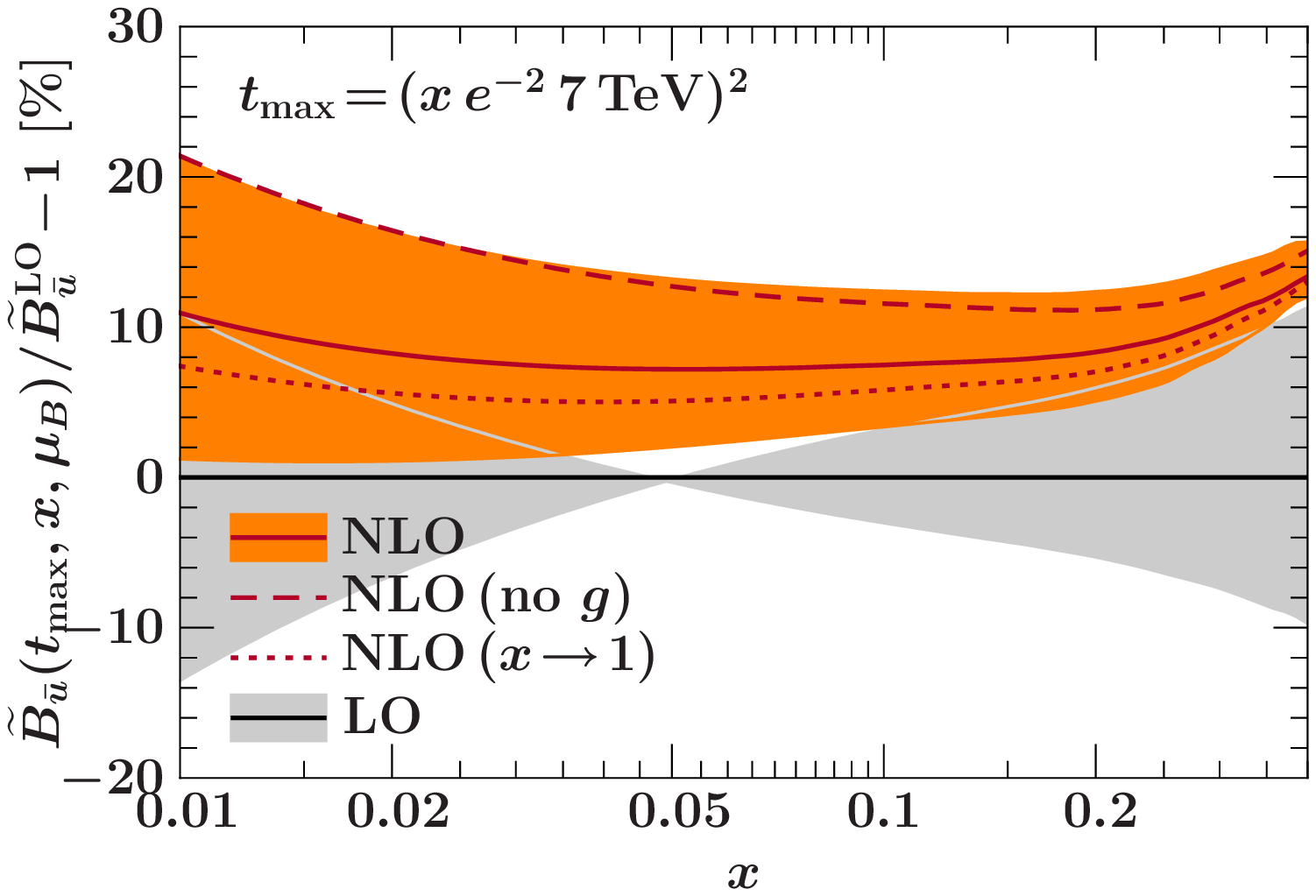}%
\hfill%
\includegraphics[width=0.49\textwidth]{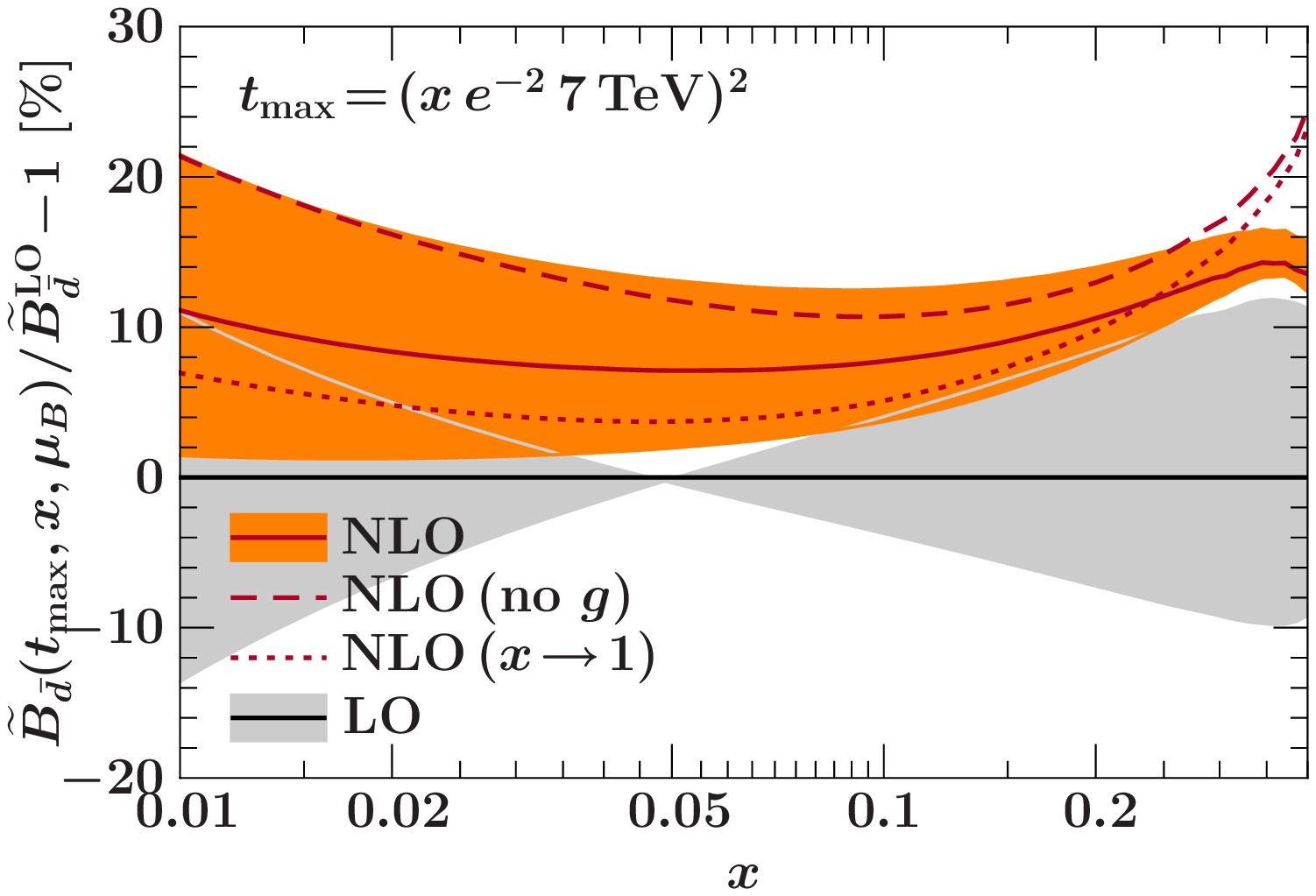}%
\caption{The $\bar u$ (left column) and $\bar d$ (right column) beam functions at the beam scale. The meaning of the curves is analogous to \fig{Bq_muB}.}
\label{fig:Bqbar_muB}}

To be able to plot the beam function as a function of the momentum fraction $x$ including the virtual terms proportional to $\delta(t)$, we consider the integral over $t$ up to some maximum $t_\max$,
\begin{equation}
 \tB_i(t_\max,x,\mu) = \int\! \df t\, B_i(t,x,\mu) \theta(t_\max - t)
\,,\end{equation}
where $B_i(t, x, \mu)$ is given by \eqs{beam_fact}{Brun}. In the plots, we
always choose $t_\max = (x e^{-2} 7 \TeV)^2$, which one should think of as
$t_\max = (e^{-y^\cut} x \Ecm)^2$. Hence, this choice of $t_\max$ corresponds to
a rapidity cut $y^\cut = 2$ for $\Ecm = 7\TeV$ or equivalently $y^\cut = 2.4$
for $\Ecm = 10\TeV$. This is motivated by the upper bound $y^\cut = y_B^\cut \pm
Y$, which follows from the factorization theorem \eq{DYbeamrun} when we
integrate $\tau_B \leq \exp(-2y_B^\cut)$.

Figure~\ref{fig:Bu_muH} shows the integrated $u$ and $\bar u$-quark beam
function $x \tB_i(t_\max,x,\mu_H)$ evaluated at the hard scale $\mu_H = Q = x\,
7 \TeV$. For the fixed-order results at LO (lowest gray band) and NLO (wide
green band), the bands are obtained by varying $\mu_H$ by factors of two, since
this is the scale at which the perturbation series for the matching coefficients
in \eq{beam_fact} is evaluated.  At LO, the resulting variation is entirely due
to the scale dependence of the PDF. At NLO the sizeable variation indicates the
presence of large single and double logarithms of $t_\max/Q^2$ when the fixed
order beam functions are evaluated at $\mu_H$.

For the resummed results at NLL (blue band) and NNLL (orange band) the beam
function OPE, \eq{beam_fact}, is evaluated at the beam scale $\mu_B^2 \simeq
t_\max$, and the beam function is then evolved to $\mu_H$ using its RGE,
\eq{Brun}. In this way, the large logarithms of $\mu_B^2/\mu_H^2 \simeq
t_\max/\mu_H^2$ are resummed. Here the bands correspond to perturbative
uncertainties evaluated by varying the matching scale $\mu_B$ while keeping
$\mu_H$ fixed. The dependence on the scale $\mu_B$ cancels between the fixed
order perturbative result for the beam function and its evolution factor, up to
higher order corrections. An estimate for these higher order corrections is
given by the NLL and NNLL bands. These uncertainty bands show the minimum and
maximum variation in the interval $\sqrt{t_\max}/2 \leq \mu_B \leq
2\sqrt{t_\max}$ (which due to the double-logarithmic series do not occur at the
edges of the interval) with the central value given by the center of the bands.
The NLL result is close to the NLO result, showing that the large logarithms
make up by far the biggest part in the NLO corrections.  Consequently, the
corrections from NLL to NNLL are of reasonable size and within the NLL
uncertainties. Hence, for the beam function at the hard scale, fixed-order
perturbation theory is not applicable. Resummed perturbation theory is
well-behaved and should be used.

To study the perturbative corrections to the beam functions in more detail, we consider them at the scale $\mu_B^2 \simeq t_\max$, where there are no large logarithms and we can use fixed-order perturbation theory. The $u$ and $d$ beam functions at LO and NLO are shown in \fig{Bq_muB}, and the $\bar u$ and $\bar d$ beam functions in \fig{Bqbar_muB}. The top rows show $x \tB_i(t_\max, x, \mu_B)$. The bottom rows show the same results but as relative corrections with respect to the LO results. At LO, the only scale variation comes from the PDFs and the minimum and maximum variations are obtained for $\mu_B = \{\sqrt{t_\max}/2, 2\sqrt{t_\max}\}$ with the central value at $\mu_B = \sqrt{t_\max}$. For the NLO results, the maximum variation in the range $\sqrt{t_\max}/2 \leq \mu_B \leq 2\sqrt{t_\max}$ is approximately attained for $\mu_B =\{0.7 \sqrt{t_\max},2.0\sqrt{t_\max} \}$ and the corresponding central value for $\mu_B = 1.4\sqrt{t_\max}$. To be consistent we use the same central value $\mu_B = 1.4\sqrt{t_\max}$ for the NLO results in the threshold limit and without gluon contribution. In all cases the NLO perturbative corrections are of $\ord{10 \%}$ and exhibit reasonable uncertainties.

The integration limits $x \leq \xi \leq 1$ in the beam function OPE, \eq{beam_fact}, force $z = x/\xi \to 1$ in the limit $x\to 1$. Hence, the threshold terms in \eq{Ithres} are expected to dominate over the non-threshold terms at large values of $x$. This can be seen in \figs{Bq_muB}{Bqbar_muB}, where  the threshold results shown by the dotted lines approach the full results towards large $x$ values where the beam functions vanish. For the quark beam functions in \fig{Bq_muB}, away from the endpoint, $x\lesssim 0.5$, the threshold corrections give a poor approximation to the full NLO corrections. For the antiquark beam functions in \fig{Bqbar_muB}, the threshold result turns out to be relatively close to the full result even for small $x$. However, the reason for this is a relatively strong cancellation between the non-threshold terms in the quark and gluon contributions $\cI_{qq}$ and $\cI_{qg}$ at one loop. As shown by the result without the gluon contribution (dashed lines) the non-threshold terms in the quark and gluon contributions each by themselves are of the same size or larger than the threshold contributions. Note also that for the $\bar d$ beam function the threshold result approaches the no-gluon result rather than the full result at large $x$. A similar but less strong cancellation can also be observed at small $x$ in the quark beam functions. These appear to be accidental cancellations, which depend on both the relative size of the (anti)quark and gluon PDFs as well as the relative size of the non-threshold terms in $\cI_{qq}$ and $\cI_{qg}$. Therefore one must be careful when applying the numerical dominance of the threshold terms to cases where it is not explicitly tested.

It has been argued~\cite{Appell:1988ie, Catani:1998tm} that the steep fall-off of the PDFs causes a systematic enhancement of the partonic threshold region $z\to 1$ even away from the hadronic threshold limit $x\to 1$. This likely explains why the threshold terms in \figs{Bq_muB}{Bqbar_muB} start to dominate already close to the $x$ values where the PDFs are close to zero, rather than strictly near $x = 1$~\cite{Becher:2007ty}. However, our results show that the same arguments do not apply in the relevant region of $x$ where the PDFs and beam functions are substantially nonzero.

\section{Conclusions}
\label{sec:conclusions}

At the LHC or Tevatron, the appropriate description of the initial state of the collision depends on the measurements made on the hadronic final state. The majority of measurements trying to identify a specific hard interaction process do so by finding a certain number of central jets, leptons, or photons that are distinguished from energetic initial-state radiation in the forward direction. These measurements effectively probe the proton at an intermediate beam scale $\mu_B \ll Q$ and the initial state is described by universal beam functions. The beam functions encode initial-state effects including both PDF effects as well as initial-state radiation forming an incoming jet around the incoming hard parton above $\mu_B$.

We have discussed in detail the field-theoretic treatment of beam functions using SCET. We discussed their renormalization properties and showed that they satisfy an RGE with the same anomalous dimension as the jet function to all orders in perturbation theory. The beam function RGE determines the evolution of the initial state above $\mu_B$. It resums a double logarithmic series associated to the virtuality $t$ of the incoming parton, while leaving the parton's identity and momentum fraction $x$ unchanged.

We gave a general discussion of the operator product expansion for the beam functions that allows us to match them onto PDFs $f_j(\xi, \mu_B)$ convoluted with matching coefficients $\cI_{ij}(t, x/\xi, \mu_B)$. The latter encode the effects of the initial-state radiation and are perturbatively calculable at the scale $\mu_B$. We performed this matching at one loop for the quark beam function onto quark and gluon PDFs. Our calculation explicitly confirms at one loop that the quark beam function contains the same IR singularities as the PDFs, and this required a proper handling of zero-bin subtractions.

In \sec{results}, we presented an explicit expression for the resummed beam thrust cross section for Drell-Yan production, $pp\to X\ell^+\ell^-$, with the necessary ingredients for its evaluation at NNLL collected in \app{pert}. An analysis of the cross section at this order is left to a separate publication~\cite{Stewart:2010pd}. Here, we discussed in detail numerical results for the quark beam function at NLO and NNLL. The gluon beam function is important for Higgs production at the LHC. The one-loop matching of the gluon beam function onto gluon and quark PDFs is discussed in a separate publication and used to calculate the Higgs production cross section for small beam thrust at NNLL~\cite{Berger:2010xi}. Another application is to define a $p_T$ dependent beam function to study the $p_T$-spectrum of the Higgs~\cite{Mantry:2009qz}.

So far, effects of strong initial-state radiation that go beyond the inclusive treatment via PDFs have only been studied using Monte Carlo methods and models for initial-state parton showers. The physical picture behind the beam function and initial-state parton showers are in fact in close correspondence. Beam functions and the beam thrust factorization theorem provide a complementary field-theoretic approach to study these effects analytically. Hence, they provide a crucial tool to obtain an accurate description of the initial state, which is mandatory to obtain precise and realistic theory predictions for the LHC.

\begin{acknowledgments}
  This work was supported in part by the Office of Nuclear Physics of the
  U.S.\ Department of Energy under the Contract DE-FG02-94ER40818, and by a
  Friedrich Wilhelm Bessel award from the Alexander von Humboldt foundation.
\end{acknowledgments}

\appendix

\section{Plus Distributions and Discontinuities}
\label{app:disc}

The standard plus distribution for some function $g(x)$ can be defined as
\begin{equation}
\bigl[\theta(x) g(x)\bigr]_+
= \lim_{\beta \to 0} \frac{\df}{\df x} \bigl[\theta(x-\beta)\, G(x) \bigr]
\qquad\text{with}\qquad
G(x) = \int_1^x\!\df x'\, g(x')
\,,\end{equation}
satisfying the boundary condition $\int_0^1 \df x\, [\theta(x) g(x)]_+ = 0$. Two special cases we need are
\begin{align} \label{eq:plusdef}
\cL_n(x)
&\equiv \biggl[ \frac{\theta(x) \ln^n x}{x}\biggr]_+
 = \lim_{\beta \to 0} \biggl[
  \frac{\theta(x- \beta)\ln^n x}{x} +
  \delta(x- \beta) \, \frac{\ln^{n+1}\!\beta}{n+1} \biggr]
\,,\nn\\
\cL^\eta(x)
&\equiv \biggl[ \frac{\theta(x)}{x^{1-\eta}}\biggr]_+
 = \lim_{\beta \to 0} \biggl[
  \frac{\theta(x - \beta)}{x^{1-\eta}} +
  \delta(x- \beta) \, \frac{x^\eta - 1}{\eta} \biggr]
\,.\end{align}
In addition, we need the identity
\begin{equation} \label{eq:distr_id}
 \frac{\theta(x)}{x^{1+\eps}} = - \frac{1}{\eps}\,\delta(x) + \cL_0(x)
  - \eps \cL_1(x) + \ord{\eps^2}
\,,\end{equation}
the Fourier transform
\begin{equation} \label{eq:distr_FT}
\cL_0(x) = - \int\!\frac{\df y}{2\pi}\,e^{\img x y}\,\ln\bigl[\img(y - \img 0) e^{\gamma_E}\bigr]
\,,\end{equation}
and the two limits
\begin{align} \label{eq:limits}
\lim_{\bt\to 0}\biggl[\frac{\theta(x-\bt)\ln(x-\bt)}{x} + \delta(x-\bt)\,\frac{1}{2}\ln^2 \bt\biggr]
&= \cL_1(x) - \frac{\pi^2}{6}\,\delta(x)
\,,\nn\\
\lim_{\bt\to 0} \frac{\theta(x-\bt)\,\bt}{x^2}
&= \delta(x)
\,.\end{align}
Away from $x=0$ these relations are straightforward, while the behavior at $x = 0$ is obtained by taking the integral of both sides. General relations for the rescaling and convolutions of $\cL_n(x)$ and $\cL^\eta(x)$ can be found in App.~B of Ref.~\cite{Ligeti:2008ac}.

The discontinuity of a function $g(x)$ is defined as
\begin{align} \label{eq:Disc_def2}
\Disc_x\, g(x) = \lim_{\beta\to 0} \bigl[ g(x + \img\beta) - g(x - \img\beta) \bigr]
\,.\end{align}
If we are only interested in the discontinuity in some interval in $x$, we simply multiply the right-hand side with the appropriate $\theta$ functions, as in \eq{Disc_def}. If $g(x)$ is real then $\Disc_x g(x) = 2\img\, \Im\, g(x+\img0)$. Two useful identities are
\begin{align} \label{eq:disc_os}
\frac{\img}{2\pi} \Disc_x\, \frac{1}{x^{n+1}}
= \frac{(-1)^n}{n!}\, \delta^{(n)}(x)
\,,\qquad
\frac{\img}{2\pi}\, \Disc_x\, (-x)^{n-\eps}
= (-1)^{n-1} \frac{\sin\pi\eps}{\pi}\, \theta(x) x^{n-\eps}
\,.\end{align}
To derive the last identity, note that
$(-x-\img 0)^{n-\eps} = \exp[ (n-\eps)\ln(-x-\img 0)] =
|x|^{n-\eps} \exp[-\img\pi (n-\eps)\theta(x)] $, so taking the imaginary part gives
$\Im (-x-\img0)^{n-\eps} =  (-1)^n \sin(\pi\eps)\, \theta(x)\, x^{n-\eps}$.

\section{Renormalization of the Beam Function}
\label{app:B_RGE}

In this appendix we derive the general structure of the beam function RGE in \eq{B_RGE} to all orders in perturbation theory. The two essential ingredients will be the known all-order renormalization properties of lightlike Wilson lines~\cite{Brandt:1981kf, Korchemsky:1987wg, Korchemskaya:1992je, Korchemsky:1992xv} and the factorization theorem for the isolated $pp\to X L$ cross section, where $X$ is the hadronic and $L$ the non-hadronic final state. In Ref.~\cite{Stewart:2009yx} we proved that to all orders in perturbation theory and leading order in the power counting this cross section factorizes as
\begin{align} \label{eq:sigBab}
\frac{\df\sigma}{\df q^2 \df Y \df B_a^+\df B_b^+}
&= \sum_{ij} H_{ij}(q^2, Y, \mu) \int\!\df k_a^+\, \df k_b^+\, S^{ij}_\hemiin(k_a^+,k_b^+, \mu)
\nn\\ & \quad\times
 q^2 B_i[\w_a(B_a^+ - k_a^+), x_a, \mu] B_j[\w_b(B_b^+ -k_b^+),x_b, \mu]
\,.\end{align}
The sum over $ij$ runs over parton species $ij = \{gg, u\bar u, \bar u u, d\bar d, \bar d d, \ldots\}$. The soft function does not depend on the quark flavor, and its superscript only refers to the color representation. The variables $q^2$ and $Y$ are the total invariant mass and rapidity of the non-hadronic system $L$, $x_{a,b} = \sqrt{q^2} e^{\pm Y}/\Ecm$ and $\w_{a,b} = x_{a,b} \Ecm$.
The hadronic variables $B_{a,b}^+$ are the hemisphere plus momenta of the hadronic final state $X$ with respect to the directions $n_a$ and $n_b$ of the incoming protons. Their precise definition will not be relevant for our discussion.

The three ingredients in \eq{sigBab} are the renormalized hard, beam, and soft functions, $H_{ij}(q^2, Y, \mu)$, $B_i(t, x, \mu)$, $S^{ij}_\hemiin(k_a^+, k_b^+, \mu)$. Their dependence on the renormalization scale $\mu$ must cancel in \eq{sigBab}, because the cross section must be $\mu$ independent. The structure of the RGE for the hard and soft functions thus uniquely determines the allowed structure of the beam function RGE.

The hard function is a contraction between the relevant leptonic matrix element squared and the square of the Wilson coefficients of the color-singlet $q\bar q$ and $gg$ local SCET currents
\begin{equation} \label{eq:Oi_SCET}
O_{q\bar{q}}^{\alpha\beta}
= \bar\chi_{n_a,-\w_a}^\alpha\,\chi_{n_b,\w_b}^\beta
\,,\qquad
O_{gg}^{\mu\nu}
= \sqrt{\w_a\,\w_b}\,\cB_{n_a,-\w_a\perp }^{\mu c}\, \cB_{\bn_b,-\w_b\perp}^{\nu c}
\,,\end{equation}
where $\alpha$ and $\beta$ are spin indices. In each collinear sector, total label momentum and fermion number for each quark flavor are conserved. Thus, the currents cannot mix with each other and are multiplicatively renormalized. Furthermore, RPI-III invariance implies that the RGE for the currents can only depend on $q^2 = \w_a \w_b$. The renormalization of these SCET currents also does not depend on their spin structure, so the RGE for the hard function must have the same structure as for the currents. Therefore, to all orders in perturbation theory we have (with no sum on $ij$)
\begin{equation} \label{eq:H_RGE}
\mu \frac{\df}{\df\mu} H_{ij}(q^2, Y, \mu) = \gamma^{ij}_H(q^2, \mu)\, H_{ij}(q^2, Y, \mu)
\,.\end{equation}

Next, the incoming hemisphere soft function, $S^{ij}_\hemiin(k_a^+, k_b^+, \mu)$, is given by the vacuum matrix element of incoming soft lightlike Wilson lines along the $n_a$ and $n_b$ directions. In position space,
\begin{equation}
\tS^{ij}_\hemiin(y_a^-, y_b^-, \mu)
= \int\! \df k_a^+ \df k_b^+\, e^{-\img (k_a^+ y_a^- + k_b^+ y_b^-)/2}\,S^{ij}_\hemiin(k_a^+, k_b^+, \mu)
\end{equation}
has two cusps, one at spacetime position $0$ and one at $y = y_a^- n_a/2 + y_b^- n_b/2$. The renormalization properties of lightlike Wilson lines with cusps~\cite{Brandt:1981kf, Korchemsky:1987wg, Korchemskaya:1992je, Korchemsky:1992xv} then imply that to all orders in perturbation theory,
\begin{align} \label{eq:tS_RGE}
\mu \frac{\df}{\df\mu} \tS^{ij}_\hemiin(y_a^-, y_b^-, \mu)
&= \tga_S^{ij}(y_a^-, y_b^-, \mu)\, \tS^{ij}_\hemiin(y_a^-, y_b^-, \mu)
\,,\\\nn
\tga_S^{ij}(y_a^-, y_b^-, \mu)
&= 2\Gamma^i_\cusp(\alpha_s) \Bigl[
- \ln\Bigl(\img \frac{y_a^-\! - \img 0}{2} \mu e^{\gamma_E}\Bigr)
- \ln\Bigl(\img \frac{y_b^-\! - \img 0}{2} \mu e^{\gamma_E}\Bigr) \Bigr]
\!+ \gamma_S^{ij}(\alpha_s)
\,,\end{align}
where $\Gamma^i_\cusp$ is the cusp anomalous dimension for quarks/antiquarks or gluons, and $\gamma_S^{ij}[\alpha_s(\mu)]$ and $\Gamma_\cusp^i[\alpha_s(\mu)]$ depend only indirectly on $\mu$ via $\alpha_s(\mu)$. Dimensional analysis and RPI-III invariance imply that the single logarithm multiplying $2\Gamma^i_\cusp$ scales like $\ln(y_a^- y_b^- \mu^2)$. (The additional dimensionless factors are chosen for convenience. Any change in them can be absorbed into $\gamma_S^{ij}(\alpha_s)$.) The correct overall sign and $\img 0$ prescription for the logarithms can be deduced from the explicitly known one-loop result~\cite{Stewart:2009yx, Schwartz:2007ib, Fleming:2007xt}.

Taking the Fourier transform of the cross section in \eq{sigBab} with respect to $B_a^+$ and $B_b^+$ and differentiating the result with respect to $\mu$ yields
\begin{align}
0 &= \mu\frac{\df}{\df\mu} \biggl[
\sum_{ij} H_{ij}(q^2, Y, \mu)
\tB_i\Bigl(\frac{y_a^-}{2\w_a}, x_a, \mu \Bigr) \tB_j\Bigl(\frac{y_b^-}{2\w_b}, x_b, \mu \Bigr)
\tS^{ij}_\hemiin(y_a^-, y_b^-, \mu) \biggr]
\nn\\
&= \sum_{ij} H_{ij}(q^2, Y, \mu) \tS^{ij}_\hemiin(y_a^-, y_b^-, \mu)
\nn\\ & \quad\times
\Bigl[ \gamma_H^{ij}(\w_a \w_b, \mu) + \tga_S^{ij}(y_a^-, y_b^-, \mu) + \mu\frac{\df}{\df\mu} \Bigr]
\tB_i\Bigl(\frac{y_a^-}{2\w_a}, x_a, \mu\Bigr) \tB_j\Bigl(\frac{y_b^-}{2\w_b}, x_b, \mu\Bigr)
\,.\end{align}
The factorization theorem for the cross section neither depends on the choice of $L$, which affects the form of $H_{ij}$ for different $ij$, nor the type of the colliding hadrons. This implies that each term in the sum over $ij$ must vanish separately. (For example, choosing Drell-Yan, $L = \ell^+\ell^-$, there is no contribution from $ij = gg$, so the quark and gluon contributions are separately zero. Then, by assigning arbitrary electroweak quark charges, the contribution from each quark flavor must vanish separately. Finally, the $ij = q\bar{q}$ and $ij = \bar{q}q$ contributions for a single quark flavor $q$ must vanish separately by choosing various different incoming hadrons.) Therefore, the RGE for the product of the two beam functions is
\begin{equation} \label{eq:tBB_RGE}
\Bigl[ \gamma_H^{ij}(\w_a \w_b, \mu) + \tga_S^{ij}(y_a^-, y_b^-, \mu) + \mu\frac{\df}{\df\mu} \Bigr]
\tB_i\Bigl(\frac{y_a^-}{2\w_a}, x_a, \mu\Bigr) \tB_j\Bigl(\frac{y_b^-}{2\w_b}, x_b, \mu\Bigr) = 0
\,,\end{equation}
which shows that the beam functions in position space renormalize multiplicatively and independently of $x_{a,b}$. The RGE for each individual beam function can only depend on the RPI-III invariant $y^-/2\w$ and obviously cannot depend on the variables of the other beam function. Hence, we find that to all orders in perturbation theory
\begin{equation} \label{eq:tB_RGE}
\mu\frac{\df}{\df\mu} \tB_i\Bigl(\frac{y-}{2\w}, x, \mu\Bigr)
= \tga_B^i\Bigl(\frac{y^-}{2\w}, \mu \Bigr) \tB_i\Bigl(\frac{y-}{2\w}, x, \mu\Bigr)
\,,\end{equation}
which is the result we set out to prove in this Appendix. Using \eq{tB_RGE} together with \eq{tBB_RGE}, the anomalous dimensions must satisfy the consistency condition
\begin{equation}
0 =
\gamma_H^{ij}(\w_a \w_b, \mu) + \tga_S^{ij}(y_a^-, y_b^-, \mu)
+ \tga_B^i\Bigl(\frac{y^-_a}{2\w_a}, \mu \Bigr) + \tga_B^j\Bigl(\frac{y^-_b}{2\w_b}, \mu \Bigr)
\,.\end{equation}
Given the form of $\tga_S^{ij}$ in \eq{tS_RGE}, it follows that the anomalous dimensions are given to all orders by
\begin{align} \label{eq:gammas}
\gamma_H^{ij}(\w_a \w_b, \mu)
&= 2\Gamma^i_\cusp(\alpha_s) \ln\frac{\w_a \w_b}{\mu^2} + \gamma_H^{ij}(\alpha_s)
\,,\nn\\
\tga_B^i\Bigl(\frac{y^-}{2\w}, \mu \Bigr)
&= 2\Gamma^i_\cusp(\alpha_s) \ln\Bigl(\img \frac{y^-\!- \img 0}{2\w}\mu^2 e^{\gamma_E}\Bigr)
+ \gamma_B^i(\alpha_s)
\,,\nn\\
\gamma_S^{ij}(\alpha_s) &= -\gamma_H^{ij}(\alpha_s) - \gamma_B^i(\alpha_s) - \gamma_B^j(\alpha_s)
\,.\end{align}
Taking the Fourier transform using \eq{distr_FT}, the momentum-space anomalous dimensions become
\begin{align} \label{eq:gammas2}
\gamma_S^{ij}(k_a^+, k_b^+, \mu)
&= 2\Gamma^i_\cusp(\alpha_s)
\biggl[\frac{1}{\mu}\cL_0\Bigl(\frac{k_a^+}{\mu}\Bigr) \delta(k_b^+) + \delta(k_a^+) \frac{1}{\mu}\cL_0\Bigl(\frac{k_b^+}{\mu}\Bigr) \biggr] + \gamma_S^{ij}(\alpha_s)\, \delta(k_a^+) \delta(k_b^+)
\,,\nn\\
\gamma_B^i(t, \mu)
&= - 2\Gamma^i_\cusp(\alpha_s)\, \frac{1}{\mu^2}\cL_0\Bigl(\frac{t}{\mu^2}\Bigr)
+ \gamma_B^i(\alpha_s)\, \delta(t)
\,.\end{align}

The same all-order structure of the soft anomalous dimension as in \eq{gammas2} was obtained in Ref.~\cite{Fleming:2007xt} for the hemisphere soft function with outgoing Wilson lines in $e^+e^-\to 2$ jets using analogous consistency conditions. In fact, the hard SCET currents here and there are the same and in \subsec{B_RGE} we proved that the anomalous dimensions for the beam and jet function are the same, $\gamma_B^i = \gamma_J^i$. Hence, the hemisphere soft functions with incoming and outgoing Wilson lines have in fact identical anomalous dimensions to all orders.

\section{Matching Calculation in Pure Dimensional Regularization}
\label{app:dimreg}

Here we repeat the NLO \SCETa to \SCETb matching calculation from \sec{oneloop}
using dimensional regularization for both the UV and IR. Since we only change
the IR regulator, the final results for the matching coefficients $\cI_{ij}(t,
z, \mu)$ should not be affected.

In pure dimensional regularization all the loop diagrams contributing to the bare matrix elements of $\oq_q$ vanish, since by dimensional analysis there is no Lorentz invariant quantity they can depend on. Hence, including the counter terms in \eq{Zpdf} to subtract the UV divergences, the renormalized matrix elements consist of pure IR divergences with opposite signs to the UV divergences,
\begin{align} \label{eq:fDR}
\Mae{q_n}{\oq_q(\w,\mu)}{q_n}^\one
&= - \frac{1}{\eps}\,\frac{\alpha_s(\mu) C_F}{2\pi}\, \theta(z) P_{qq}(z)
\,,\nn\\
\Mae{g_n}{\oq_q(\w,\mu)}{g_n}^\one
&=  - \frac{1}{\eps}\,\frac{\alpha_s(\mu) T_F}{2\pi}\, \theta(z) P_{qg}(z)
\,.\end{align}
This shows explicitly that the conventional $\overline{\rm MS}$ definition of the PDFs in QCD, which also yields \eq{fDR}, is indeed identical to the SCET definition used in our OPE for the beam function.

Considering the beam function matrix elements, the bare results for \figs{Bone_c}{Bone_d} now vanish, because their loop integrals are again scaleless. For the remaining diagrams we can reuse the intermediate results from \subsec{NLO_B} before carrying out the Feynman parameter integrals and taking the discontinuity. Setting $t' = 0$ the denominator in the Feynman parameter integrals in \eq{I12} becomes
$(1-\alpha)A - \alpha B = t(1 - \alpha/z)$. In this case it easier to carry out the integral after taking the discontinuity. The discontinuity we need is
\begin{equation}
\frac{\img}{2\pi}\text{Disc}_{t>0} \Bigl[\Bigl(1-\frac{\alpha}{z}\Bigr)t\Bigr]^{-1-\eps}
=  \frac{\sin \pi\eps}{\pi}\, \frac{\theta(t)}{t^{1+\eps}}\, \theta\Bigl(\frac{\alpha}{z}-1\Bigr) \Bigl(\frac{\alpha}{z}-1\Bigr)^{-1-\eps}
\,,\end{equation}
where we used \eq{disc_os}. Since we require $z > 0$, the first $\theta$ function becomes $\theta(\alpha - z)$, and so we have
\begin{align} \label{eq:discI12}
-\theta(z)\, \frac{\img}{2\pi}\Disc_{t>0}\, I_1(A, B, \eps)
&= -\theta(z)\,\frac{\sin \pi\eps}{\pi}\, \frac{\theta(t)}{t^{1+\eps}}
\int_0^1\!\df\alpha\, \theta(\alpha - z) \Bigl(\frac{\alpha}{z}-1\Bigr)^{-1-\eps}
\nn\\
&= \theta(z)\,\frac{\sin\pi\eps}{\pi\eps}\, \frac{\theta(t)}{t^{1+\eps}}\, \theta(1-z) z^{1+\eps}(1-z)^{-\eps}
\,,\nn\\
-\theta(z)\, \frac{\img}{2\pi}\Disc_{t>0}\, I_2(A, B, \eps)
&= \theta(z)\,\frac{\sin\pi\eps}{\pi\eps(1-\eps)}\, \frac{\theta(t)}{t^{1+\eps}}\, \theta(1-z)z^{1+\eps}(1-z)^{1-\eps}
\,.\end{align}
For \fig{Bone_a}, using \eqs{Ba}{discI12} we obtain
\begin{align} \label{eq:BaDR}
& \Mae{q_n}{\theta(\w)\op^\bare_q(t,\w)}{q_n}^{(a)}
\nn\\ & \qquad
= \frac{\alpha_s(\mu)C_F}{2\pi}\,\frac{\theta(z)}{z}\, \Gamma(1+\eps) (e^{\gamma_E} \mu^2)^\eps (1-\eps)^2
  \Bigl[-\frac{\img}{2\pi}\Disc_{t>0}\, I_2(A, B, \eps) \Bigr]
\nn\\ & \qquad
= \frac{\alpha_s(\mu)C_F}{2\pi}\,\theta(z)\theta(1-z)(1-z) \Gamma(1+\eps) (e^{\gamma_E} \mu^2)^\eps (1-\eps)
\frac{\sin\pi\eps}{\pi\eps}\, \frac{\theta(t)}{t^{1+\eps}}\, \Bigl(\frac{z}{1-z}\Bigr)^{\eps}
\nn \\ & \qquad
=  \frac{\alpha_s(\mu)C_F}{2\pi}\, \theta(z) \theta(1-z) (1-z) \biggl\{
\frac{1}{\mu^2} \cL_0\Bigl(\frac{t}{\mu^2}\Bigr)
+ \delta(t) \Bigl(-\frac{1}{\eps} + \ln \frac{1-z}{z} + 1\Bigr)
\biggr\}
\,,\end{align}
where in the last step we used \eq{distr_id} to expand in $\eps$. For \fig{Bone_b}, we start from the third line in \eq{Bb1} and using \eqs{discI12}{distr_id} we get
\begin{align} \label{eq:BbDR}
 & \Mae{q_n}{\theta(\w)\op_q^\bare(t,\w)}{q_n}^{(b)}
\nn\\ & \qquad
= \frac{\alpha_s(\mu)C_F}{\pi}\, \frac{\theta(z)}{1 -z}\, \Gamma(1+\eps)(e^{\gamma_E} \mu^2)^\eps
  \Bigl[-\frac{\img}{2\pi}\Disc_{t>0}\, I_1(A, B, \eps) \Bigr]
\nn\\ & \qquad
= \frac{\alpha_s(\mu)C_F}{\pi}\, \theta(z) \Gamma(1+\eps)(e^{\gamma_E} \mu^2)^\eps
\frac{\sin\pi\eps}{\pi\eps}\, \frac{\theta(t)}{t^{1+\eps}}\, \frac{\theta(1-z)z^{1+\eps} }{(1-z)^{1+\eps}}
\nn\\ & \qquad
= \frac{\alpha_s(\mu) C_F}{\pi}\, \theta(z) \biggl\{
\biggl[- \frac{1}{\eps}\,\delta(t) + \frac{1}{\mu^2} \cL_0\Bigl(\frac{t}{\mu^2}\Bigr) \biggr]
\Bigl[- \frac{1}{\eps}\,\delta(1-z) + \cL_0(1-z)z \Bigr]
\nn \\ & \qquad\quad
+ \frac{1}{\mu^2} \cL_1\Bigl(\frac{t}{\mu^2}\Bigr) \delta(1 - z)
+  \delta(t) \Bigl[\cL_1(1 - z)z  - \cL_0(1 - z) z \ln z -  \frac{\pi^2}{12} \delta(1 - z) \Bigr]
 \biggr\}
\,.\end{align}
Adding up \eqs{BaDR}{BbDR}, the bare quark matrix element in pure dimensional regularization becomes
\begin{align} \label{eq:BqbareDR}
&\Mae{q_n}{\theta(\w)\op_q^\bare(t,\w)}{q_n}^\one
\nn\\ & \qquad
= \frac{\alpha_s(\mu) C_F}{2\pi}\,\theta(z) \biggl\{
   \biggl[ \delta(t) \Bigl(\frac{2}{\eps^2} +
    \frac{3}{2\eps} \Bigr) -
    \frac{2}{\eps}\, \frac{1}{\mu^2} \cL_0\Bigl(\frac{t}{\mu^2}\Bigr)
  \biggr]\delta(1-z)
  - \frac{1}{\eps}\,\delta(t) P_{qq}(z)
\nn \\ & \qquad\quad
  +\frac{2}{\mu^2} \cL_1\Bigl(\frac{t}{\mu^2}\Bigr)\delta(1-z) +
  \frac{1}{\mu^2} \cL_0\Bigl(\frac{t}{\mu^2}\Bigr)\cL_0(1-z)(1 + z^2)
\nn\\ & \qquad\quad
  + \delta(t) \biggl[
  \cL_1(1 - z)(1 + z^2)
  -\frac{\pi^2}{6}\, \delta(1 - z)
  + \theta(1 - z)\Bigl(1 - z - \frac{1 + z^2}{1 - z}\ln z \Bigr) \biggr]
  \biggr\}
\,.\end{align}
We can now proceed in two ways to obtain the matching coefficient $\cI_{qq}(t, z, \mu)$.

First, we can subtract $\delta(t)$ times \eq{fDR} from \eq{BqbareDR} to obtain the bare matching coefficient. This simply removes the $(1/\eps)\delta(t) P_{qq}(z)$ in the first line of \eq{BqbareDR}. Assuming that the IR divergences between the PDF and beam function cancel (and including the vanishing zero-bin) the remaining poles in the first line are of UV origin and determine the necessary $\overline{\mathrm{MS}}$ counter term, reproducing our previous result for $Z_B^q(t, \mu)$ in \eq{ZB}.

Alternatively, we can use our general result that the beam function has the same renormalization as the jet function. In this case, we subtract the one-loop counter term for $\op_q^\bare$ in \eq{ZB} (which is already known from the jet function's renormalization) from \eq{BqbareDR} to obtain the renormalized quark matrix element, which equals \eq{BqbareDR} without the $[...]\delta(1-z)$ term in the first line. The remaining $1/\eps$ pole must then be of IR origin, so we again have an explicit check that the IR divergences in the beam function match those of the PDF in \eq{fDR}. Either way, the finite terms in the last two lines of \eq{BqbareDR} determine the renormalized matching coefficient $\cI_{qq}(t, z, \mu)$, which agrees with our previous result in \eq{Iresult}.

For the gluon matrix element, \fig{Bone_f} again does not contribute. For \fig{Bone_e}, starting from the third line of \eq{Be}, we find
\begin{align}  \label{eq:BgbareDR}
 & \Mae{g_n}{\theta(\w)\op_q^\bare(t,\w)}{g_n}^\one
\\ & \qquad
= \frac{\alpha_s(\mu)T_F}{2\pi}\,\frac{\theta(z)}{z}\,
 \Gamma(1+\eps) (e^{\gamma_E} \mu^2)^\eps \Bigl(\frac{1-\eps}{1-z} - 2z\Bigr)
  \Bigl[-\frac{\img}{2\pi}\Disc_{t>0}\, I_2(A, B, \eps) \Bigr]
\nn\\ & \qquad
= \frac{\alpha_s(\mu)T_F}{2\pi}\,\theta(z)\theta(1-z)
 \Gamma(1+\eps) (e^{\gamma_E} \mu^2)^\eps (1 - 2z + 2z^2 -\eps)
\frac{\sin\pi\eps}{\pi\eps(1-\eps)}\, \frac{\theta(t)}{t^{1+\eps}}\, \Bigl(\frac{z}{1-z}\Bigr)^\eps
\nn\\\nn & \qquad
 = \frac{ \alpha_s(\mu) T_F }{2\pi}\, \theta(z) \biggl\{
\frac{1}{\mu^2} \cL_0\Bigl(\frac{t}{\mu^2}\Bigr) P_{qg}(z)
 + \delta(t) \biggl[P_{qg}(z)\Bigl(-\frac{1}{\eps} + \ln\frac{1-z}{z} - 1\Bigr) +  \theta(1-z) \biggr]
\biggr\}
\,.\end{align}
The same discussion as for the quark matrix element above can be repeated for the gluon matrix element. The $(1/\eps)\delta(t)P_{qg}(z)$ term matches the IR divergence in the PDF in \eq{fDR}. Since there are no further poles, no UV renormalization is required and the quark and gluon operators do not mix. The finite terms in \eq{BgbareDR} then determine the matching coefficient $\cI_{qg}(t, z, \mu)$, reproducing our previous result in \eq{Iresult}.

\section{Perturbative Results}
\label{app:pert}

In this appendix we collect perturbative results relevant for the Drell-Yan beam thrust cross section
in \eq{DYbeamrun}.

\subsection{Fixed-Order Results}
\label{app:HandS}

The one-loop Wilson coefficient from matching the quark current from QCD onto SCET was computed in Refs.~\cite{Manohar:2003vb, Bauer:2003di},
\begin{equation}
C(q^2, \mu) = 1 + \frac{\alpha_s(\mu)\,C_F}{4\pi} \biggl[-\ln^2 \Bigl(\frac{-q^2-\img 0}{\mu^2}\Bigr) + 3 \ln \Bigl(\frac{-q^2-\img 0}{\mu^2}\Bigr) - 8 + \frac{\pi^2}{6} \biggr]
\,,\end{equation}
in agreement with the one-loop quark form factors. The hard function is given by the square of the Wilson coefficient~\cite{Stewart:2009yx}
\begin{equation} \label{eq:H_oneloop}
H_{q\bar q}(q^2, \mu) = H_{\bar q q}(q^2, \mu)
= \biggl[ Q_q^2 + \frac{(v_q^2 + a_q^2) (v_\ell^2+a_\ell^2) - 2 Q_q v_q v_\ell (1-m_Z^2/q^2)}
{(1-m_Z^2/q^2)^2 + m_Z^2 \Gamma_Z^2/q^4} \biggr] \Abs{C(q^2, \mu)}^2
\,,\end{equation}
where we included the prefactor from the leptonic matrix element, $Q_q$ is the quark charge in units of $\abs{e}$, $v_{\ell,q}$ and $a_{\ell,q}$ are the standard vector and axial couplings of the leptons and quarks, and $m_Z$ and $\Gamma_Z$ are the mass and width of the $Z$ boson.

As discussed in Ref.~\cite{Stewart:2009yx}, the one-loop result for the beam thrust soft function can be extracted from the one-loop incoming hemisphere soft function~\cite{Schwartz:2007ib, Fleming:2007xt}, yielding
\begin{equation}
S_B(k^+,\mu) = \delta(k^+) + \frac{\alpha_s(\mu)\,C_F}{2\pi} \biggl[
-\frac{8}{\mu} \cL_1\Bigl(\frac{k^+}{\mu}\Bigr) + \frac{\pi^2}{6}\, \delta(k^+) \biggr]
\,.\end{equation}

Our one-loop results for the matching coefficients in the beam function OPE in \eq{beam_fact} are given in \eq{Iresult}.

\subsection{Renormalization Group Evolution}
\label{app:rge}

The RGE and anomalous dimension for the hard Wilson coefficients are~\cite{Manohar:2003vb, Bauer:2003di}
\begin{equation} \label{eq:C_RGE}
\mu \frac{\df}{\df\mu} C(q^2, \mu) = \gamma_H^q(q^2, \mu)\, C(q^2, \mu)
\,,\quad
\gamma_H^q(q^2, \mu) =
\Gamma_\cusp^q(\alpha_s) \ln\frac{- q^2 - \img 0}{\mu^2} + \gamma_H^q(\alpha_s)
\,.\end{equation}
The anomalous dimension for the $q\bar{q}$ hard function in \eqs{H_RGE}{gammas} is given by
$\gamma_H^{q\bar{q}}(q^2, \mu) = 2\mathrm{Re}[\gamma_H^q(q^2, \mu)]$. The expansion coefficients of $\Gamma_\cusp^q(\alpha_s)$ and $\gamma_H^q(\alpha_s)$ are given below in \eqs{Gacuspexp}{gaHexp}. The
solution of the RGE in \eq{C_RGE} yields for the evolution of the hard function
\begin{align} \label{eq:Hrun}
H_{q\bar q}(q^2, \mu) &= H_{q\bar q}(q^2, \mu_0)\, U_H(q^2, \mu_0, \mu)
\,,\qquad
U_H(q^2, \mu_0, \mu)
= \Bigl\lvert e^{K_H(\mu_0, \mu)} \Bigl(\frac{- q^2 - \img 0}{\mu_0^2}\Bigr)^{\eta_H(\mu_0, \mu)} \Bigr\rvert^2
\,,\nn \\
K_H(\mu_0,\mu) &= -2 K^q_\Gamma(\mu_0,\mu) + K_{\gamma_H^q}(\mu_0,\mu)
\,, \qquad
\eta_H(\mu_0,\mu) = \eta_\Gamma^q(\mu_0,\mu)
\,,\end{align}
where the functions $K_\Gamma^i(\mu_0, \mu)$, $\eta_\Gamma^i(\mu_0, \mu)$ and $K_\gamma$ are given below in \eq{Keta_def}.

The beam function RGE is [see \eqs{B_RGE}{gaB_gen}]
\begin{align}
\mu \frac{\df}{\df \mu} B_i(t, x, \mu) &= \int\! \df t'\, \gamma_B^i(t-t',\mu)\, B_i(t', x, \mu)
\,,\nn\\
\gamma_B^i(t, \mu)
&= -2 \Gamma^i_{\cusp}(\alpha_s)\,\frac{1}{\mu^2}\cL_0\Bigl(\frac{t}{\mu^2}\Bigr) + \gamma_B^i(\alpha_s)\,\delta(t)
\,,\end{align}
and its solution is~\cite{Balzereit:1998yf, Neubert:2004dd, Fleming:2007xt, Ligeti:2008ac} [see \eq{Brun}]
\begin{align} \label{eq:Brun_full}
B_i(t,x,\mu) & =  \int\! \df t'\, B_i(t - t',x,\mu_0)\, U_B^i(t',\mu_0, \mu)
\,, \nn \\
U_B^i(t, \mu_0, \mu) &= \frac{e^{K_B^i -\gamma_E\, \eta_B^i}}{\Gamma(1 + \eta_B^i)}\,
\biggl[\frac{\eta_B^i}{\mu_0^2} \cL^{\eta_B^i} \Bigl( \frac{t}{\mu_0^2} \Bigr) + \delta(t) \biggr]
\,, \nn \\
K_B^i(\mu_0,\mu) &= 4 K^i_\Gamma(\mu_0,\mu) + K_{\gamma_B^i}(\mu_0,\mu)
\,, \qquad
\eta_B^i(\mu_0,\mu) = -2\eta^i_{\Gamma}(\mu_0,\mu)
\,.\end{align}

The beam thrust soft function is given in terms of $S_{\hemiin}$ by
\begin{equation} \label{eq:SB}
S_B(k^+, \mu)
= \!\int\!\df k_a^+ \df k_b^+\, S_\hemiin(k_a^+, k_b^+, \mu )\,\delta(k^+\! - k_a^+ - k_b^+)
\,.\end{equation}
Its RGE is easily obtained by integrating \eqs{tS_RGE}{gammas2},
\begin{align} \label{eq:SB_RGE}
\mu\frac{\df}{\df\mu} S_B(k^+, \mu)
&= \int\! \df \ell^+\, \gamma_S(k^+\! - \ell^+, \mu)\, S_B(\ell^+, \mu)
\,,\\\nn
\gamma_S(k^+, \mu)
&= 4\,\Gamma_\cusp^q(\alpha_s)\, \frac{1}{\mu} \cL_0\Big(\frac{k^+}{\mu}\Big) +
\gamma_S(\alpha_s)\, \delta(k^+)
\,, \qquad
\gamma_S(\alpha_s) = - 2\gamma_H^q(\alpha_s) - 2\gamma_B^q(\alpha_s)
\,,\end{align}
whose solution is completely analogous to \eq{Brun_full},
\begin{align} \label{eq:SBrun}
S_B(k^+,\mu) & =  \int\! \df \ell^+\, S(k^+\! - \ell^+,\mu_0)\, U_S(\ell^+,\mu_0,\mu)
\,, \nn \\
U_S(k^+, \mu_0, \mu) & = \frac{e^{K_S -\gamma_E\, \eta_S}}{\Gamma(1 + \eta_S)}\,
\biggl[\frac{\eta_S}{\mu_0} \cL^{\eta_S} \Big( \frac{k^+}{\mu_0} \Big) + \delta(k^+) \biggr]
\,, \nn \\
K_S(\mu_0,\mu) &= -4K_\Gamma^q(\mu_0,\mu) + K_{\gamma_S}(\mu_0,\mu)
\,, \qquad
\eta_S(\mu_0,\mu) = 4\eta_\Gamma^q(\mu_0,\mu)
\,.\end{align}

The functions $K_\Gamma^i(\mu_0, \mu)$, $\eta_\Gamma^i(\mu_0, \mu)$, $K_\gamma(\mu_0, \mu)$ in the above RGE solutions are defined as
\begin{align} \label{eq:Keta_def}
K_\Gamma^i(\mu_0, \mu)
& = \int_{\alpha_s(\mu_0)}^{\alpha_s(\mu)}\!\frac{\df\alpha_s}{\beta(\alpha_s)}\,
\Gamma_\cusp^i(\alpha_s) \int_{\alpha_s(\mu_0)}^{\alpha_s} \frac{\df \alpha_s'}{\beta(\alpha_s')}
\,,\qquad
\eta_\Gamma^i(\mu_0, \mu)
= \int_{\alpha_s(\mu_0)}^{\alpha_s(\mu)}\!\frac{\df\alpha_s}{\beta(\alpha_s)}\, \Gamma_\cusp^i(\alpha_s)
\,,\nn \\
K_\gamma(\mu_0, \mu)
& = \int_{\alpha_s(\mu_0)}^{\alpha_s(\mu)}\!\frac{\df\alpha_s}{\beta(\alpha_s)}\, \gamma(\alpha_s)
\,.\end{align}
Expanding the beta function and anomalous dimensions in powers of $\alpha_s$,
\begin{align}
\beta(\alpha_s) &=
- 2 \alpha_s \sum_{n=0}^\infty \beta_n\Bigl(\frac{\alpha_s}{4\pi}\Bigr)^{n+1}
\,, \quad
\Gamma^i_\cusp(\alpha_s) = \sum_{n=0}^\infty \Gamma^i_n \Bigl(\frac{\alpha_s}{4\pi}\Bigr)^{n+1}
\,, \quad
\gamma(\alpha_s) = \sum_{n=0}^\infty \gamma_n \Bigl(\frac{\alpha_s}{4\pi}\Bigr)^{n+1}
\,,\end{align}
their explicit expressions at NNLL are (suppressing the superscript $i$ on $K_\Ga^i$, $\eta_\Ga^i$ and $\Ga^i_n$),
\begin{align} \label{eq:Keta}
K_\Gamma(\mu_0, \mu) &= -\frac{\Gamma_0}{4\beta_0^2}\,
\biggl\{ \frac{4\pi}{\alpha_s(\mu_0)}\, \Bigl(1 - \frac{1}{r} - \ln r\Bigr)
   + \biggl(\frac{\Gamma_1 }{\Gamma_0 } - \frac{\beta_1}{\beta_0}\biggr) (1-r+\ln r)
   + \frac{\beta_1}{2\beta_0} \ln^2 r
\nn\\ & \hspace{10ex}
+ \frac{\alpha_s(\mu_0)}{4\pi}\, \biggl[
  \biggl(\frac{\beta_1^2}{\beta_0^2} - \frac{\beta_2}{\beta_0} \biggr) \Bigl(\frac{1 - r^2}{2} + \ln r\Bigr)
  + \biggl(\frac{\beta_1\Gamma_1 }{\beta_0 \Gamma_0 } - \frac{\beta_1^2}{\beta_0^2} \biggr) (1- r+ r\ln r)
\nn\\ & \hspace{10ex}
  - \biggl(\frac{\Gamma_2 }{\Gamma_0} - \frac{\beta_1\Gamma_1}{\beta_0\Gamma_0} \biggr) \frac{(1- r)^2}{2}
     \biggr] \biggr\}
\,, \nn\\
\eta_\Gamma(\mu_0, \mu) &=
 - \frac{\Gamma_0}{2\beta_0}\, \biggl[ \ln r
 + \frac{\alpha_s(\mu_0)}{4\pi}\, \biggl(\frac{\Gamma_1 }{\Gamma_0 }
 - \frac{\beta_1}{\beta_0}\biggr)(r-1)
\nn \\ & \hspace{10ex}
 + \frac{\alpha_s^2(\mu_0)}{16\pi^2} \biggl(
    \frac{\Gamma_2 }{\Gamma_0 } - \frac{\beta_1\Gamma_1 }{\beta_0 \Gamma_0 }
      + \frac{\beta_1^2}{\beta_0^2} -\frac{\beta_2}{\beta_0} \biggr) \frac{r^2-1}{2}
    \biggr]
\,, \nn\\
K_\gamma(\mu_0, \mu) &=
 - \frac{\gamma_0}{2\beta_0}\, \biggl[ \ln r
 + \frac{\alpha_s(\mu_0)}{4\pi}\, \biggl(\frac{\gamma_1 }{\gamma_0 }
 - \frac{\beta_1}{\beta_0}\biggr)(r-1) \biggr]
\,.\end{align}
Here, $r = \alpha_s(\mu)/\alpha_s(\mu_0)$ and the running coupling is given by the three-loop expression
\begin{equation} \label{eq:alphas}
\frac{1}{\alpha_s(\mu)} = \frac{X}{\alpha_s(\mu_0)}
  +\frac{\beta_1}{4\pi\beta_0}  \ln X
  + \frac{\alpha_s(\mu_0)}{16\pi^2} \biggr[
  \frac{\beta_2}{\beta_0} \Bigl(1-\frac{1}{X}\Bigr)
  + \frac{\beta_1^2}{\beta_0^2} \Bigl( \frac{\ln X}{X} +\frac{1}{X} -1\Bigr) \biggl]
\,,\end{equation}
where $X\equiv 1+\alpha_s(\mu_0)\beta_0 \ln(\mu/\mu_0)/(2\pi)$.
As discussed in \sec{results}, in our numerical analysis we use the full NNLL expressions in \eq{Keta}, but to be consistent with the NLO PDFs we only use the two-loop expression to obtain numerical values for $\alpha_s(\mu)$, hence dropping the $\beta_2$ and $\beta_1^2$ terms in \eq{alphas}. (The numerical difference between using the two-loop and three-loop $\alpha_s$ is numerically very small and well within our theory uncertainties.) Up to three loops, the coefficients of the beta function~\cite{Tarasov:1980au, Larin:1993tp} and cusp anomalous dimension~\cite{Korchemsky:1987wg, Moch:2004pa} in $\overline{\mathrm{MS}}$ are
\begin{align} \label{eq:Gacuspexp}
\beta_0 &= \frac{11}{3}\,C_A -\frac{4}{3}\,T_F\,n_f
\,,\nn\\
\beta_1 &= \frac{34}{3}\,C_A^2  - \Bigl(\frac{20}{3}\,C_A\, + 4 C_F\Bigr)\, T_F\,n_f
\,, \nn\\
\beta_2 &=
\frac{2857}{54}\,C_A^3 + \Bigl(C_F^2 - \frac{205}{18}\,C_F C_A
 - \frac{1415}{54}\,C_A^2 \Bigr)\, 2T_F\,n_f
 + \Bigl(\frac{11}{9}\, C_F + \frac{79}{54}\, C_A \Bigr)\, 4T_F^2\,n_f^2
\\[2ex]
\Gamma^q_0 &= 4C_F
\,,\nn\\
\Gamma^q_1 &= 4C_F \Bigl[\Bigl( \frac{67}{9} -\frac{\pi^2}{3} \Bigr)\,C_A  -
   \frac{20}{9}\,T_F\, n_f \Bigr]
\,,\nn\\
\Gamma^q_2 &= 4C_F \Bigl[
\Bigl(\frac{245}{6} -\frac{134 \pi^2}{27} + \frac{11 \pi ^4}{45}
  + \frac{22 \zeta_3}{3}\Bigr)C_A^2
  + \Bigl(- \frac{418}{27} + \frac{40 \pi^2}{27}  - \frac{56 \zeta_3}{3} \Bigr)C_A\, T_F\,n_f
\nn\\* & \hspace{8ex}
  + \Bigl(- \frac{55}{3} + 16 \zeta_3 \Bigr) C_F\, T_F\,n_f
  - \frac{16}{27}\,T_F^2\, n_f^2 \Bigr]
\,.\end{align}

The $\overline{\mathrm{MS}}$ anomalous dimension for the hard function can be obtained~\cite{Idilbi:2006dg, Becher:2006mr} from the IR divergences of the on-shell massless quark form factor which are known to three loops~\cite{Moch:2005id},
\begin{align} \label{eq:gaHexp}
\gamma^q_{H\,0} &= -6 C_F
\,,\nn\\
\gamma^q_{H\,1}
&= - C_F \Bigl[
  \Bigl(\frac{82}{9} - 52 \zeta_3\Bigr) C_A
+ (3 - 4 \pi^2 + 48 \zeta_3) C_F
+ \Bigl(\frac{65}{9} + \pi^2 \Bigr) \beta_0 \Bigr]
\,,\nn\\
\gamma^q_{H\,2}
&= -2C_F \Bigl[
  \Bigl(\frac{66167}{324} - \frac{686 \pi^2}{81} - \frac{302 \pi^4}{135} - \frac{782 \zeta_3}{9} + \frac{44\pi^2 \zeta_3}{9} + 136 \zeta_5\Bigr) C_A^2
\nn\\ & \qquad\hspace{6ex}
+ \Bigl(\frac{151}{4} - \frac{205 \pi^2}{9} - \frac{247 \pi^4}{135} + \frac{844 \zeta_3}{3} + \frac{8 \pi^2 \zeta_3}{3} + 120 \zeta_5\Bigr) C_F C_A
\nn\\ & \qquad\hspace{6ex}
+ \Bigl(\frac{29}{2} + 3 \pi^2 + \frac{8\pi^4}{5} + 68 \zeta_3 - \frac{16\pi^2 \zeta_3}{3} - 240 \zeta_5\Bigr) C_F^2
\nn\\ & \qquad\hspace{6ex}
+ \Bigl(-\frac{10781}{108} + \frac{446 \pi^2}{81} + \frac{449 \pi^4}{270} - \frac{1166 \zeta_3}{9} \Bigr) C_A \beta_0
\nn\\ & \qquad\hspace{6ex}
+ \Bigl(\frac{2953}{108} - \frac{13 \pi^2}{18} - \frac{7 \pi^4 }{27} + \frac{128 \zeta_3}{9}\Bigr)\beta_1
+ \Bigl(-\frac{2417}{324} + \frac{5 \pi^2}{6} + \frac{2 \zeta_3}{3}\Bigr)\beta_0^2
\Bigr]
\,.\end{align}
Denoting $\gamma_f^q$ the coefficient of the $\delta(1-z)$ in the quark PDF anomalous dimension, \eq{PDF_RGE} (which gives the non-cusp part of the anomalous dimension in the threshold limit $z\to 1$), the factorization theorem for DIS at threshold implies that $2\gamma_H^q(\alpha_s) + \gamma_J^q(\alpha_s) + \gamma_f^q(\alpha_s) = 0$, which was used in Ref.~\cite{Becher:2006mr} to obtain $\gamma_J^q$ at three loops from the known three-loop result for $\gamma_f^q$~\cite{Moch:2004pa}. As we showed in \subsec{B_RGE}, the anomalous dimension for the beam function equals that of the jet function, $\gamma_B^q = \gamma_J^q$, so the three-loop result for $\gamma_f^q$ together with \eq{gaHexp} yields the non-cusp three-loop anomalous dimension for the beam function,
\begin{align}
\gamma_{B\,0}^q &= 6 C_F
\,,\nn\\
\gamma_{B\,1}^q
&= C_F \Bigl[
  \Bigl(\frac{146}{9} - 80 \zeta_3\Bigr) C_A
+ (3 - 4 \pi^2 + 48 \zeta_3) C_F
+ \Bigl(\frac{121}{9} + \frac{2\pi^2}{3} \Bigr) \beta_0 \Bigr]
\,,\nn\\
\gamma_{B\,2}^q
&= 2 C_F \Bigl[
  \Bigl(\frac{52019}{162} - \frac{841\pi^2}{81} - \frac{82\pi^4}{27} -\frac{2056\zeta_3}{9} + \frac{88\pi^2 \zeta_3}{9} + 232 \zeta_5\Bigr)C_A^2
\nn\\ & \quad\hspace{6ex}
+ \Bigl(\frac{151}{4} - \frac{205\pi^2}{9} - \frac{247\pi^4}{135} + \frac{844\zeta_3}{3} + \frac{8\pi^2 \zeta_3}{3} + 120 \zeta_5\Bigr) C_A C_F
\nn\\ & \quad\hspace{6ex}
+ \Bigl(\frac{29}{2} + 3 \pi^2 + \frac{8\pi^4}{5} + 68 \zeta_3 - \frac{16\pi^2 \zeta_3}{3} - 240 \zeta_5\Bigr) C_F^2
\nn\\ & \quad\hspace{6ex}
+ \Bigl(-\frac{7739}{54} + \frac{325}{81} \pi^2 + \frac{617 \pi^4}{270} - \frac{1276\zeta_3}{9} \Bigr) C_A\beta_0
\nn\\ & \quad\hspace{6ex}
+ \Bigl(-\frac{3457}{324} + \frac{5\pi^2}{9} + \frac{16 \zeta_3}{3} \Bigr) \beta_0^2
+ \Bigl(\frac{1166}{27} - \frac{8 \pi^2}{9} - \frac{41 \pi^4}{135} + \frac{52 \zeta_3}{9}\Bigr) \beta_1
\Bigr]
\,.\end{align}
At NNLL, we only need the one- and two-loop coefficients of $\gamma_B^q$ and $\gamma_H^q$. The three-loop coefficients, $\gamma^q_{H\,2}$ and $\gamma^q_{B\,2}$, are given here for completeness. They are required for the resummation at N$^3$LL, where one would also need the four-loop beta function and cusp anomalous dimension, the latter of which is has not been calculated so far. In addition, the full N$^3$LL would also require the two-loop fixed-order corrections, which are known for the hard function, but not yet for the beam and soft functions.

\bibliographystyle{../jhep}
\bibliography{../pp.bib}

\providecommand{\href}[2]{#2}\begingroup\raggedright\begin{thebibliography}{10}

\bibitem{Stewart:2009yx}
I.~W. Stewart, F.~J. Tackmann, and W.~J. Waalewijn, {\it {Factorization at the
  LHC: From PDFs to Initial State Jets}},  {\em Phys. Rev. D} {\bf 81} (2010)
  094035, [\href{http://arXiv.org/abs/0910.0467}{{\tt arXiv:0910.0467}}].

\bibitem{Gribov:1972ri}
V.~N. Gribov and L.~N. Lipatov, {\it {Deep inelastic $ep$ scattering in
  perturbation theory}},  {\em Sov. J. Nucl. Phys.} {\bf 15} (1972) 438--450.

\bibitem{Georgi:1951sr}
H.~Georgi and H.~D. Politzer, {\it {Electroproduction scaling in an
  asymptotically free theory of strong interactions}},  {\em Phys. Rev. D} {\bf
  9} (1974) 416--420.

\bibitem{Gross:1974cs}
D.~J. Gross and F.~Wilczek, {\it {Asymptotically free gauge theories. II}},
  {\em Phys. Rev. D} {\bf 9} (1974) 980--993.

\bibitem{Altarelli:1977zs}
G.~Altarelli and G.~Parisi, {\it {Asymptotic Freedom in Parton Language}},
  {\em Nucl. Phys. B} {\bf 126} (1977) 298.

\bibitem{Dokshitzer:1977sg}
Y.~L. Dokshitzer, {\it {Calculation of the Structure Functions for Deep
  Inelastic Scattering and $e^+ e^-$ Annihilation by Perturbation Theory in
  Quantum Chromodynamics. (In Russian)}},  {\em Sov. Phys. JETP} {\bf 46}
  (1977) 641--653.

\bibitem{Ellis:2009wj}
S.~D. Ellis, A.~Hornig, C.~Lee, C.~K. Vermilion, and J.~R. Walsh, {\it
  {Consistent Factorization of Jet Observables in Exclusive Multijet
  Cross-Sections}},  {\em Phys. Lett. B} {\bf 689} (2010) 82--89,
  [\href{http://arXiv.org/abs/0912.0262}{{\tt arXiv:0912.0262}}].

\bibitem{Jouttenus:2009ns}
T.~T. Jouttenus, {\it {Jet Function with a Jet Algorithm in SCET}},  {\em Phys.
  Rev. D} {\bf 81} (2010) 094017, [\href{http://arXiv.org/abs/0912.5509}{{\tt
  arXiv:0912.5509}}].

\bibitem{Ellis:2010rwa}
S.~D. Ellis, C.~K. Vermilion, J.~R. Walsh, A.~Hornig, and C.~Lee, {\it {Jet
  Shapes and Jet Algorithms in SCET}},  {\em JHEP} {\bf 11} (2010) 101,
  [\href{http://arXiv.org/abs/1001.0014}{{\tt arXiv:1001.0014}}].

\bibitem{Bauer:2000ew}
C.~W. Bauer, S.~Fleming, and M.~E. Luke, {\it {Summing Sudakov logarithms in $B
  \to X_s\gamma$ in effective field theory}},  {\em Phys. Rev. D} {\bf 63}
  (2000) 014006, [\href{http://arXiv.org/abs/hep-ph/0005275}{{\tt
  hep-ph/0005275}}].

\bibitem{Bauer:2000yr}
C.~W. Bauer, S.~Fleming, D.~Pirjol, and I.~W. Stewart, {\it An effective field
  theory for collinear and soft gluons: Heavy to light decays},  {\em Phys.
  Rev. D} {\bf 63} (2001) 114020,
  [\href{http://arXiv.org/abs/hep-ph/0011336}{{\tt hep-ph/0011336}}].

\bibitem{Bauer:2001ct}
C.~W. Bauer and I.~W. Stewart, {\it Invariant operators in collinear effective
  theory},  {\em Phys. Lett. B} {\bf 516} (2001) 134--142,
  [\href{http://arXiv.org/abs/hep-ph/0107001}{{\tt hep-ph/0107001}}].

\bibitem{Bauer:2001yt}
C.~W. Bauer, D.~Pirjol, and I.~W. Stewart, {\it Soft-collinear factorization in
  effective field theory},  {\em Phys. Rev. D} {\bf 65} (2002) 054022,
  [\href{http://arXiv.org/abs/hep-ph/0109045}{{\tt hep-ph/0109045}}].

\bibitem{Collins:1988ig}
J.~C. Collins, D.~E. Soper, and G.~Sterman, {\it Soft gluons and
  factorization},  {\em Nucl. Phys. B} {\bf 308} (1988) 833.

\bibitem{Aybat:2008ct}
S.~M. Aybat and G.~Sterman, {\it {Soft-Gluon Cancellation, Phases and
  Factorization with Initial-State Partons}},  {\em Phys. Lett. B} {\bf 671}
  (2009) 46--50, [\href{http://arXiv.org/abs/0811.0246}{{\tt
  arXiv:0811.0246}}].

\bibitem{Manohar:2006nz}
A.~V. Manohar and I.~W. Stewart, {\it {The zero-bin and mode factorization in
  quantum field theory}},  {\em Phys. Rev. D} {\bf 76} (2007) 074002,
  [\href{http://arXiv.org/abs/hep-ph/0605001}{{\tt hep-ph/0605001}}].

\bibitem{Collins:1999dz}
J.~C. Collins and F.~Hautmann, {\it {Infrared divergences and non-lightlike
  eikonal lines in Sudakov processes}},  {\em Phys. Lett. B} {\bf 472} (2000)
  129--134, [\href{http://arXiv.org/abs/hep-ph/9908467}{{\tt hep-ph/9908467}}].

\bibitem{Lee:2006nr}
C.~Lee and G.~Sterman, {\it Momentum flow correlations from event shapes:
  Factorized soft gluons and soft-collinear effective theory},  {\em Phys. Rev.
  D} {\bf 75} (2007) 014022, [\href{http://arXiv.org/abs/hep-ph/0611061}{{\tt
  hep-ph/0611061}}].

\bibitem{Idilbi:2007ff}
A.~Idilbi and T.~Mehen, {\it {On the equivalence of soft and zero-bin
  subtractions}},  {\em Phys. Rev. D} {\bf 75} (2007) 114017,
  [\href{http://arXiv.org/abs/hep-ph/0702022}{{\tt hep-ph/0702022}}].

\bibitem{Chay:2002vy}
J.~Chay and C.~Kim, {\it {Collinear effective theory at subleading order and
  its application to heavy-light currents}},  {\em Phys. Rev. D} {\bf 65}
  (2002) 114016, [\href{http://arXiv.org/abs/hep-ph/0201197}{{\tt
  hep-ph/0201197}}].

\bibitem{Manohar:2002fd}
A.~V. Manohar, T.~Mehen, D.~Pirjol, and I.~W. Stewart, {\it Reparameterization
  invariance for collinear operators},  {\em Phys. Lett. B} {\bf 539} (2002)
  59--66, [\href{http://arXiv.org/abs/hep-ph/0204229}{{\tt hep-ph/0204229}}].

\bibitem{Bauer:2002nz}
C.~W. Bauer, S.~Fleming, D.~Pirjol, I.~Z. Rothstein, and I.~W. Stewart, {\it
  Hard scattering factorization from effective field theory},  {\em Phys. Rev.
  D} {\bf 66} (2002) 014017, [\href{http://arXiv.org/abs/hep-ph/0202088}{{\tt
  hep-ph/0202088}}].

\bibitem{Collins:1981uw}
J.~C. Collins and D.~E. Soper, {\it {Parton Distribution and Decay Functions}},
   {\em Nucl. Phys. B} {\bf 194} (1982) 445.

\bibitem{Balzereit:1998yf}
C.~Balzereit, T.~Mannel, and W.~Kilian, {\it {Evolution of the light-cone
  distribution function for a heavy quark}},  {\em Phys. Rev. D} {\bf 58}
  (1998) 114029, [\href{http://arXiv.org/abs/hep-ph/9805297}{{\tt
  hep-ph/9805297}}].

\bibitem{Neubert:2004dd}
M.~Neubert, {\it {Renormalization-group improved calculation of the $B \to
  X_s\gamma$ branching ratio}},  {\em Eur. Phys. J. C} {\bf 40} (2005)
  165--186, [\href{http://arXiv.org/abs/hep-ph/0408179}{{\tt hep-ph/0408179}}].

\bibitem{Fleming:2007xt}
S.~Fleming, A.~H. Hoang, S.~Mantry, and I.~W. Stewart, {\it {Top Jets in the
  Peak Region: Factorization Analysis with NLL Resummation}},  {\em Phys. Rev.
  D} {\bf 77} (2008) 114003, [\href{http://arXiv.org/abs/0711.2079}{{\tt
  arXiv:0711.2079}}].

\bibitem{Ligeti:2008ac}
Z.~Ligeti, I.~W. Stewart, and F.~J. Tackmann, {\it {Treating the b quark
  distribution function with reliable uncertainties}},  {\em Phys. Rev. D} {\bf
  78} (2008) 114014, [\href{http://arXiv.org/abs/0807.1926}{{\tt
  arXiv:0807.1926}}].

\bibitem{Fleming:2003gt}
S.~Fleming, A.~K. Leibovich, and T.~Mehen, {\it {Resumming the color-octet
  contribution to $e^+e^- \to J/\psi + X$}},  {\em Phys. Rev. D} {\bf 68}
  (2003) 094011, [\href{http://arXiv.org/abs/hep-ph/0306139}{{\tt
  hep-ph/0306139}}].

\bibitem{Moch:2004pa}
S.~Moch, J.~A.~M. Vermaseren, and A.~Vogt, {\it {The three-loop splitting
  functions in QCD: The non-singlet case}},  {\em Nucl. Phys. B} {\bf 688}
  (2004) 101--134, [\href{http://arXiv.org/abs/hep-ph/0403192}{{\tt
  hep-ph/0403192}}].

\bibitem{Moch:2005id}
S.~Moch, J.~A.~M. Vermaseren, and A.~Vogt, {\it {The quark form factor at
  higher orders}},  {\em JHEP} {\bf 08} (2005) 049,
  [\href{http://arXiv.org/abs/hep-ph/0507039}{{\tt hep-ph/0507039}}].

\bibitem{Fleming:2006cd}
S.~Fleming, A.~K. Leibovich, and T.~Mehen, {\it {Resummation of Large Endpoint
  Corrections to Color-Octet $J/\psi$ Photoproduction}},  {\em Phys. Rev. D}
  {\bf 74} (2006) 114004, [\href{http://arXiv.org/abs/hep-ph/0607121}{{\tt
  hep-ph/0607121}}].

\bibitem{Martin:2009bu}
A.~D. Martin, W.~J. Stirling, R.~S. Thorne, and G.~Watt, {\it {Uncertainties on
  $\alpha_s$ in global PDF analyses and implications for predicted hadronic
  cross sections}},  {\em Eur. Phys. J. C} {\bf 64} (2009) 653--680,
  [\href{http://arXiv.org/abs/0905.3531}{{\tt arXiv:0905.3531}}].

\bibitem{Appell:1988ie}
D.~Appell, G.~Sterman, and P.~B. Mackenzie, {\it {Soft Gluons and the
  Normalization of the Drell-Yan Cross Section}},  {\em Nucl. Phys. B} {\bf
  309} (1988) 259.

\bibitem{Catani:1998tm}
S.~Catani, M.~L. Mangano, and P.~Nason, {\it {Sudakov resummation for prompt
  photon production in hadron collisions}},  {\em JHEP} {\bf 07} (1998) 024,
  [\href{http://arXiv.org/abs/hep-ph/9806484}{{\tt hep-ph/9806484}}].

\bibitem{Becher:2007ty}
T.~Becher, M.~Neubert, and G.~Xu, {\it {Dynamical Threshold Enhancement and
  Resummation in Drell-Yan Production}},  {\em JHEP} {\bf 07} (2008) 030,
  [\href{http://arXiv.org/abs/0710.0680}{{\tt arXiv:0710.0680}}].

\bibitem{Stewart:2010pd}
I.~W. Stewart, F.~J. Tackmann, and W.~J. Waalewijn, {\it {The Beam Thrust Cross
  Section for Drell-Yan at NNLL Order}},
  \href{http://arXiv.org/abs/1005.4060}{{\tt arXiv:1005.4060}}.

\bibitem{Berger:2010xi}
C.~F. Berger, C.~Marcantonini, I.~W. Stewart, F.~J. Tackmann, and W.~J.
  Waalewijn, {\it {Higgs Production with a Central Jet Veto at NNLL+NNLO}},
  \href{http://arXiv.org/abs/1012.4480}{{\tt arXiv:1012.4480}}.

\bibitem{Mantry:2009qz}
S.~Mantry and F.~Petriello, {\it {Factorization and Resummation of Higgs Boson
  Differential Distributions in Soft-Collinear Effective Theory}},  {\em Phys.
  Rev. D} {\bf 81} (2010) 093007, [\href{http://arXiv.org/abs/0911.4135}{{\tt
  arXiv:0911.4135}}].

\bibitem{Brandt:1981kf}
R.~A. Brandt, F.~Neri, and M.~aki Sato, {\it {Renormalization of Loop Functions
  for All Loops}},  {\em Phys. Rev. D} {\bf 24} (1981) 879.

\bibitem{Korchemsky:1987wg}
G.~P. Korchemsky and A.~V. Radyushkin, {\it {Renormalization of the Wilson
  Loops Beyond the Leading Order}},  {\em Nucl. Phys. B} {\bf 283} (1987)
  342--364.

\bibitem{Korchemskaya:1992je}
I.~A. Korchemskaya and G.~P. Korchemsky, {\it {On lightlike Wilson loops}},
  {\em Phys. Lett. B} {\bf 287} (1992) 169--175.

\bibitem{Korchemsky:1992xv}
G.~P. Korchemsky and G.~Marchesini, {\it {Structure function for large $x$ and
  renormalization of Wilson loop}},  {\em Nucl. Phys. B} {\bf 406} (1993)
  225--258.

\bibitem{Schwartz:2007ib}
M.~D. Schwartz, {\it {Resummation and NLO Matching of Event Shapes with
  Effective Field Theory}},  {\em Phys. Rev. D} {\bf 77} (2008) 014026,
  [\href{http://arXiv.org/abs/0709.2709}{{\tt arXiv:0709.2709}}].

\bibitem{Manohar:2003vb}
A.~V. Manohar, {\it {Deep inelastic scattering as $x \to 1$ using
  soft-collinear effective theory}},  {\em Phys. Rev. D} {\bf 68} (2003)
  114019, [\href{http://arXiv.org/abs/hep-ph/0309176}{{\tt hep-ph/0309176}}].

\bibitem{Bauer:2003di}
C.~W. Bauer, C.~Lee, A.~V. Manohar, and M.~B. Wise, {\it {Enhanced
  nonperturbative effects in Z decays to hadrons}},  {\em Phys. Rev. D} {\bf
  70} (2004) 034014, [\href{http://arXiv.org/abs/hep-ph/0309278}{{\tt
  hep-ph/0309278}}].

\bibitem{Tarasov:1980au}
O.~V. Tarasov, A.~A. Vladimirov, and A.~Y. Zharkov, {\it {The Gell-Mann-Low
  Function of QCD in the Three Loop Approximation}},  {\em Phys. Lett. B} {\bf
  93} (1980) 429--432.

\bibitem{Larin:1993tp}
S.~A. Larin and J.~A.~M. Vermaseren, {\it {The three-loop QCD $\beta$ function
  and anomalous dimensions}},  {\em Phys. Lett. B} {\bf 303} (1993) 334--336,
  [\href{http://arXiv.org/abs/hep-ph/9302208}{{\tt hep-ph/9302208}}].

\bibitem{Idilbi:2006dg}
A.~Idilbi, X.~dong Ji, and F.~Yuan, {\it {Resummation of Threshold Logarithms
  in Effective Field Theory For DIS, Drell-Yan and Higgs Production}},  {\em
  Nucl. Phys. B} {\bf 753} (2006) 42--68,
  [\href{http://arXiv.org/abs/hep-ph/0605068}{{\tt hep-ph/0605068}}].

\bibitem{Becher:2006mr}
T.~Becher, M.~Neubert, and B.~D. Pecjak, {\it {Factorization and momentum-space
  resummation in deep-inelastic scattering}},  {\em JHEP} {\bf 01} (2007) 076,
  [\href{http://arXiv.org/abs/hep-ph/0607228}{{\tt hep-ph/0607228}}].

\end{thebibliography}\endgroup

\end{document}